\documentclass[twocolumn]{aastex7}

\usepackage{mhchem}
\usepackage{subcaption}

\defcitealias{Teske2025AJ....169..249T}{Teske, Batalha et al.\ 2025}
\defcitealias{Teske2025AJ_paren}{Teske, Batalha et al.\ (2025)}

\newcommand{\ltt}{LTT 1445A~b}
\newcommand{\jedi}{\texttt{ExoTiC-JEDI}}
\newcommand{\eureka}{\texttt{\texttt{Eureka!}}}
\graphicspath{{figures/}}

\begin{document}

\title{JWST COMPASS Program: The 3--5$\mu$m  transmission spectrum of LTT 1445 A b}

% tier 1 direct data/models/text for paper 

\author[0000-0003-1240-6844]{Natasha E. Batalha}
\affiliation{NASA Ames Research Center, Moffett Field, CA 94035, USA}
\email{natasha.e.batalha@nasa.gov}

\author[0000-0003-0354-0187]{Nicole Wallack}
\affiliation{Earth and Planets Laboratory, Carnegie Institution for Science, 5241 Broad Branch Road, NW, Washington, DC 20015, USA}
\email{nwallack@carnegiescience.edu}

\author{Tyler Gordon}
\affiliation{Department of Astronomy and Astrophysics, University of California, Santa Cruz, CA 95064, USA}
\email{tygordon@ucsc.edu}

\author[0000-0002-0413-3308]{Nicholas F. Wogan}
\affiliation{SETI Institute, Mountain View, CA 94043}
\affiliation{NASA Ames Research Center, Moffett Field, CA 94035}
\email{nicholas.f.wogan@nasa.gov}

\author[0000-0002-9030-0132]{Katherine A. Bennett}
\affiliation{Department of Earth \& Planetary Sciences, Johns Hopkins University, Baltimore, MD 21218, USA}
\email{kbenne50@jhu.edu}

% tier 2 LOTS of work pre paper to get this target actually successfully observed (contributed to TA measurements, APT, TTRBs, etc) 

\author[0000-0002-4489-3168]{Jea Adams Redai} 
\affiliation{Center for Astrophysics ${\rm \mid}$ Harvard {\rm \&} Smithsonian, 60 Garden St, Cambridge, MA 02138, USA}
\email{jea.adams@cfa.harvard.edu}

\author[0000-0003-3204-8183]{Mercedes L\'opez-Morales} 
\affiliation{Center for Astrophysics ${\rm \mid}$ Harvard {\rm \&} Smithsonian, 60 Garden St, Cambridge, MA 02138, USA}
\email{mlopez-morales@stsci.edu}

\author[0009-0008-2801-5040]{Johanna Teske} 
\affiliation{Earth and Planets Laboratory, Carnegie Institution for Science, 5241 Broad Branch Road, NW, Washington, DC 20015, USA}
\affiliation{The Observatories of the Carnegie Institution for Science, 813 Santa Barbara St., Pasadena, CA 91101, USA}
\email{jteske@carnegiescience.edu}

\author[0000-0003-3305-6281]{Jeff Valenti} 
\affiliation{Space Telescope Science Institute, 3700 San Martin Drive, Baltimore, MD 21218, USA}
\email{valenti@stsci.edu}
%Team members that read the paper and contributed to comments 

\author[0000-0003-4157-832X]{Munazza K. Alam}
\affiliation{Space Telescope Science Institute, 3700 San Martin Drive, Baltimore, MD 21218, USA}
\email{malam@stsci.edu}

\author[0000-0001-8703-7751]{Lili Alderson} 
\affiliation{Department of Astronomy, Cornell University, 122 Sciences Drive, Ithaca, NY 14853, USA}
\email{lili.alderson@cornell.edu}

\author[0000-0002-8949-5956]{Artyom Aguichine}
\affiliation{Instituto de Astronomía, Universidad Nacional Autónoma de México, Apartado Postal 106, CP 22800 Ensenada, Baja California, México}
\email{aaguichi@ucsc.edu}

\author[0000-0002-7030-9519]{Natalie M. Batalha}
\affiliation{Department of Astronomy and Astrophysics, University of California, Santa Cruz, CA 95064, USA}
\email{nabatalh@ucsc.edu}

\author[0009-0003-2576-9422]{Anna Gagnebin}
\affiliation{Department of Astronomy and Astrophysics, University of California, Santa Cruz, CA 95064, USA}
\email{akgagneb@ucsc.edu}

\author[0000-0002-8518-9601]{Peter Gao} 
\affiliation{Earth and Planets Laboratory, Carnegie Institution for Science, 5241 Broad Branch Road, NW, Washington, DC 20015, USA}
\email{pgao@carnegiescience.edu}

\author[0000-0002-7500-7173]{Annabella Meech}
\affiliation{Space Telescope Science Institute, 3700 San Martin Drive, Baltimore, MD 21218, USA}
\email{annabella.meech@cfa.harvard.edu}

\author[0000-0002-6721-3284]{Sarah E. Moran}
\altaffiliation{NHFP Sagan Fellow}
\affiliation{Space Telescope Science Institute, 3700 San Martin Drive, Baltimore, MD 21218, USA}
\email{semoran@umd.edu}

\author[0000-0003-4328-3867]{Hannah R. Wakeford} 
\affiliation{School of Physics, University of Bristol, HH Wills Physics Laboratory, Tyndall Avenue, Bristol BS8 1TL, UK}
\email{hannah.wakeford@bristol.ac.uk}

\author[0000-0003-2862-6278]{Angie Wolfgang}
\affiliation{Eureka Scientific Inc., 2452 Delmer Street Suite 100, Oakland, CA 94602-3017}
\email{dawolfgang@gmail.com}

%opt out for LTT 
%\author[0000-0003-3623-7280]{Nicholas Scarsdale} 
%\affiliation{Department of Astronomy and Astrophysics, University of California, Santa Cruz, CA 95064, USA}

\begin{abstract}

The search for an atmosphere on the closest rocky M dwarf planet, \ltt, has been the subject of intense investigation from both the ground and space. Here, we present the first JWST transmission spectrum of \ltt ~using a single visit spanning 3-5~$\mu$m using NIRSpec/G395H. We conduct two independent reductions of the data using both the \texttt{Eureka!} and \texttt{ExoTiC-JEDI} pipelines. Overall, we measure the NRS1 transit depths to a median precision of $\sim23$~ppm in 41 spectroscopic channels with uniform widths of 30 pixels ($\sim$ 0.02 $\mu$m), and the NRS2 transit depths to $\sim36$~ppm precision in 65 spectroscopic channels, also with uniform widths of 30 pixels. We rule out any statistically significant spectral features at this precision and place limits on atmospheric metallicity using a grid of chemical equilibrium models with grey opaque clouds. Using NIRSpec/G395H alone, we can place limits on the atmospheric metallicity of $\gtrsim350~\times$ Solar when the opaque pressure level is greater than 0.01~bars. We also conduct a combined analysis of JWST/NIRSpec and HST/WFC3 transmission data and find our atmospheric limits can be extended $\gtrsim500~\times$ Solar when considering both datasets. Future analyses both in transit and emission will uncover whether there are detectable atmospheric features. 

\end{abstract}

% Reference Management 
% Public ADS LIbrary Link 
% https://ui.adsabs.harvard.edu/public-libraries/aHGglSvkQl-tJ7378bbvzg

%Classical Novae (251) --- Ultraviolet astronomy(1736) --- History of astronomy(1868) --- Interdisciplinary astronomy(804)}

%%%%%%%%%%%%%%%%%%%%%%%%%%%%%%%%%%%%%%%%%%%%%%%%%%%%%%%%%%%
\section{Introduction} 
\label{sec:intro}
%%%%%%%%%%%%%%%%%%%%%%%%%%%%%%%%%%%%%%%%%%%%%%%%%%%%%%%%%%%
%intro 
Rocky planets, defined here as  $\lesssim$ 1.8~R$_{\oplus}$, which orbit M dwarfs have been a key area of focus for atmospheric reconnaissance with both the Hubble Space Telescope (HST) and the James Webb Space Telescope (JWST). M dwarfs are approximately three times more common in our stellar neighborhood than solar-type stars \citep{Henry2024}, and their smaller size allows us to push atmospheric observations towards smaller and cooler rocky worlds, some of which could reside in the habitable zone of their host star \citep{Kopparapu2013ApJ...767L...8K}. Both JWST and HST have devoted several hundred hours of observational time targeting these worlds. However, thus far it has been difficult to determine whether or not rocky planets orbiting M dwarfs  have observable atmospheres.

With HST, ten planets across four M dwarf systems have been targeted with WFC3/G141 (1.1-1.65~$\mu$m) transmission spectroscopy: TRAPPIST-1 b, c, d, e, f, g, and h, \citep{dewit2016combined, 2018AJ....156..252M,2019AJ....157...11W, zhang2018near, garcia2022hst,Gressier2022T1h} L98-59~b \citep{damiano2022transmission,Barclay2025L9859c}, GJ 1132~b \citep{libby2022featureless}, and most recently, LTT 1445A~b \citep{Bennett2025AJ....169..111B}. Ultimately, these observations demonstrated that these planets do not carry large \ce{H2}-envelopes.  At the same time, none yielded robust atmospheric constraints because the data were void of any observable spectral features. 

JWST has also tried to tackle the question of the presence of atmospheres around M dwarf hosted rocky planets. However, to date, atmospheric signals from transmission spectroscopy have remained elusive. Some transit spectra have yielded tentative signs of an atmosphere, for example, hints of gaseous sulfur species have been suggested on both L 98-59~b \citep{BelloArufe2025ApJ...980L..26B} and L 98-59~d \citep{Gressier2024ApJ...975L..10G}, GJ 486 b \citep{2023ApJ...948L..11M} and TOI-270 b \citep{2025AJ....170..226C} have hints of water feature that could either be of planetary or stellar contamination origin. However, the majority of rocky planets around M dwarfs show no strong atmospheric spectral signals, including in transmission spectroscopy across ten planets: 
TRAPPIST-1~c \citep{2025ApJ...979L...5R};
TRAPPIST-1~e \citep{2025ApJ...990L..52E,2025ApJ...990L..53G, 2026AJ....171..105A};
L 98-58~c \citep{Scarsdale2024AJ....168..276S};  
GJ 341~b \citep{Kirk2024AJ....167...90K}; 
LHS 475~b \citep{Lustig2023NatAs...7.1317L}; 
GJ 1132~b \citep{May2023ApJ...959L...9M, Bennett2025AJ....170..205B}; 
GJ 357~b \citep{Taylor2025MNRAS.540.3677T,Adams2025AJ....170..219A}; 
L 168-9~b \citep{Alam2025AJ....169...15A}; 
TOI-1685~b \citep{2025AJ....170...49L,Fisher2026MNRAS.545f2187F}; 
TOI-776~b \citep{Alderson2025AJ....169..142A}. 

%what emission spectra has shown
Transmission spectroscopy has not been the only avenue for exploration with JWST. Other small planets that orbit M stars have also been explored with emission spectroscopy and/or photometry. Though not an exhaustive list, some have been: GJ 367~b \citep{Zhang2024ApJ...961L..44Z}, TRAPPIST-1 b and c \citep{Zieba2023Natur.620..746Z,Greene2023Natur.618...39G,2025arXiv250902128G}, LHS 1140 b and c \citep{Fortune2025A&A...701A..25F,Damiano2024ApJ...968L..22D}, and GJ 486~b \citep{Weiner2024ApJ...975L..22W}. So far these investigations involving rocky planets around M dwarfs show emission spectra that are broadly consistent with a bare rock. 

%why more observations, why ltt
The LTT 1445 stellar system, a hierarchical triple of mid-to-late M dwarfs (T$_\mathrm{eff}$=3562~K for LTT 1445 A), is of particular interest because it is only 6.9 pc away, making it the nearest known M dwarf system with transiting rocky planets. \ltt\,, first reported in  \citet{Winters2019}, has a radius of 1.304$\pm$0.063 R$_\oplus$, a mass of 2.87$\pm$0.25 M$_\oplus$, and an equilibrium temperature T(A$_b$=0)=424~K. A few years later, its transiting sibling LTT 1445A~c, (R$_\oplus$ (min) = 1.15, M$_\oplus$=1.54$\pm$0.20, T(A$_b$=0)=508~K) was discovered by \citet{Winters2022AJ....163..168W}.
Planet parameters for this system have also been refined in \citet{Pass2023AJ....166..171P}. The combination of \ltt's escape velocity, bolometric stellar flux, and host star luminosity makes it amenable to retaining an atmosphere based on  the purported ``cosmic shoreline'' \citep{Zahnle2017ApJ...843..122Z,BertaThompson2025arXiv250702136B}, the theoretical line delineating planets that could versus could not have atmospheres. 

\ltt\ was previously observed via low-resolution 0.6--1 $\mu$m transmission spectroscopy with Magellan II/LDSS3C from the ground \citep{Diamond2023AJ....165..169D} and from space with HST \citep{Bennett2025AJ....169..111B}. These observations confidently rule out $\le$100$\times$solar metallicity atmospheres. \ltt\ has also been studied in emission with MIRI/LRS spectroscopy (5--12$\mu$m); \citet{Wachiraphan2025AJ....169..311W} inferred from those observations that although the planet likely lacks a thick \ce{CO2} atmosphere, it could potentially have a thin atmosphere similar to Mars, Titan or Earth. Secondary eclipse photometry at 15$\mu$m using MIRI will be taken via the Rocky Worlds Directors Discretionary Program, which deemed \ltt\ a high community interest target. Taken as a whole, \ltt\, is one of the most well-studied rocky planets orbiting an M-dwarf  to date and offers us a unique opportunity to address overarching questions surrounding the ability of rocky M-dwarf planets to retain atmospheres. However, it is clear from these previous investigations that to fully understand the nature of rocky M dwarf worlds, multiple follow-up observations are needed to refine atmospheric spectral precision, expand wavelength coverage, and assess multiple planetary viewing geometries across both transmission and emission. This is especially true for targets that are of high community interest, such as \ltt. 

To that end, here we present the first transmission spectrum of \ltt\ taken with JWST via the COMPASS (Compositions of Mini-Planet Atmospheres for Statistical Study) Program  (PID \# 2512, \citealt{2021jwst.prop.2512B}). Broadly speaking, COMPASS has an overarching goal of obtaining NIRSpec/G395H transmission spectra of eleven 1-3 R$_\oplus$ planets. The main driver of the program is to kickstart observations of small planets and build a link between atmospheric characterization and planetary demographics of super-Earths and sub-Neptunes. We have taken both an individual, ``deep-dive'' approach for each planet (\citealt{Wallack2024AJ....168...77W, Alderson2024AJ....167..216A, Scarsdale2024AJ....168..276S, Alam2025AJ....169...15A, Adams2025AJ....170..219A, Alderson2025AJ....169..142A}, \citetalias{Teske2025AJ....169..249T}, \citealt{Meech2026,Wallack2026}, Gagnebin et al. submitted), as well as a broader population-level look at our sample (for the first seven, \citealt{Gordon2026}) and how it compares to other JWST-observed small planets. The investigation published here on \ltt\ will round out the sample as one of the last of the eleven planets to be observed by the COMPASS program. 
%. COMPASS has also just released the first preliminary population-level look at our sample, as well as how it compares to other JWST-observed small planets %at the entire population 
%\citep{Gordon2025arXiv251118196G}. The investigation in this paper, however, only focuses on the last of the eleven planets, \ltt.  

In what follows, we describe our observations and data reduction procedures, including light curve fitting, in \S \ref{sec:data_reduction}. We present the transmission spectrum of LTT 1445A~b in \S \ref{sec:spectrum} including basic feature detection analysis and atmospheric modeling. Lastly, we interpret our results in the context of other transmission and emission observations in \S \ref{sec:discussion}. We also take a forward look at what the future of \ltt\ might be after several more JWST transmission observation campaigns. We summarize our results in \S \ref{sec:summary}.

%%%%%%%%%%%%%%%%%%%%%%%%%%%%%%%%%%%%%%%%%%%%%%%%%%%%%%%%%%%
\section{Observations \& Data Reduction} 
\label{sec:data_reduction}
%%%%%%%%%%%%%%%%%%%%%%%%%%%%%%%%%%%%%%%%%%%%%%%%%%%%%%%%%%%

\subsection{NIRSpec/G395H}
We observed a single transit of \ltt\ with JWST/NIRSpec using the high-resolution ($R\sim$2700) G395H grating, which provides spectroscopy between 2.87--5.14\,$\mu$m across the NRS1 and NRS2 detectors (with a $\sim$0.1\,$\mu$m detector gap between 3.72--3.82\,$\mu$m). The observations were taken with the NIRSpec Bright Object Time Series (BOTS) mode using the SUB2048 subarray, the F290LP filter, the S1600A1 slit, and the NRSRAPID readout pattern. The 4.5\,hr exposure (beginning at Sep 1, 2025 22:37:46 UT and ending at Sep 2, 2025 04:43:42 UT) consisted of 4468 integrations with 3 groups per integration, and was designed to be centered on the transit event (0.5042~hrs) with sufficient out-of-transit baseline.

\begin{table*}[]
    \centering
    \caption{Fitted parameters from the \texttt{\texttt{Eureka!}} and \texttt{ExoTiC-JEDI} white light curves.}
    \begin{tabular}{c c c c c c c}
    \hline
    \hline
         Pipeline & Detector & $R_p/R_*$ & $T_0$\,[MJD] & $i$\,[$^\circ$] & $a/R_*$  \\
         \hline

        \eureka   & NRS1 & 0.044916$\pm$0.000065 &60920.11176$\pm$0.000017 &89.62$\pm$ 0.14 &30.72$\pm$0.42\\
         &           NRS2 &0.044476$\pm$0.000067&60920.11170$\pm$0.000020  &89.62$\pm$ 0.16 &30.77$\pm$0.49\\

        \hline
        \jedi & NRS1 & 0.044229$\pm$0.000088 & 60920.11182$\pm$0.000018 & 89.57$\pm$0.18 & 30.48$\pm$0.53 \\
        &  NRS2 & 0.044151$\pm$0.000081 & 60920.11172$\pm$0.000020 & 89.44$\pm$0.13 & 29.98$\pm$0.55 \\

    \end{tabular}

    \label{tab:bestfit}
\end{table*}    

%%FIGURE
\begin{figure*}
    \centering
    \includegraphics[width=\linewidth]{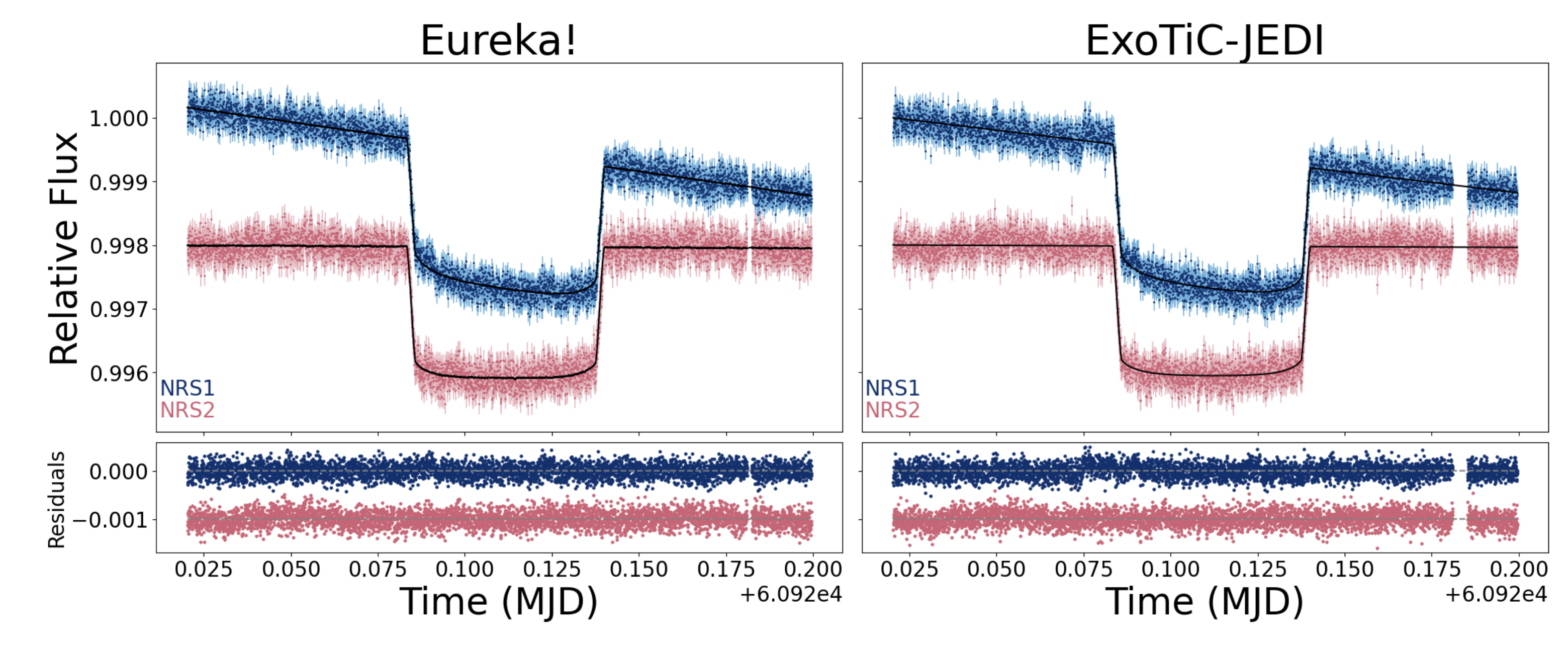}
    \caption{White light curves and associated residuals for both reduction methods with NRS1 in blue and NRS2 in pink. The \eureka\ data reduction is on the left and \jedi\ data reduction is on the right. The points missing in the post-transit baseline were removed automatically using the outlier rejection for \eureka\ and manually for \jedi\ due to a high gain antenna move.}
    \label{fig:wlc}
\end{figure*}

%%FIGURE
\begin{figure*}
    \centering
    \includegraphics[width=\linewidth]{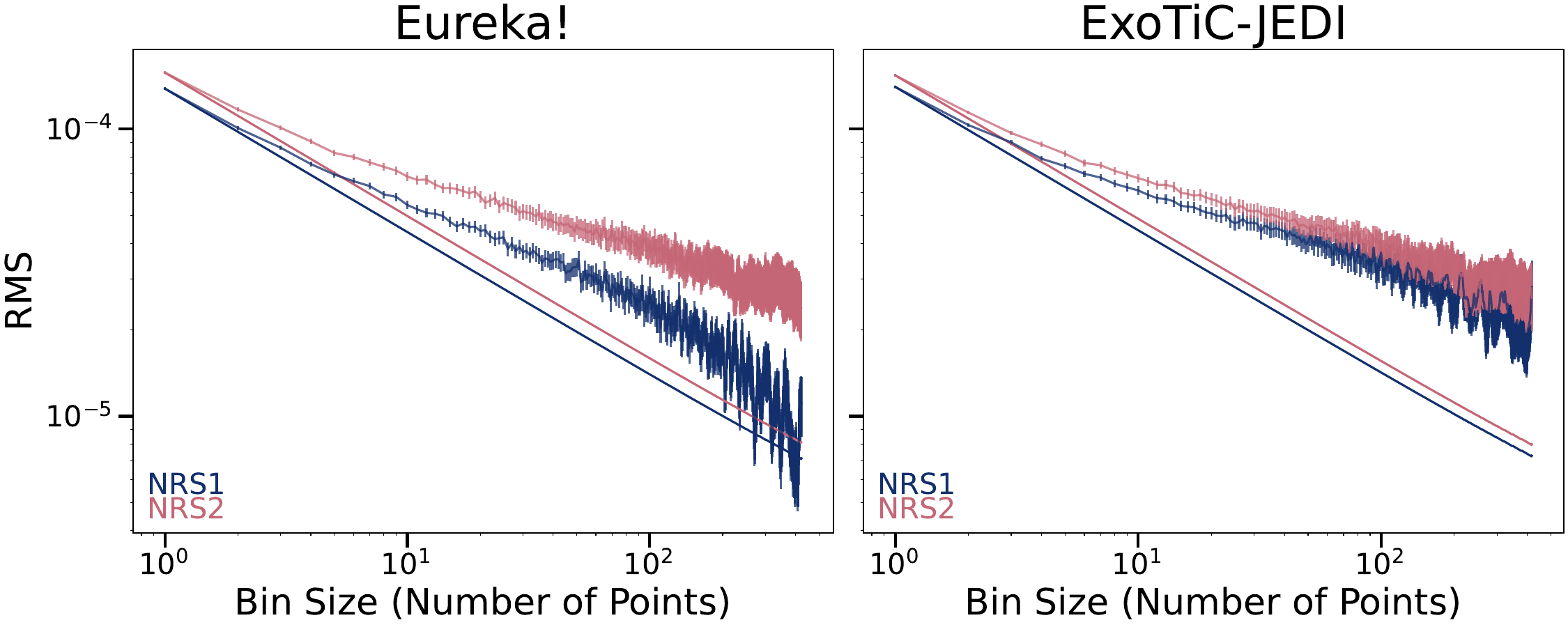}
    \caption{RMS versus bin size for the white light curves of NRS1 in purple and NRS2 in pink. The residuals would follow the solid lines in the absence of red noise. }
    \label{fig:rms}
\end{figure*}

\subsubsection{\texttt{Eureka!}}
\label{sec:eureka}
%Citations: \citep{2022JOSS....7.4503B}
We reduce the observation first with the \texttt{\texttt{Eureka!}} \citep{2022JOSS....7.4503B} pipeline in a similar manner as we have done in previous COMPASS papers \citepalias[e.g.,][]{Teske2025AJ....169..249T}. \texttt{\texttt{Eureka!}} is an end-to-end pipeline for the reduction of JWST and HST data, acting as a wrapper for the \texttt{jwst} pipeline and employing additional reduction steps. Overall, we utilize \texttt{\texttt{Eureka!}} version 0.13 and \texttt{jwst} version 1.18.0 with context map jwst$\_$1364.pmap. For Stages 1 and 2 we primarily use the default \texttt{\texttt{Eureka!}} reduction steps (with the exception of a 15$\sigma$ jump-detection threshold), which in addition to the default \texttt{jwst} steps includes a group-level background subtraction to account for the 1/f noise noise in the observation by removing the median of each column after masking the trace.  In Stage 3, we optimize the extraction aperture, background aperture, polynomial order for the background subtraction, and sigma threshold for the outlier rejection of the optimal extraction, selecting the version of the reduction that minimizes the median absolute deviation of each white light curve (generated by summing over 2.863--3.714 microns for NRS1 and 3.820--5.082 microns for NRS2). We consider extraction aperture half-widths of 4-8 pixels, background aperture half-widths of 8-11 pixels, either a full frame or additional column-by-column background subtraction, and a sigma threshold of 10 or 60 for the outlier rejection during the optimal extraction to allow for an approximate standard extraction. The result of this optimization is as follows: extraction aperture half-widths of 5 pixels and 8 pixel backgrounds (as measured from the center of the trace to the edge of the detector) are preferred for both NRS1 and NRS2 respectively, while an additional full frame background subtraction and a sigma threshold of 10 is preferred for NRS1, and an additional column-by-column background subtraction and a sigma threshold of 60 is preferred for NRS2. We then generate 30 pixel spectroscopic binned light curves from these optimal white light curves.

We then implement a custom light curve fitting code to fit both the white light curves and spectroscopic light curves from \texttt{\texttt{Eureka!}} as we have done in previous COMPASS papers. We first iteratively trim 3$\sigma$ outliers from a 50
point rolling median three times and trim the initial 200 points ($\sim$ 12 minutes) from the light curves to remove any initial ramps in the observation. We then fit the white light curve with a combination of an astrophysical model using \texttt{batman} \citep{Kreidberg2015} with transit depth ($R_p/R_{*}$), time of transit ($T_0$), inclination ($i$), and ratio of the semi-major axis to the stellar radius ($a/R_{*}$) as free parameters, and a systematic model of the form 

\begin{equation}
S= p_{1} + p_{2}\times T+ p_{3}\times X + p_{4}\times Y , 
\label{eq:1}
\end{equation}

\noindent where  $p_{N}$ are free parameters, $T$ is the array of times, and $X$ and $Y$ are arrays of the positions of the trace to be consistent with the simplest noise model that we have utilized in previous COMPASS papers that have used \eureka\ \citep[e.g.,][]{Wallack2026}. We also fit for an additional per-point error added in quadrature to the measurement errors. We fix the quadratic limb darkening to the theoretical values from Set One of the MPS-ATLAS models using the stellar parameters from \cite{Winters2019} (T$_{\rm eff}$= 3337 K, [Fe/H]= -0.340, log(g)= 4.967) with {\tt ExoTiC-LD} \citep{Grant2024} and assume a circular orbit. Next, we first fit our white light curves using a Levenberg-Marquardt minimization where we minimize the log likelihood. To these best-fit parameters we then initialize 3$\times$ the number of free parameters as walkers for a Markov chain Monte Carlo (MCMC) fit using the package \texttt{emcee} \citep{Foreman-Mackey2013}. We run 100,000 steps, discarding the first 50,000 steps as burn-in. We show our best fit parameters from these white light curves in Table~\ref{tab:bestfit}, the white light curves themselves in Figure~\ref{fig:wlc}, and our RMS versus bin plots in Figure~\ref{fig:rms}.

We fit the 30 pixel spectroscopically binned spectra in a similar manner as for the white light curves, fixing $T_0$, $i$, and $a/R_{*}$ to the best fit value from that detector's white light curve. We present the resulting transmission spectrum in Figure~\ref{fig:spectra}. We also test the effect of fixing and fitting the limb darkening. When we allow the limb darkening to vary freely without a prior, the spectrum is in good agreement (at the 1.1 $\sigma$ level or better per point) with the version of the spectrum where we fix the limb darkening, with median differences of 14 ppm. Therefore, we opt to consider the fixed limb darkening version when determining model constraints from our spectrum. 

\subsubsection{ExoTiC-JEDI}
We use the \texttt{ExoTiC-JEDI} pipeline \citep{2022zndo...7185855A} to carry out a second reduction following the same setup as described in \cite{Gordon2026}. We use the standard \texttt{jwst} pipeline \cite[version 1.18.1, context map 1364;][]{jwst_pipeline} for the first stage with the addition of the custom bias subtraction routine implemented in \cite{2023Natur.614..664A} and a group-level column-by-column background subtraction routine. We then flag pixels with data quality bit values indicating ``do not use'', ``saturated'', ``dead'', ``hot'', ``low quantum efficiency'', and ``no gain value'' and replace them with the median of the neighboring four pixels. Pixels exceeding 10$\sigma$ in the spatial dimension and 6$\sigma$ in the time dimension are flagged as outliers and replaced with the median of the nearest 10 pixels. The spectrum is extracted using the intrapixel aperture extraction routine implemented in \texttt{ExoTiC-JEDI} with an aperture of 7 pixel FWHM, which is the value that minimizes the median absolute deviation of the resulting white lightcurve. 

After extracting the spectral timeseries, the white lightcurves for each detector were independently fitted with a custom light curve fitting code. First, we apply a Gaussian filter with a standard deviation of 50 points to the white lightcurve. We then trim flux measurements that deviate more than $4\sigma$ from the filtered light curves, which excludes 9 data points in NRS1 and 12 in NRS2. Finally, we trim the first 200 integrations 12.1 minutes of observation to avoid any effects of instrument settling. We model the light curve as the product of a transit model computed using \texttt{batman} \citep{Kreidberg2015} and a systematics model of the form 
\begin{equation}
    S = p_1 + p_2 \times T
\end{equation}
where $p_1$ and $p_2$ are free parameters and $T$ is the array of times. Note that we omit the x- and y-positions of the trace used in the systematics model for the \texttt{\texttt{Eureka!}} reduction, as they did not appear to correlate strongly with the red noise in the light curves based on a visual inspection. The transit parameters we fit for are the planet-star radius ratio ($R_p/R_*$), the semimajor axis in units of stellar radii ($a/R_*$), the time of transit ($T_0$), the inclination ($i$). As for the \texttt{Eureka!} reduction, we assume a circular orbit. We use \texttt{ExoTiC-LD} \citep{Grant2022} to compute priors for the quadratic limb-darkening parameters using the values from Set One of the MPS-ATLAS models with the stellar parameters from \cite{Winters2019}. Finally, we include an additional error term that is added in quadrature to the errors output by the \texttt{ExoTiC-JEDI} pipeline. 

To infer the parameters of our model we initialize 22 walkers (twice the number of model parameters) and run them for 100,000 steps each using \texttt{emcee}\citep{Foreman-Mackey2013}, discarding the first 50,000 steps as burn-in. The best fit parameters from this white light curve fitting procedure are given in Table \ref{tab:bestfit}, with the white light curves themselves and best fit models in Figure \ref{fig:wlc} and the RMS vs. bin size plots for the light curves minus the model shown in Figure \ref{fig:rms}. 

\begin{figure*}
    \centering
    \includegraphics[width=\linewidth]{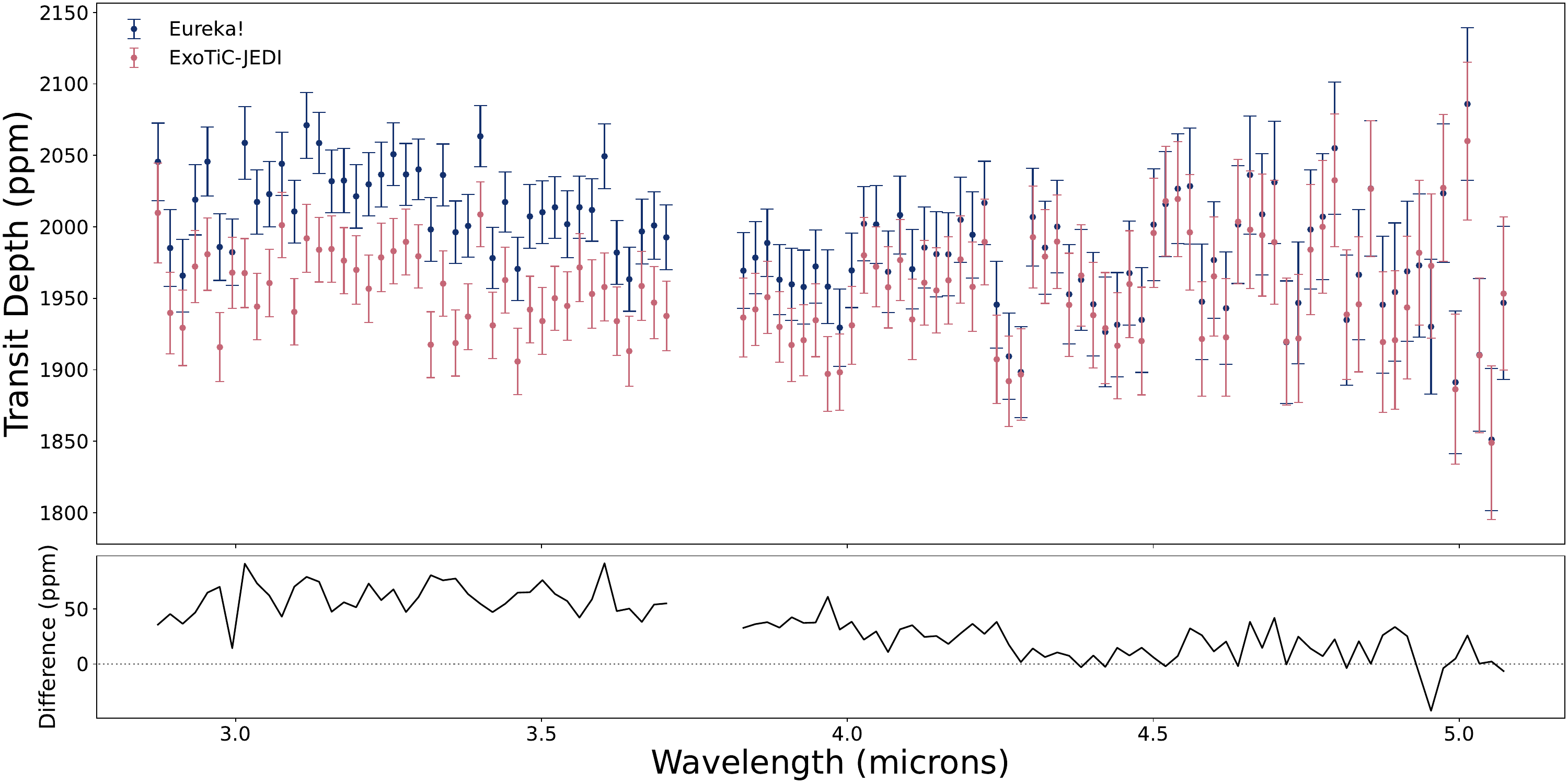}
    \caption{Spectrum of LTT~1445A~b from both the \texttt{ExoTiC-JEDI} and \texttt{\texttt{Eureka!}} reductions. While there is an offset in the baselines for the two reductions in NRS1, the overall shapes of the spectra are in good agreement.}
    \label{fig:spectra}
\end{figure*}

Following white light curve fitting we bin the spectral timeseries into 30 pixel spectroscopic bins with the same binning scheme as was used for the \texttt{\texttt{Eureka!}} reduction. We fit these binned light curves with the same systematics and transit models as for the white light curves. We fix all transit parameters except the limb-darkening parameters and $R_p/R_*$ to the best-fit values from the white light curves for each respective detector. For the limb-darkening parameters we again use \texttt{ExoTiC-LD} with the same stellar parameters and MPS-ATLAS models to compute priors for each wavelength bin. During the fitting procedure for each spectral light curve the two parameters of the systematics model are allowed to vary, as is the additional noise term. We initialize 12 MCMC chains for each light curve and run 10,000 steps of MCMC, discarding the first 5,000 steps as burn-in. Figure \ref{fig:spectra} shows the resulting transmission spectrum.

%As for the \texttt{Eureka} reduction, we trim the initial 200 points to remove an initial ramp seen in the NRS1 light curve. 

%%FIGURE

%%FIGURE
\begin{figure}
    \centering
    \includegraphics[width=\linewidth]{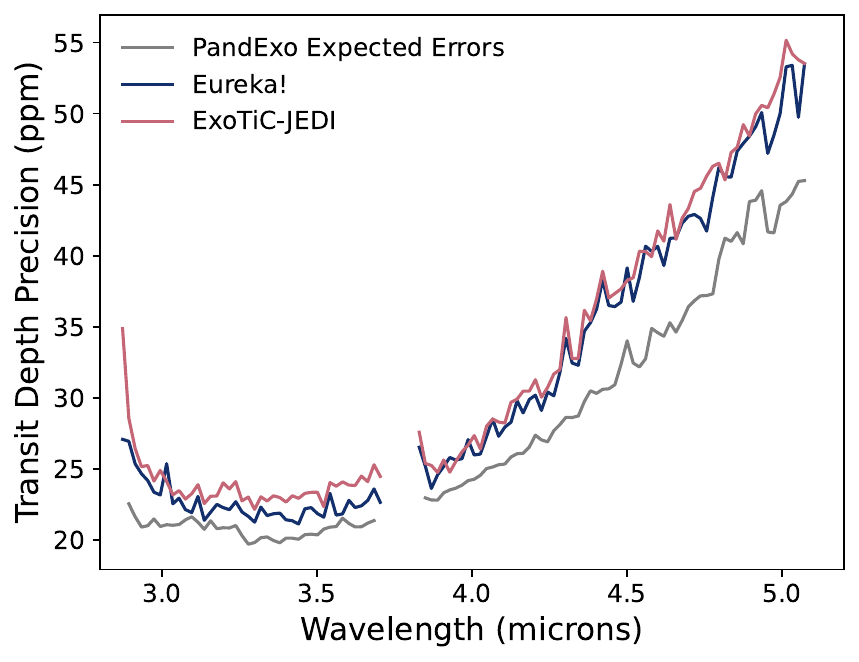}
    \caption{The expected errors on the transmission spectrum from \texttt{PandExo} \citep{Batalha2017}, compared to the measured errors from our two reductions.}
    \label{fig:pandexo}
\end{figure}

\subsection{Overall quality and comparison of data reduction results}
\label{sec:offsets}

As shown in Figure \ref{fig:spectra}, the overall structure of the spectra agree well. Additionally, Figure \ref{fig:pandexo} shows how each data reduction compares to \texttt{PandExo} \citep{Batalha2017} predicted simulations of \ltt. In NRS1, the median \texttt{PandExo} precision is 20~ppm compared to the \eureka\ and \jedi\ median precision of 22.2~ppm and 23.6~ppm, respectively. In NRS2 the median \texttt{PandExo} precision is 32~ppm compared to the \eureka\ median precision of 36.4~ppm and \jedi\ median precision of 36.7~ppm. Therefore, overall the quality of the spectral precision is in good alignment with predictions made prior to taking the observations, although slightly poorer in NRS2 as shown in previous COMPASS analysis \citep[e.g.,][]{Wallack2024AJ....168...77W, Alderson2024AJ....167..216A}. There is a slight offset between the NRS1 spectra between the two reductions (Figure \ref{fig:spectra}). We investigated the cause of the offsets and have ruled out, for example, the differences in fitting versus fixing the limb darkening as the cause. Moreover, differences of such magnitudes have been seen in previous NIRSpec/G395H observations \citep[e.g.,][]{Alam2025AJ....169...15A}. Additionally, there is potentially a linear slope offset between the data reductions in NRS2. This has been observed in other data analyses comparisons as well \citep{2026MNRAS.546ag143C,2025MNRAS.537.3027K}. \citet{2025MNRAS.537.3027K} suggested the treatment of LD parameters as a possible cause. To investigate this ( as described in \S \ref{sec:eureka}) we fix and fit for the LD parameters and find it does not strongly affect the data reduction. We leave further exploration of this to future work. Ultimately, we conduct our theoretical interpretation on both data reductions and give special attention to how we treat for the offsets between the detectors to ensure there these differences does not impact our derived atmospheric inferences. 

%The largest difference between the two data reductions is the NRS1/NRS2 offset. The size of the offset in the \texttt{ExoTiC-JEDI} reduction was observed to depend on the width of the spectral extraction aperture over several trial reductions. The final choice of aperture was determined by minimization of the median absolute difference (MAD). For this aperture the offset is insensitive to different choices of outlier rejection thresholds in the time and spatial dimensions during extraction. We therefore elected to accept the \texttt{ExoTiC-JEDI} with a large offset rather than attempting to minimize it. In section \ref{sec:spect_np} we address the modeling of the detector offsets, and in section \ref{sec:spect_mh} we discuss the impact of the offset on the interpretation of the transmission spectrum.

%%%%%%%%%%%%%%%%%%%%%%%%%%%%%%%%%%%%%%%%%%%%%%%%%%%%%%%%%%%

\section{Interpretation of Planet's Transmission Spectrum} 
\label{sec:spectrum}
%%%%%%%%%%%%%%%%%%%%%%%%%%%%%%%%%%%%%%%%%%%%%%%%%%%%%%%%%%%
Our strategy to interpret the planet's transmission spectrum is two-fold and based on the other published works in COMPASS. First, we apply a purely parametric, non-physical approach to assess the statistical significance of any structure in our spectrum. By structure, we mean slopes, offsets, or any Gaussian-like features that could resemble astrophysical signals. These tests are separately applied to both data reductions, allowing us to determine any differences between them. Second, we apply a more physically informed set of models assuming either chemical equilibrium or a simple three-gas component (e.g., \ce{H2}/He background with \ce{CO2}). These physically informed models allow us to test the bounds of atmospheric metallicities and mean molecular weights that can be statistically ruled out by the data. More details on our methods as well as the results are outlined below.  

\subsection{Spectral Feature Detection}
\label{sec:spect_np}

%%FIGURE
\begin{figure*}
    \centering
    \includegraphics[width=\linewidth]{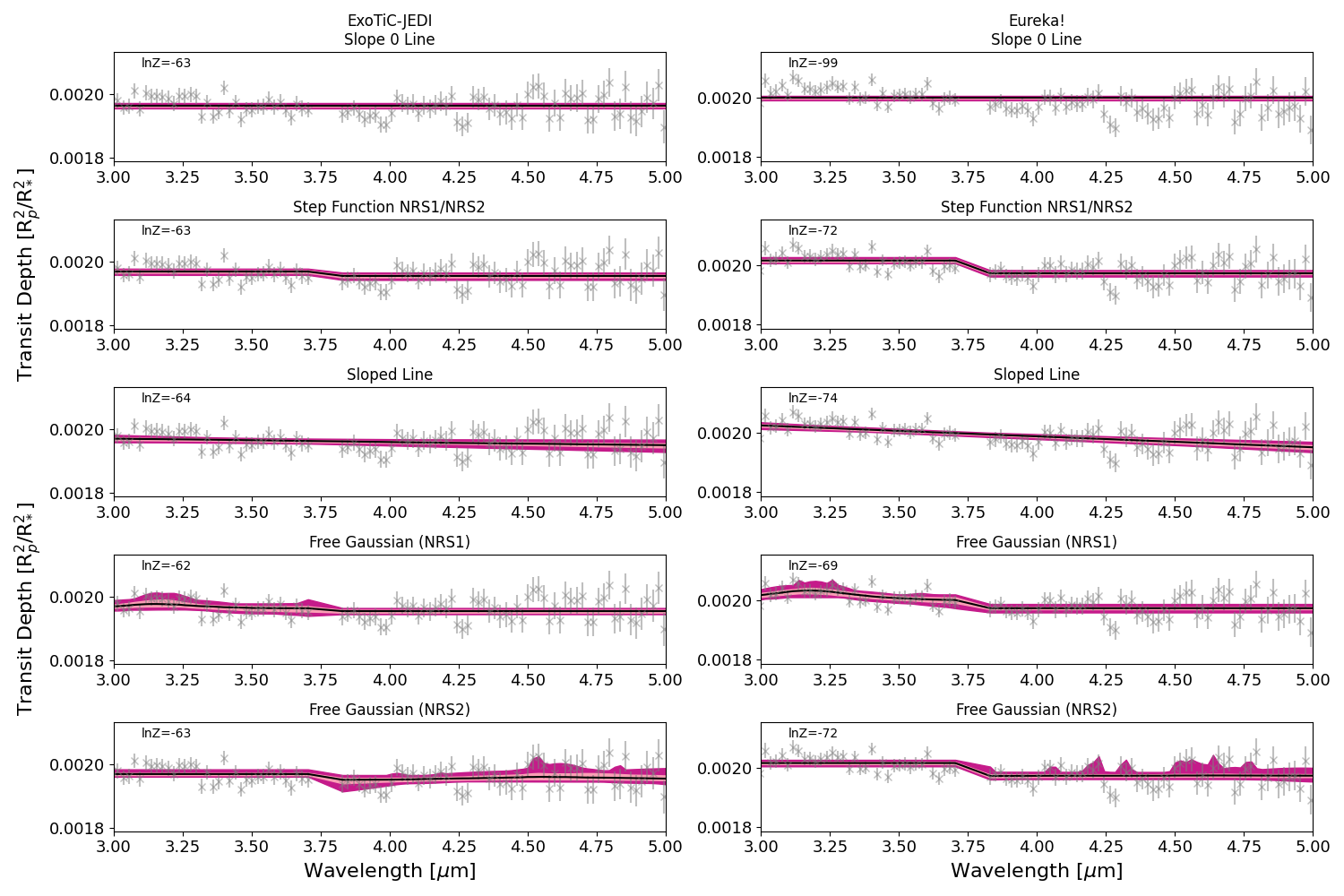}
    \caption{Non-physical model interpretation of \jedi\ (left) and \eureka\ (right) reductions, showing structural differences between the two. Each column contains a different model (outlined in \S\ref{sec:spect_np}) and is roughly ordered by increasing complexity. In individual figures the black line represents the median retrieved model, light pink the 1$\sigma$-banded model, and dark pink the 3$\sigma$-banded model. Statistical quantities, including model evidence, are included in Table \ref{tab:spect_np}. An offset between NRS1/NRS2 (row 2) is the only statistically significant feature in the \eureka\ dataset. The offset fit for both reductions differs. Overall, there are no astrophysical spectral features in the first visit of \ltt. }
    \label{fig:spect_np}
\end{figure*}

Our spectral feature detection technique is described in many other COMPASS works (\citealt[e.g.,][]{Wallack2024AJ....168...77W, Alderson2024AJ....167..216A}). Briefly, we use the MultiNest fitting code, \texttt{Ultranest} \citep{UltraNest} to fit a set of five model types, ranging in parameter complexity: 1) a zero-slope line (1 parameter), 2) a step function to represent two zero-sloped lines with an offset between NRS1 and NRS2 (2 parameter), 3) a sloped-line (2 parameter), 4) a Gaussian-feature constrained to be within the NRS1 wavelength range, plus an NRS1/NRS2 offset (5 parameter), and 5) same as \#4 but in NRS2. 

\begin{table*}[ht]
\centering
\caption{Results of synthetic fits to both \texttt{ExoTiC-JEDI} and \texttt{Eureka!} data reductions.}
\begin{tabular}{llccc}
\hline \hline
\textbf{Model Type} & \textbf{Parameter (Units)} & $\ln Z$ & $\chi^2/N$ & \textbf{Value} \\
\hline
\multicolumn{5}{c}{\textbf{Eureka!}} \\
\hline
Slope 0 Line & Baseline ($(R_p/R_*)^2$ ppm) & -99.0 & 1.8 & $1999 \pm 3$ \\
Step Function NRS1/NRS2 & Offset ($(R_p/R_*)^2$ ppm) & -72.0 & 1.21 & $-43 \pm 5$ \\
Sloped Line & Gradient (ppm/$\mu$m) & -74.0 & 1.25 & $-37 \pm 5$ \\
Free Gaussian (NRS1) & Center $\lambda$ ($\mu$m) & -69.0 & 1.07 & $3.2 \pm 0.1$ \\
Free Gaussian (NRS2) & Center $\lambda$ ($\mu$m) & -72.0 & 1.21 & $4.5^{+0.5}_{-0.4}$ \\
\hline
\multicolumn{5}{c}{\textbf{ExoTiC-JEDI}} \\
\hline
Slope 0 Line & Baseline ($(R_p/R_*)^2$ ppm) & -63.0 & 1.11 & $1962 \pm 3$ \\
Step Function NRS1/NRS2 & Offset ($(R_p/R_*)^2$ ppm) & -63.0 & 1.05 & $-15 \pm 6$ \\
Sloped Line & Gradient (ppm/$\mu$m) & -64.0 & 1.07 & $-10 \pm 5$ \\
Free Gaussian (NRS1) & Center $\lambda$ ($\mu$m) & -62.0 & 0.97 & $3.2 \pm 0.2$ \\
Free Gaussian (NRS2) & Center $\lambda$ ($\mu$m) & -63.0 & 1.02 & $4.6^{+0.4}_{-0.3}$ \\
\hline \hline
\end{tabular}
\label{tab:spect_np}
\end{table*}

 The results of this analysis are shown in Table \ref{tab:spect_np} and corresponding Figure \ref{fig:spect_np}. Neither data reduction prefers a zero sloped (flat) line. \eureka\  strongly prefers ($\Delta \ln$Z = 27) an offset between NRS1 and NRS2.  The \eureka\ data reduction results in an offset of $-43 \pm 5$~ppm (NRS2 baseline below NRS1), and the \jedi\ data reduction results in an offset of $-15 \pm 6$~ppm. %The overall $(R_p/R_*)^2$ agrees to 1$\sigma$ only when comparing \eureka\ NRS2 to \jedi\ NRS1.  
The \eureka\ NRS1 and NRS2 baselines are 2016.8 $\pm$ 3.5~ppm and 1973 $\pm$ 4~ppm, respectively, and the \jedi\ NRS1 and NRS2 baselines are 1969 $\pm$ 4~ppm and 1954 $\pm$ 4~ppm, respectively. 
 
 None of the other models we tested were a better match to the data. However, both \eureka\ and \jedi\ Free Gaussian (NRS1) model tests indicate some structure (weak preference $\Delta \ln \ge$ 1) at $\sim$3.1~$\mu$m. We mention this because weak structure around 3~$\mu$m has been seen in other COMPASS datasets; it was most prominently discussed in TOI-260~b \citep{Meech2026} but also appeared in TOI-836~c \citep{Wallack2024AJ....168...77W}. However, \citet{Gordon2026} found that there was no common spectral structure (being of either astrophysical or systematic in nature) between the first seven COMPASS datasets.

\subsection{Physical Atmospheric Modeling Methodology}
\label{sec:spect_mh}
We take two different physical modeling approaches to analyzing the \ltt\ spectra. The first is based on pure chemical equilibrium, in which we parameterize the atmosphere by the atmospheric metallicity ($\log$ M/H) and other driving parameters such as cloud top pressure. This method allows us to place limits on atmospheric metallicity (a proxy for mean molecular weight) while assessing any degeneracies that exist with the presence of clouds. Given \ltt's mass and radius, as well as insights from previous observations of this planet, it likely does not possess a large H$_2$/He envelope that is governed by chemical equilibrium. Therefore, to complement the limits we obtain on $\log$ M/H, we also parameterize the atmosphere with simple three-gas mixture models, considering cases of a \ce{H2}/He-background with varying quantities of \ce{CH4}, \ce{CO2}, and \ce{H2O}. This allows us to determine how our mean molecular weight constraints might be affected by the assumption of chemical equilibrium. Below, we further detail these two methodologies, then describe the results of each. 

For our chemical equilibrium-based grid we compute the spectral models as a function of metallicity ($\log$ M/H), carbon-to-oxygen ratio (C/O), and cloud top pressure. Our pressure-temperature (PT) profile is computed via the \citet{2010A&A...520A..27G} parameterization, which for weakly irradiated atmospheres approximates to $T(P)^4 \approx \frac{3}{4}T_\mathrm{eq}^4(P+\frac{2}{3})$, where $T$ is the pressure ($P$) dependent profile. Given this P-T profile we compute altitude-dependent abundances using the equilibrium chemistry solver in the \texttt{photochem} package \citep{photochem2025PSJ.....6..256W} for 20 evenly log-spaced metallicities from 1 -- 1000 $\times$ Solar. We choose two carbon-to-oxygen ratios -- Solar and 0.5 $\times$ Solar, assuming \citet{2009LanB...4B..712L} abundances -- solely to ensure our results are robust against these values. To approximate the effect of clouds, we include a wavelength-independent (grey) opacity source of $\tau=10$ at 5 varying pressures. Though this likely oversimplifies cloud structure in small planets like \ltt\ \citep[e.g.,][]{2025NatAs.tmp..256R} we use it to simply probe the degeneracy that could exist between increased metallicity (or mean molecular weight) and muted features from an aerosol obscuring opacity. Beyond our own COMPASS survey papers, this technique has been used in many analyses of small planets with JWST and HST \citep[e.g.,][]{2017PASP..129d4402K, 2018AJ....156..252M, Lustig2023NatAs...7.1317L}. We note that this technique also mimics the effect of varying surface pressures. For example, \citet{Lustig2023NatAs...7.1317L} referred to this as an ``apparent surface''. Here and in other works (e.g., \citetalias{Teske2025AJ....169..249T}) we refer to this as an ``opaque pressure level''. 

Given the P-T profile, chemistry, and opaque pressure level, we compute spectra using the open source spectra and climate code \texttt{PICASO-v4} \citep{2019ApJ...878...70B,Mang2026}. \texttt{PICASO} enables the computation of transmission, emission, and reflected light spectroscopy for a wide range of substellar atmospheres, including Brown Dwarfs. It also computes radiative-convective equilibrium climate models, though that functionality is not used here. It relies on an opacity database, which  is available at \citet{natasha_batalha_2025_14861730} and includes opacity sources from \ce{C2H2} \citep{hitran2012}, \ce{CH4} \citep{2020ApJS..247...55H},  \ce{CO} \citep{li15rovibrational}, \ce{CO2} \citep{HUANG2014reliable}, \ce{H2O} \citep{Polyansky2018H2O, GharibNezhad2021}, \ce{H2S} \citep{azzam16exomol}, \ce{K} \citep{Allard2019}, \ce{Na} \citep{Allard2019}, \ce{OCS} \citep{HITRAN2016}, \ce{SO2} \citep{underwood2016exomol}, and collision induced opacity from \ce{H2-H2} \citep{Saumon12,Lenzuni1991h2h2} and \ce{H2-He} \citep{Saumon12}. The opacity is resampled from line-by-line calculations at roughly $R\sim$ 1e6 down to $R$ = 15,000. Though this is lower resolution than is sometimes used in analysis of NIRSpec/G395H data \citep[e.g., $R$ = 60,000 in][]{2023Natur.614..664A} it is sufficient for this work, as we do not observe any spectral features.  

We fit the data to the grid of models in two ways to ensure our method of handling the NRS1/NRS2 offset is robust. In our first method, we correct the offset in the data using the fitted value shown in Table \ref{tab:spect_np}. Then, we use \texttt{scipy.curve\_fit} to fit the overall model to the entire offset-corrected dataset. We proceed by computing the chi-squared statistic per data point, $\chi^2/N$, and follow through by converting it to a p-value and $\sigma$-confidence, assuming the errors are normally distributed. In our second method, we leave the offset as a free parameter within the fit. We compute the offset by using \texttt{scipy.curve\_fit} to fit the spectra to NRS1 and NRS2 spectral ranges independently. Then we compute the ultimate  $\chi^2/N$ and $\sigma$-confidence similar to our first method. 

For our three-gas component model fits we use identical methodology as our metallicity grid exercise for creating a pressure-temperature profile, modeling clouds, and computing spectra with \texttt{PICASO}. We make two main changes: 1) we swap chemical equilibrium for flexible abundances, and 2) we swap the grid fitting method with the \texttt{Ultranest} Bayesian fitting algorithm to directly fit for the free parameters of interest. For the three-gas component model we fix an \ce{H2}/He fraction to a Solar value (5.93, \citealt{2009LanB...4B..712L}) then fit freely for the fraction of $X_i$/$\left( H_2+He \right)$ where $X_i$ is either \ce{H2O}, \ce{CH4}, or \ce{CO2}. We choose these three gases because they have the strongest absorption bands from 3-5~$\mu$m and because they are all naturally generated from simple chemical equilibrium processes \citep{photochem2025PSJ.....6..256W}. Section \S\ref{sec:spect_np} clearly demonstrates no evidence for spectral features, therefore there is no physical motivation for more exotic molecular species. In addition to fitting for the abundance ratio, we also fit for an NRS1/NRS2 offset, an opaque pressure level, and an error inflation term (see Eqn. 2 \& 3 in \citetalias{Teske2025AJ....169..249T}) -- a total of five free parameters per run. The analysis in \citetalias{Teske2025AJ_paren} used this retrieval methodology and demonstrated its importance in determining how limits on mean molecular weight depend on the model setup. 

\begin{figure*}[b!]
    \centering
    \begin{subfigure}[b]{0.48\textwidth}   
        \centering
        \includegraphics[width=\linewidth]{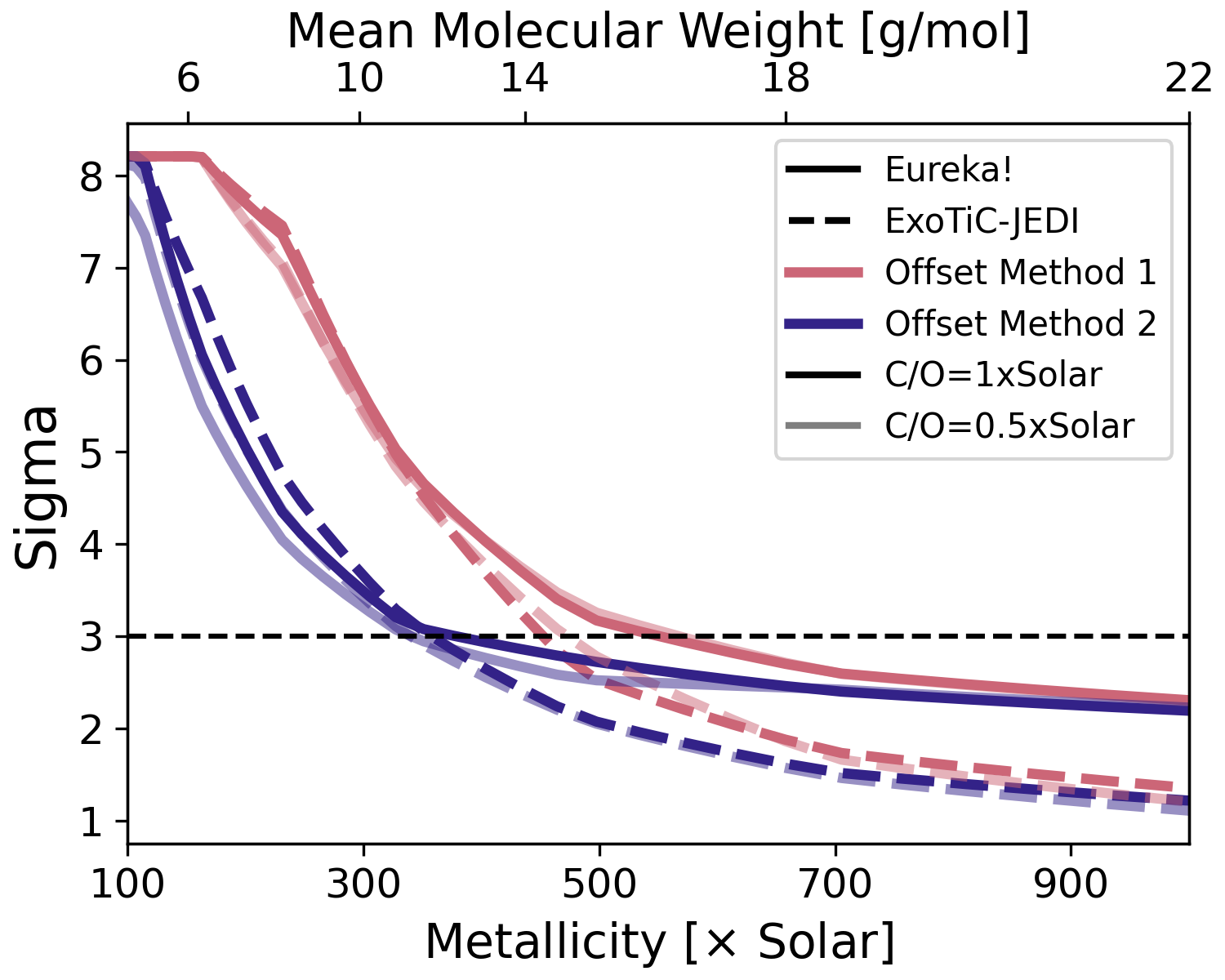}
        \caption{How atmospheric metallicity limits depend on offset fitting method}
        \label{fig:spect_mhA}
    \end{subfigure}
    \begin{subfigure}[b]{0.48\textwidth}   
        \centering
        \includegraphics[width=\linewidth]{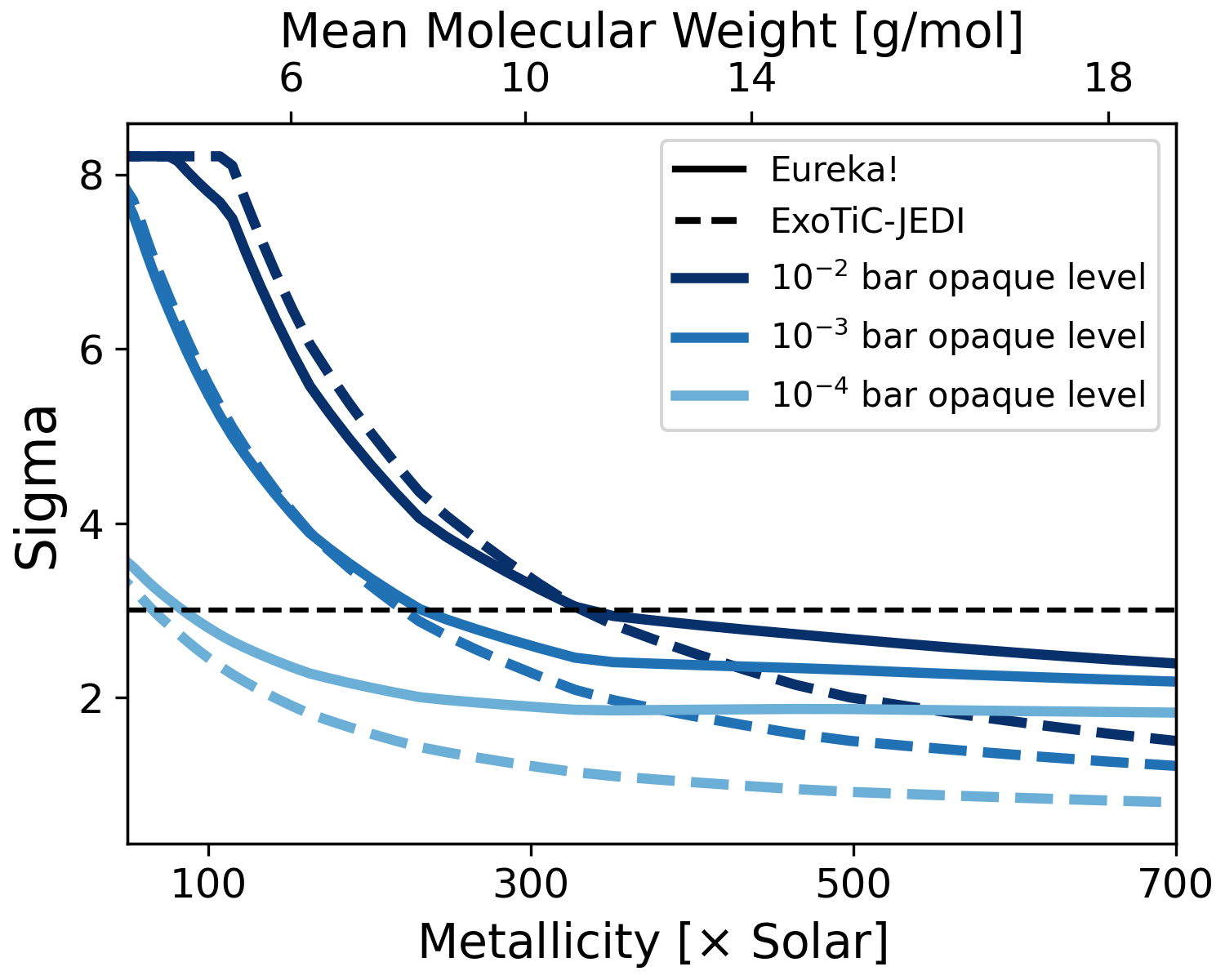}
        \caption{How atmospheric metallicity limits depend on opaque pressure levels}
        \label{fig:spect_mhB}
    \end{subfigure}
    \caption{Curves demonstrate how well the \ltt\ data can rule out atmospheric metallicity (and corresponding mean molecular weight). Solid and dashed lines represent the \eureka\ and \jedi\ reductions, respectively. In the left figure (a) all curves are computed with no opaque pressure level (``cloud-free''). Additionally, in (a) the dark lines show results for solar C/O (=0.458) and transparent lines show corresponding results for 0.5 $\times$ Solar (=0.229), demonstrating insensitivity of results to C/O. In the right figure (b) we only show  solar C/O  results since they are qualitatively the same as the 0.5 $\times$ Solar C/O case, as shown in (a). Also on the right, we only show offset method 2, since this is where the two reductions show agreement at the 3$\sigma$-level. }
    \label{fig:spect_mh}
\end{figure*}

\subsection{Physical Atmospheric Modeling Results}
\label{sec:spect_results}
Figure \ref{fig:spect_mh} shows the results of our atmospheric metallicity analysis. We first show how our metallicity limits are sensitive to the method of including offsets in the fit (pink vs.\ purple in Figure \ref{fig:spect_mhA}). When considering method \#1 (offset is corrected before fit), the \eureka\ and \jedi\ metallicity limits show slight disagreement at the 3$\sigma$-level, with 450 $\times$ Solar and 550 $\times$ Solar limits, respectively. When considering method \#2 (offset is a free parameter in the fit), the results of both the \eureka\ and \jedi\ reductions become more conservative toward lower limits in metallicity, ruling out metallicities less than 380 $\times$ Solar and 360 $\times$ Solar, respectively. Because offset method \#2 allows both NRS1 and NRS2 to be treated independently, it is intuitive that they provide more conservative limits and a closer agreement with respect to the data reduction method used. Therefore, we proceed with this method for our final conclusions. However, we urge future studies to consider multiple techniques for accounting for offsets when using grids to fit spectral data. In Figure \ref{fig:spect_mhA} we also demonstrate that atmospheric C/O does not play a large part in determining how well we can place limits on atmospheric metallicity (solid vs. transparent lines). 

In Figure \ref{fig:spect_mhB}, we show the well-known degeneracy between atmospheric metallicity and opaque pressure level (cloud or surface) as applied to this planet's spectra. %As expected, we show how an opaque pressure level (cloud or surface) affects our ability to put limits on atmospheric metallicity in Figure \ref{fig:spect_mhB}. 
With a fully opaque cloud deck at 10$^{-2}$~bar, our 3$\sigma$ results change only slightly to  rule out atmospheric metallicities of M/H $<330\ \times$ Solar. An opaque pressure level of 10$^{-3}$~bar and  10$^{-4}$~bar (e.g., clouds higher in the atmosphere) would shift our inference down to $<220\ \times$ Solar and $<100\ \times$ Solar, respectively. We briefly discuss the possibility of forming low pressure aerosols in the atmosphere of \ltt\ in \S\ref{sec:clouds}.

\begin{figure*}[b!]
    \centering
    \begin{subfigure}[b]{0.32\textwidth}   
        \centering
        \includegraphics[width=\linewidth]{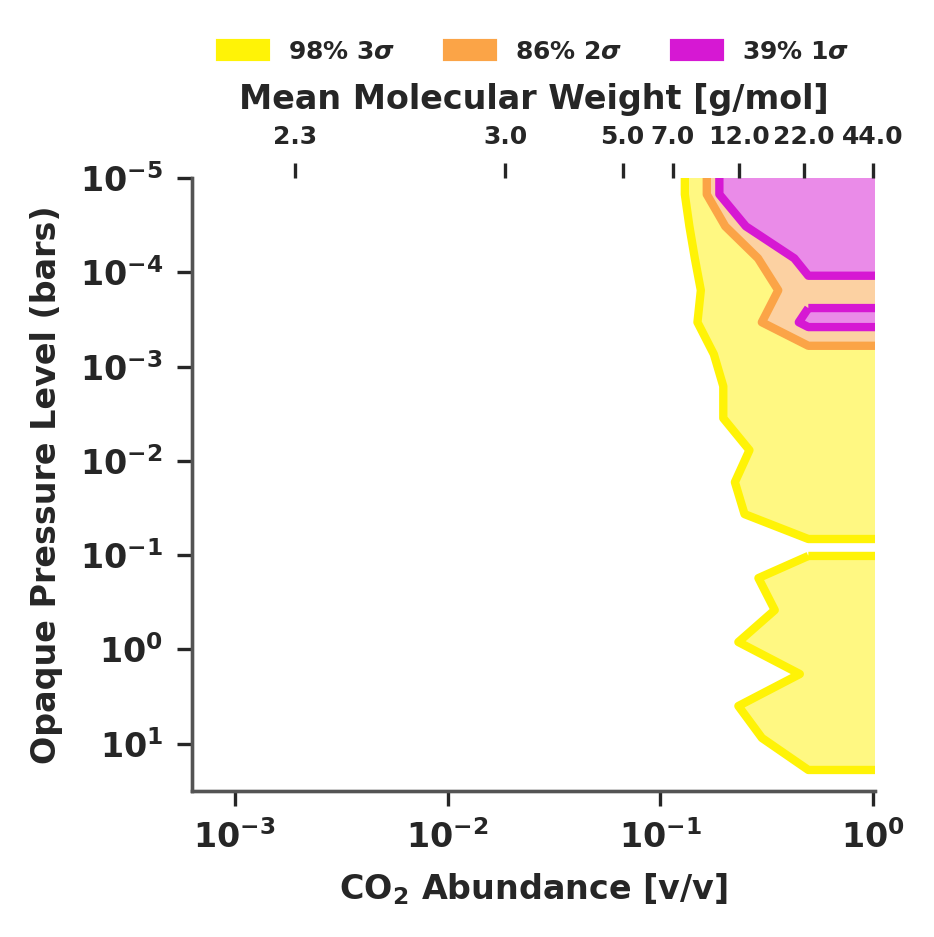}
        \caption{Ruling out \ce{CO2} abundances}
        \label{fig:spect_molA}
    \end{subfigure}
    \begin{subfigure}[b]{0.32\textwidth}   
        \centering
        \includegraphics[width=\linewidth]{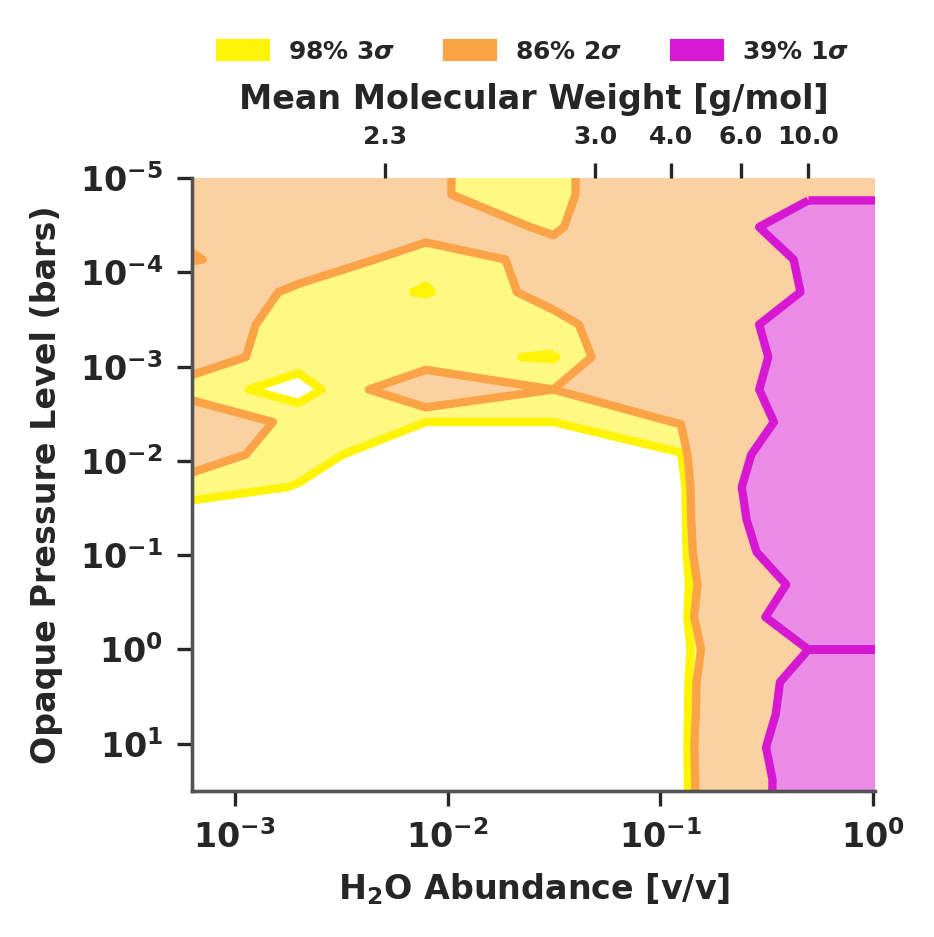}
        \caption{Ruling out \ce{H2O} abundances}
        \label{fig:spect_molB}
    \end{subfigure}
    \begin{subfigure}[b]{0.32\textwidth}   
        \centering
        \includegraphics[width=\linewidth]{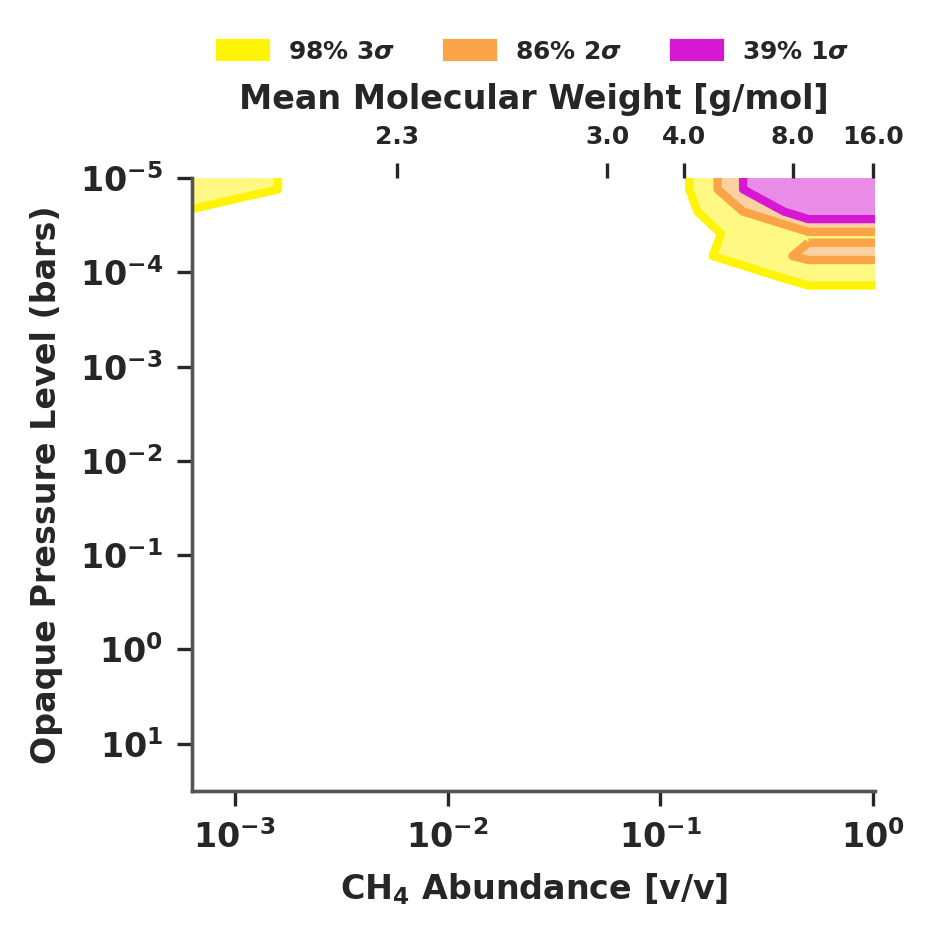}
        \caption{Ruling out \ce{CH4} abundances}
        \label{fig:spect_molC}
    \end{subfigure}
    \caption{Posterior probability distributions for the atmospheric abundances of main (either \ce{CO2}, \ce{H2O}, or \ce{CH4}) gas species versus the opaque pressure level in our three-gas model fits. Each of these models are computed with \ce{H2}/He backgrounds. The two-dimensional marginal distributions are shown with shaded contours representing the 1$\sigma$ (39.3\%), 2$\sigma$ (86.5\%), and 3$\sigma$ (98.9\%) credible regions. White space represents regions which are ruled out to greater than 3$\sigma$. Results are shown for \eureka\ data reductions but are qualitatively consistent with \jedi.}    \label{fig:spect_mol}
\end{figure*}

\begin{figure*}
    \centering
    \includegraphics[width=\textwidth]{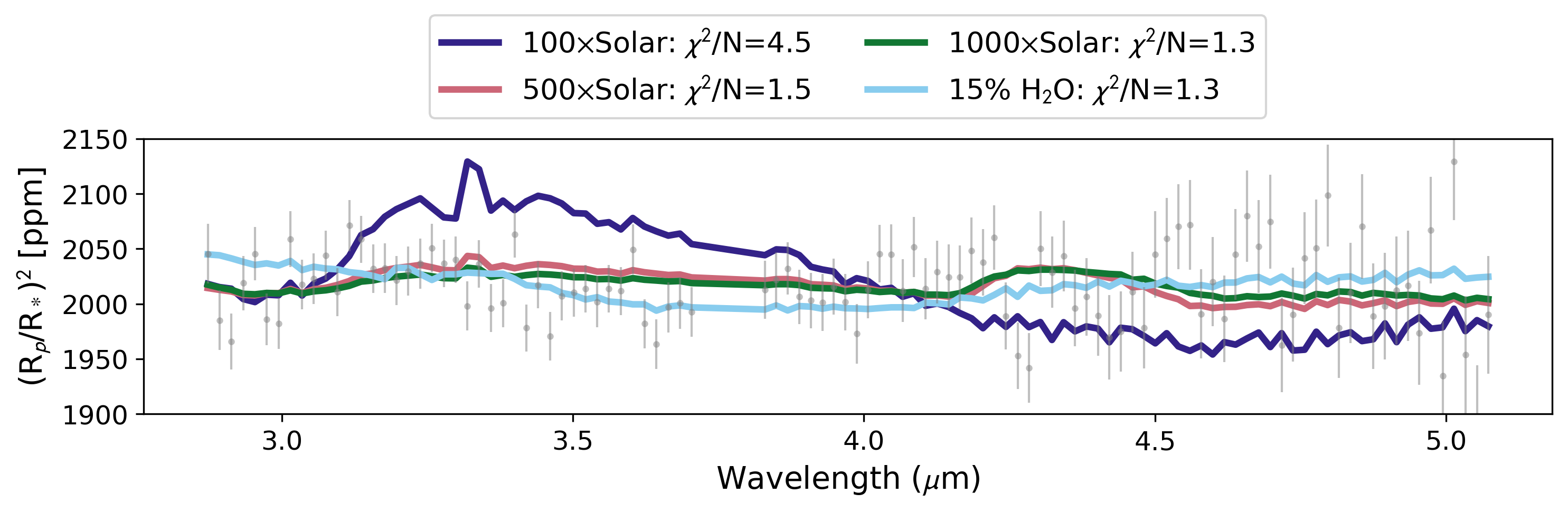}
    \caption{Spectra showing \eureka\ data reduction with four representative models to contextualize the fitting results discussed in \S\ref{sec:spect_results}. We can rule out 100 $\times$ Solar metallicities with high confidence because of the large \ce{CH4} features present at 3.3~$\mu$m. Toward high metallicity ($>300\ \times$ Solar), a smaller amplitude \ce{CO2} feature becomes visible from 4.2-4.5$\mu$m, which is also not present in our observation. Lastly, we show a 15\% \ce{H2O} case from our three-gas component model fit to demonstrate why pure steam atmospheres are difficult to rule out with NIRSpec/G395H data alone. All the $\chi$/N statistics shown use the offset-corrected data from the ``method \#1'' fitting procedure and all models are shown without the effect of an added opaque pressure level. }
    \label{fig:models}
\end{figure*}

Figure \ref{fig:spect_mol} shows the results of the three-gas fits for \ce{CO2}, \ce{H2O}, and \ce{CH4} for the \jedi\ reduction. The 3$\sigma$ contour lines of the \eureka\ reduction agree with the \jedi\ reductions, similar to the metallicity grid fitting results shown in Figure \ref{fig:spect_mhA}. Assuming \ce{H2}/He backgrounds we can rule out approximately $X_i<$ 10\% abundances of \ce{CO2}. For \ce{H2O} we can also approximately rule out $X_i<$ 10\% abundances when the opaque pressure level is higher than roughly 0.01~bar. Our data is more sensitive to \ce{CO2} given the strong spectral feature at 4.3~$\mu$m.  The \ce{H2O} opacity only presents itself from 3-5~$\mu$m as a ``U''-continuum shape (see Figure \ref{fig:models}). Our \ce{H2O} model setup is also more sensitive to the presence of clouds, which dampen this ``U''-shape. Lastly, all models are generally insensitive to abundances lower than $X_i\lesssim$ 1$\times10^{-3}$. For reference, at solar M/H and solar C/O the abundance of H$_2$O is 1$\times10^{-3.04}$ \citep{2010ApJ...716.1060V}.  Toward higher abundances we lose the ability to detect a molecule (or atmosphere) due to the smaller scale height. Toward low abundances we lose the ability to detect a molecule (or atmosphere) since the spectrum approaches the flat background continuum of \ce{H2}/He. Here, abundances lower than one solar metallicity are not relevant to the small nature of \ltt. 

The \ce{CH4} posterior probability distributions exhibit a different behavior (Figure \ref{fig:spect_molC}). We can rule out up to 100\% \ce{CH4} atmospheres for pressures $>10^{-4}$~bar and are only insensitive to the below solar metallicity abundances when the \ce{CH4} feature ($\sim$3.4~$\mu$m) is overpowered by the collision induced absorption opacity. These \ce{CH4} model results are perhaps the most optimistic in terms of the maximum mean molecular weight we can rule out (16~g/mol). However, as pointed out in \citetalias{Teske2025AJ_paren}, mean molecular weight constraints are largely driven by the particular model setup and assumption of dominant background gas. For example, given our \ce{H2O} three-gas component model we can rule out mean molecular weights of about $<4$~g/mol when the opaque pressure level is greater than 1$\times10^{-2}$ bar. However, given our \ce{CO2} results we can instead rule out mean molecular weights of about $<7$~g/mol.  

Finally, Figure \ref{fig:models} shows modeled spectra that can be used to contextualize the two fitting results we have just described. We show spectra at chemical equilibrium values of $100-1000\ \times$ Solar along with a \ce{H2O}+\ce{H2}+He model. We do not show \ce{CO2} or \ce{CH4} models because those spectral features are represented in the chemical equilibrium models. Namely, at lower metallicity (toward 100~$\times$~Solar),  \ce{CH4} is the dominant spectral feature (3.3~$\mu$m) at the temperatures of \ltt\ and subsequently is the reason we can rule out those scenarios. Toward higher metallicity, \ce{CO2} also shows up toward 4.3~$\mu$m, albeit at smaller spectral significance. The \ce{H2O} model shown demonstrates the difficulty of ruling out pure steam models with NIRSpec/G395H. The peak of the \ce{H2O} bands occur just outside the wavelength range of NIRSpec/G395H on both the blue and red edges. Therefore, the effect of \ce{H2O} is just the subtle ``U-shape'' shown. 

%%%%%%%%%%%%%%%%%%%%%%%%%%%%%%%%%%%%%%%%%%%%%%%%%%%%%%%%%%%
\section{Discussion} 
\label{sec:discussion}
%%%%%%%%%%%%%%%%%%%%%%%%%%%%%%%%%%%%%%%%%%%%%%%%%%%%%%%%%%%
\subsection{Are aerosols likely present at pressures less than 10$^{-2}$~bar on \ltt?}
\label{sec:clouds}
Our reported metallicity limits of $>500~\times$~Solar become dependent on the opaque pressure level when this level becomes lower than 0.01~bar (see Figure \ref{fig:spect_mhB}). Therefore, it is worth considering whether an opaque pressure level $<0.01$~bar is plausible. Full scale cloud modeling of this system is beyond the scope of this analysis but we can hypothesize potential aerosols that could exist on \ltt\ given its equilibrium temperature. Our parametrized pressure-temperature profile for \ltt\ gives it a temperature of 365~K in the upper atmosphere (p $<$ 0.01~bar), and a temperature of 700~K at depth (p=200~bar). These values give us an approximate temperature range in which to hypothesize possible aerosol species. 

The most straightforward species to consider are condensible species, which we determine based on the open source atmospheric cloud code \texttt{Virga} \citep{2026AJ....171...98B}. % we determine which condensible species are most likely to exist in the atmosphere of \ltt. 
Of the canonical condensibles species available in \texttt{Virga}, the two that have condensation temperatures closest to that of \ltt's atmosphere are \ce{H2O} and KCl. Condensation temperatures are generally dependent on metallicity and mean molecular weight, with higher metallicities raising the point of condensation. For example, in a 1 $\times$ Solar metallicity atmosphere, the condensation temperature of \ce{H2O} is 210~K at 0.01~bar. In a 500 $\times$ Solar metallicity atmosphere, the condensation temperature of \ce{H2O} rises to 270~K at 0.01~bar. However, this is still well-below the temperature expected at that pressure for \ltt\ \citep{2026AJ....171...98B}. Therefore \ce{H2O} is unlikely to be in a condensed form. The condensation temperature of KCl at 1 $\times$ Solar metallicity is 889~K at 0.01~bar. In order for KCl to condense on \ltt\ it would need temperatures to cross this threshold. However, the temperature of \ltt's atmosphere at these pressures is likely too cold and would not cross the threshold for condensation of KCl to occur either. Other more exotic aerosols could also be considered as viable aerosols. 

%The internal temperature of \ltt\ is an unconstrained quantity. However, given its density [mention density of ltt if we add this text back in], it is not likely this system is analogous to planets like GJ~1214~b, for which KCl has been suggested a viable condensable \citep[e.g.,][]{2018ApJ...863..165G, 2018ApJ...859...34O,2025ApJ...994..116M}. %For example, many studies have considered whether or not lofted KCl particles are responsible for the muted features on GJ~1214~b ($T\mathrm{eq}$=566~K) \citep[e.g.,][]{2018ApJ...863..165G, 2018ApJ...859...34O,2025ApJ...994..116M}. They concluded that depending on the porosity of the KCl particles, it would be possible to flatten the spectrum of a planet like GJ~1214~b if it had a highly metal-rich atmosphere and strong vertical mixing. GJ~1214~b is $\sim100$~K warmer and more than 2$\times$ larger in radius than \ltt. Therefore, KCl also does not present such a straightforward option for a condensible on \ltt. 

Photochemically produced hydrocarbon organic hazes could also be an avenue to create an effective opaque pressure level at lower pressures. The most likely pathway to create hydrocarbon haze requires \ce{CH4} \citep{doi:10.1073/pnas.0608561103}. However, abundant H-bearing species like \ce{CH4} are unlikely to be stable over geologic time on warm and small rocky planets like {\ltt} because of rampant hydrogen escape to space \citep[e.g.,][]{Adams2025AJ....170..219A}. Self-consistent simulations of rocky planets orbiting M stars \citep[e.g.,][]{KirssansenTotton2024} predict that billions of years of escape should most often yield hydrogen poor atmospheres rich in \ce{CO2} or \ce{O2}. On such oxidizing atmospheres, significant \ce{CH4} would not be thermodynamically stable, and thus hydrocarbon hazes from methane photochemistry seem improbable. Additionally, hazes derived from CO or CO$_2$ photochemistry would be much less abundant and unlikely to build up to opaque layers high in the atmosphere \citep{Horst2018,Moran2022}. 

Exotic sulfur-bearing condensates such as \ce{H2SO4} or \ce{S8} could also be considered for \ltt. \ce{H2SO4} clouds are feasible in relatively oxidizing atmospheres (e.g. Venus or modern Earth) that are rich in \ce{H2O} and \ce{SO2}, as both molecules can be photochemically processed to \ce{H2SO4} \cite[e.g.,][]{photochem2025PSJ.....6..256W}. Assuming {\ltt} has a 0.01 bar temperature of $\sim 400$ K, then gas-phase \ce{H2SO4} volume mixing ratios above 0.01 would be super-saturated and would condense to form clouds \citep{Dai2022}. Determining whether such concentrations are achievable on {\ltt} would require photochemical simulations that are beyond the scope of this study. If the atmosphere was instead relatively reducing in its bulk content, \ce{S8} vapor would be more likely than \ce{H2SO4} \citep[e.g., \ce{H2}-rich;][]{Gao2017}. \ce{S8} saturation vapor pressure is very similar to \ce{H2SO4} \citep{Lyons2008}, so a similar $\sim1$\% concentration at 0.01 bar would be necessary for \ce{S8} to condense. Again, whether such a concentration is attainable on {\ltt} could be investigated by future studies that employ photochemical models.

Overall, there are several plausible pathways that could lead to aerosol formation in \ltt's atmosphere at lower pressures. However, each presents a unique challenge.  Increasing the signal-to-noise of \ltt\ observations combined with more detailed cloud and photochemical modeling would enable a clearer determination of whether or not these processes are in fact viable.

\subsection{Combining HST and JWST Data}
\label{sec:hst}

\begin{figure*}[b!]
    \centering
    \begin{subfigure}[b]{0.55\textwidth}   
        \centering
        \includegraphics[width=\linewidth]{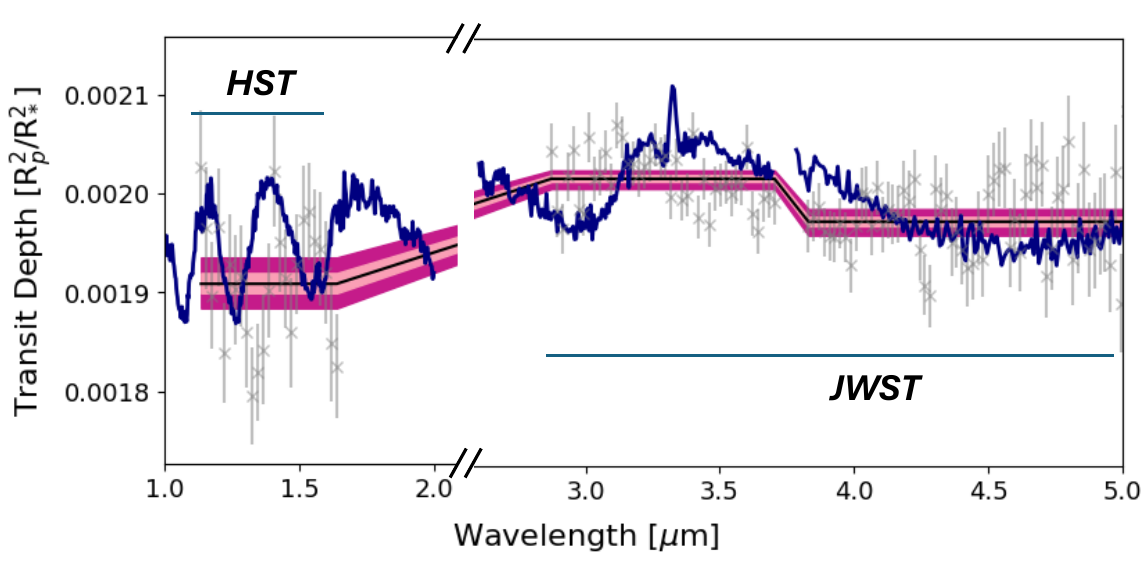}
        \caption{HST and JWST data shown together}
        \label{fig:hstA}
    \end{subfigure}
    \begin{subfigure}[b]{0.35\textwidth}   
        \centering
        \includegraphics[width=\linewidth]{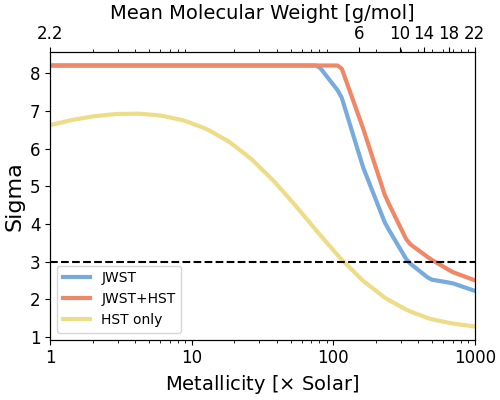}
        \caption{Ability to rule out atmospheric metallicities}
        \label{fig:hstB}
    \end{subfigure}
    \caption{On the left, we show our data in comparison with HST data from \citep{Bennett2025AJ....169..111B}. Along with the data we show the 2$\sigma$ (light pink) and 3$\sigma$ (dark pink) widths for the retrieved transit depth baseline of each dataset. In dark blue we show a 100 $\times$ Solar metallicity model purely for reference. On the right, we show our ability to infer atmospheric metallicity when considering the data from \citet{Bennett2025AJ....169..111B} and the \eureka\ data reduction presented here. When adding HST, our atmospheric metallicity constraints improve from $\sim$ 380 $\times$ Solar to $\sim$ 520 $\times$ Solar.}
    \label{fig:hst}
\end{figure*}

%\begin{figure}
%    \centering
%    \includegraphics[width=\linewidth]{HST_JWST_Comparison.png}
%    \caption{Ability to infer atmospheric metallicity when considering data from \citet{Bennett2025AJ....169..111B} and the data presented here (\texttt{Eureka!} reduction, method \#2 for offset correction, or the dark purple line in \ref{fig:spect_mhA}). When adding HST, our atmospheric metallicity constraints improve from 350$\times$Solar to 500$\times$Solar.} 
%    \label{fig:hst}
%\end{figure}
We use our chemical equilibrium grid fitting method from \S \ref{sec:spect_mh} to understand how and if adding the HST data from \citet{Bennett2025AJ....169..111B} aids in our derived metallicity constraints. Introducing the HST data adds another free parameter as there is an inherent offset between HST, NRS1, and NRS2. We show how the HST data compare to the \eureka\ reduction presented here in Figure \ref{fig:hstA}. Overall there is an offset between the HST data and the NRS1/NRS2 data and the HST data has generally lower SNR compared to the data presented here. 

In \S \ref{sec:spect_mh} we show that offset method \#2 (freely fitting each individual spectrum section to the data rather than pre-correcting the data for all offsets) is more robust to differing data baselines. Therefore we opt to use that method for this joint HST and JWST analysis.  %two data reduction methods agree when we correct the offset between NRS1 and NRS2 using our Bayesian fitting technique, then proceed with fitting the spectrum to data. Applying this same method (referred to as method \#1) to the offset between both HST and JWST produces metallicity constraints that are too optimistic, likely because the continuum matching between HST and JWST data is less straightforward than the slight shift between NRS1 and NRS2. 
When applying this method \#2 and fitting each section of the spectra individually, we can get an estimate for how the addition of HST improves our atmospheric constraints, shown in Figure \ref{fig:hstB}. With HST only we report a 3$\sigma$ limit of 120$\times$Solar (nearly identical to the 100$\times$Solar limit reported in \citet{Bennett2025AJ....169..111B}), a reassuring check that our modeling framework is reproducible. When we add the \eureka\ reduction we increase our 3$\sigma$ JWST-only limit from 380$\times$Solar to 520$\times$Solar. %Note, this ``JWST only'' limit derived with the \eureka\ reduction of 300$\times$Solar is our most conservative estimate. 
Overall, the broader wavelength coverage from HST data tightens the overall limits we can place on \ltt's atmosphere, assuming chemical equilibrium. %provide a statistically significant increase to our overall metallicity constraints, likely as a result of the lower SNR compared to the data reported here. 

\begin{figure*}[b!]
    \centering
    \includegraphics[width=1\textwidth]{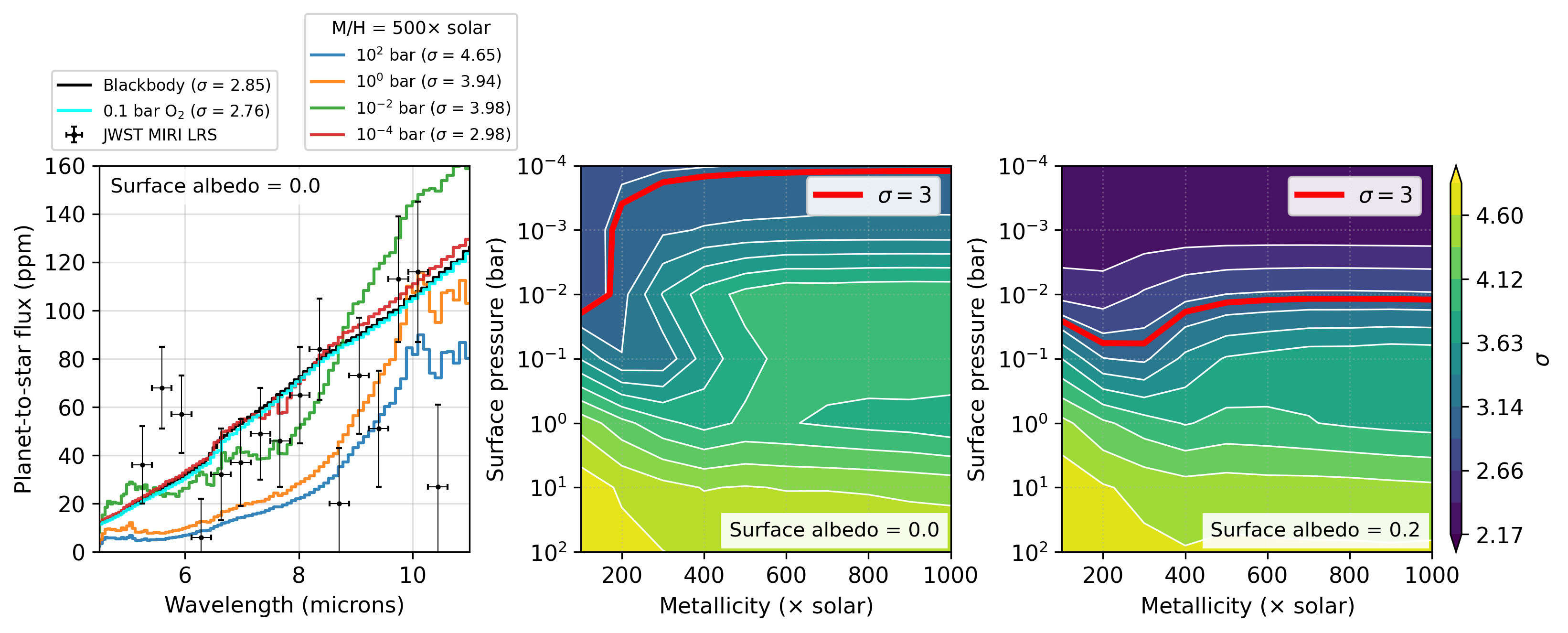}
    \caption{Left: Simulated emission spectra for various atmospheric compositions and thicknesses (all with zero surface albedo) compared to the MIRI LRS observations published in \citet{Wachiraphan2025AJ....169..311W}. The black line is a zero-albedo blackbody (bare rock) and the cyan line is a 0.1 bar pure \ce{O2} atmosphere. Blue, orange, green and red lines are 500 $\times$ Solar metallicity atmospheres at chemical equilibrium. As a reminder, 500 $\times$ Solar is approximately the M/H we can rule out with the combined JWST+HST transmission analysis presented in Section \ref{sec:hst}. Center: the degree to which the MIRI LRS observations rule out atmospheres of various metallicity and surface pressure for a surface albedo of 0. For metallicities of 200 - 1000 $\times$ Solar, the LRS data rule out surface pressures $\gtrsim10^{-4}$ bar. Right: Similar plot as the center panel, except the surface albedo is set to 0.2. In this case, for metallicities of 100 - 1000 $\times$ Solar, the LRS data rule out surface pressures $\gtrsim10^{-1}$ to $10^{-2}$ bar.}
    \label{fig:emis}
\end{figure*}

\subsection{Combining transmission with emission data from JWST}

Here, we show how our transmission-based constraints on {\ltt}'s atmosphere compare to the emission spectrum of the planet published by \citet{Wachiraphan2025AJ....169..311W}.

As described previously (Section \ref{sec:spectrum}), we interpret our featureless transmission spectrum of {\ltt} by ruling out certain atmospheric metallicities (as a proxy for mean molecular weight) based on fully chemical equilibrium atmospheres  and opaque pressure level (as a proxy for muting by aerosols, though it could also be in reference to a surface pressure). Therefore, to compare our transmission-based results to the emission spectrum, we analyze the emission observations using the same metallicity-based parameterization for atmospheric composition. We note that previous studies of rocky planet emission spectra have taken a different approach: typically, eclipse data are compared to simulated atmospheres with ranging mixes of simple compositions (e.g. 100\% \ce{CO2}, \ce{O2} or \ce{H2O}, or some combination thereof) at various surface pressures \citep[e.g.,][]{Wachiraphan2025AJ....169..311W}. Some exoplanet analyses with both emission and transmission data have also attempted to place each other in context. For example, the JWST transmission spectrum analysis of GJ 1132~b from \citet{Bennett2025AJ....170..205B} compared 2$\sigma$-limits on atmospheric pressure and mean molecular weight obtained from the emission analysis \citep{2024ApJ...973L...8X} to their own 2$\sigma$-limits on the same parameters. They also considered if the constraints on atmospheric pressure in a pure \ce{H2O} atmosphere from transmission were physically plausible considering the nightside temperature derived from the emission results.  Such approaches are valid, and perhaps even preferable in some cases, but we do not adopt them here as we aim to maintain a consistent, directly comparable framework between our emission and transmission analyses.

To interpret the JWST emission spectrum of {\ltt} we use the climate model and equilibrium chemistry solver in the \texttt{Photochem} software package \citep{photochem2025PSJ.....6..256W} to simulate atmospheres at radiative-convective and chemical equilibrium for various metallicities and surface pressures. We use the \citet{Koll2022ApJ...924..134K} parameterization to capture the effects of day-to-night heat redistribution. We use this code to compute a grid of atmospheres for surface albedos of 0 and 0.2, metallicities between 100 and $1000 \times$ Solar, and pressure from 100 to $10^{-4}$ bar. For each calculation, we perform a $\chi^2$-test between the simulated emission spectrum and the 16-bin MIRI LRS spectrum from \citet{Wachiraphan2025AJ....169..311W}, following the grid fitting procedure described in Section \ref{sec:spect_mh}. Note this is not meant to be an exhaustive modeling analysis of the emission data.

Figure \ref{fig:emis} shows results for a surface albedo of 0 (center panel) and 0.2 (right panel). For a surface albedo of 0, the MIRI LRS data is consistent with atmospheres at $\gtrsim200~\times$ Solar metallicity and $\lesssim10^{-4}$ bar or atmospheres between $100$ and $200~\times$ Solar metallicity and $\lesssim10^{-2}$ bar. For a surface albedo of 0.2, the LRS observations permit any modeled metallicity for surface pressures $\lesssim10^{-2}$ bar. For comparison, we found in Section \ref{sec:hst} that our combined JWST+HST transmission spectrum is compatible with atmospheres $\gtrsim500~\times$ Solar metallicity for opaque pressure level $\lesssim10^{-2}$ bar. %Taken together, the transmission and emission constraints jointly favor a high–mean molecular weight atmosphere ($\gtrsim200\times$ solar metallicity) with an opaque pressure level at $\lesssim10^{-4}$ bar.

It is important to note that this constraint on surface pressure is only valid for atmospheres based on metallicity at chemical equilibrium. For example, a pure 0.1 bar \ce{O2} atmosphere (zero surface albedo) is consistent with the emission observation to within $3\sigma$ (Figure \ref{fig:emis}, left). Further modeling work must be done to consider best strategies to extract information from combined emission and transmission spectra as the community builds a set of rocky planets with both (e.g., {\ltt}, GJ 1132 b, GJ 486b).

\subsection{A look ahead at future observations}
\label{sec:future}
This paper presents the second of many JWST analyses of the LTT 1445A system (and the first transmission spectroscopy analysis). Including our own program, there are 12 independent datasets of LTT-1445A~b planned as part of  JWST General Observer programs (including Cycles 1-5). One additional NIRSpec/G395H (3-5~$\mu$m) transit will be taken under program \#7073 (PI:Justig-Yaeger), which will also obtain two additional NIRISS/SOSS transits. With respect to transmission spectra, there are 4 transits using NIRCam F322W2 (2.5-4~$\mu$m) and 4 transits using F444W (4-5~$\mu$m) via program \#7251 PI: Bennett. This program will also  test  the new NIRCam DHS mode (1-2~$\mu$m) \citep{2017PASP..129a5001S}. Both LTT 1445A b (program \# is not yet available) and c (program \# 9234) are also part of the Rocky Worlds DDT program\footnote{https://rockyworlds.stsci.edu/}, which will gather secondary eclipse emission photometry at 15~$\mu$m. 

\begin{figure*}
    \centering
    \includegraphics[width=\linewidth]{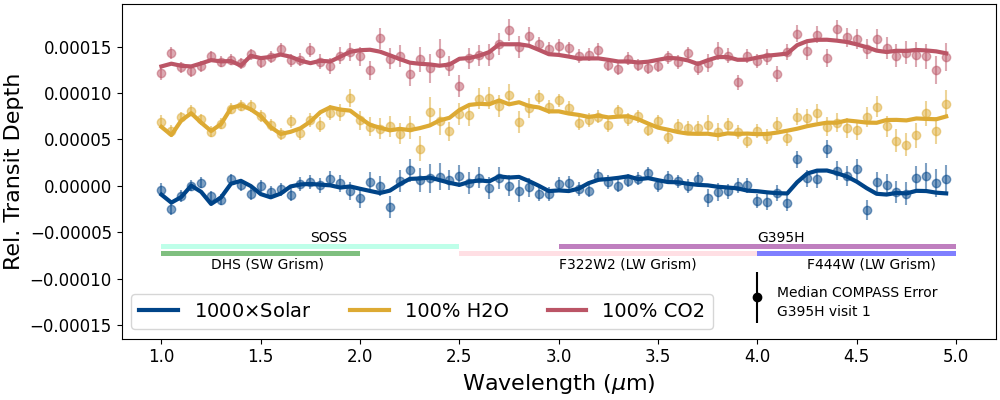}
    \caption{Simulated NIRISS, NIRCam, and NIRSpec observations of all 12 currently planned transits allocated for \ltt. In total this includes 2 visits of NIRSS/SOSS, 2 visits of NIRSpec/G395H, 4 visits of NIRCam/F322W2, and 4 visits of NIRCam/F444W from Programs \#2512, \#7073, and \#7251. The NIRCam/DHS mode, which provides shortwave coverage, will also be utilized in \#7251. All band wavelength ranges are shown for reference. We also display the median error bar from the \eureka\ reduction presented here (median from \jedi\ is only 1ppm higher).  All simulated visits are combined and binned to a resolving power of $R<100$. }
    \label{fig:future}
\end{figure*}
Focusing specifically on the transmission data, we show a fully simulated 1--5~$\mu$m spectrum of \ltt\ given the 12 transits allocated thus far. We use the JWST ETC and \texttt{PandExo} \citep{2017PASP..129f4501B} to produce simulated error bars for NIRISS/SOSS, NIRCam F322W2, NIRCam F444W, and NIRSpec/G395H using the public details of each program outlined above. Then we combine all the datasets assuming that binning reduces the error by exactly the square root of the number of combined bins, for both repeated visits and overlapping wavelength regions. We also include no hypothetical offsets between data. Note that this is an optimistic outlook on what the full dataset will look like since previous analyses have shown that stacking data does not reduce error bars as expected from photon noise alone \citep[e.g.,][]{Alderson2024AJ....167..216A}. Figure \ref{fig:future} shows a 1000 $\times$ Solar metallicity model alongside a 100\% H$_2$O and 100\% \ce{CO2} model binned to a resolving power of $R<100$. All models are effectively cloud free. The exact $\sigma$-detection level of each of these is dependent on random noise realizations. For example, the 1000 $\times$ Solar metallicity model can be distinguished from a flat line at 4.6 $\pm$ 1.1 $\sigma$, given 1000 random realizations of the data. The \ce{H2O} case will be the easiest to distinguish from a flat line, at 6.8 $\pm$ 1.0 $\sigma$. At 3.8 $\pm$ 1.1 $\sigma$, the \ce{CO2} case will be the most challenging to distinguish from a flat line. Future work should investigate in more detail how well this combined dataset could enable constrained chemical abundances, temperature structures, and opaque pressure levels.

%1000$\times$Solar 4.607996641273872 1.1127199108851618 100% H2O 6.8389296777672115 1.005857666155605 100% CO2 3.755973293124418 1.1212189101089927

%%%%%%%%%%%%%%%%%%%%%%%%%%%%%%%%%%%%%%%%%%%%%%%%%%%%%%%%%%%
\section{Summary \& Conclusions} 
\label{sec:summary}

Here we presented the first JWST transmission spectrum of the rocky planet, \ltt. It is of particular interest because: 1) it belongs to the nearest known M-dwarf system with rocky transiting planets, 
%s radius (1.31 R$_\oplus$), mass (2.87 $M_\oplus$) and equilibrium temperature (T(A$_b$=0)=424~K) make it the nearest rocky transiting M-dwarf system, 
and 2) its approximate atmospheric escape velocity, bolometric stellar flux, and host star luminosity make it amenable to retaining an atmosphere, when considering the purported ``cosmic shoreline'' \citep{BertaThompson2025arXiv250702136B}. \ltt\, has already been the subject of both ground- and space-based campaigns and has a long list of future JWST observations that will continue to try and characterize its atmosphere. 

For our JWST transmission spectrum we utilize NIRSpec/G395H, which is dispersed across the NRS1 and NRS2 detectors. In NRS1, using uniform widths of 30 pixels that produce 41 spectroscopic channels, we achieve a spectral precision of 22.3~ppm and 23.6~ppm for our two independent \eureka\ and \jedi\ reductions. We achieve 36.5 and 36.7~ppm precision, respectively, for 65 similar width spectroscopic channels in NRS2.  For comparison, \texttt{PandExo}'s median precision for the equivalent wavelength bins is 20~ppm and 32~ppm for NRS1 and NRS2, respectively. This indicates that our data closely match expectations (see Figure \ref{fig:pandexo}). 

Our reduced data show no statistically significant features besides a detector offset between NRS1 and NRS2. Specifically, we test and rule out any sloped structure or Gaussian-like features that could be indicative of molecular absorption, aerosol opacity, or even stellar activity. Both the \eureka\ and \jedi\ reductions produce offsets between NRS1 and NRS2. %The median transit depth agrees well for NRS2 wavelength ranges but deviate in NRS1. In NRS1 the \eureka\ reduction produces a higher transit depth, whereas the \jedi\ reduction produces a lower transit depth. The large offset in the \jedi\ reduction was found to depend on the width of the spectral extraction and we elected to keep the offset in our final reduction as discussed in \S\ref{sec:offsets}.

Given our featureless spectra, we conduct a series of atmospheric modeling investigations to determine what scenarios can be statistically ruled out.   Specifically, we: 1) create a grid of chemical equilibrium-based models as a function of metallicity and opaque pressure level to understand what limits can be placed on these parameters, and 2) run retrievals on simple three-gas component models (e.g., \ce{h2}/He background with either \ce{H2O}, \ce{CH4}, or \ce{CO2} to place approximate limits on the mean molecular weights that could be ruled out. Our chemical equilibrium models assume a 5-parameter parameterized pressure-temperature profile, grey optical thick clouds. Our three-gas component models make the same assumptions and also use constant-with-altitude abundances.  %We determine a ruled-out range of atmospheric metallicities given different assumptions in offset fitting method, atmospheric carbon-to-oxygen ratio, and opaque pressure level. 
Overall we find that we can rule out atmospheric metallicities greater than 350 $\times$ Solar, when the opaque pressure level is at pressures higher than 0.01~bar. This number varies slightly depending on the data reduction method used (360 $\times$ Solar for \jedi\ and 380 $\times$ Solar for \eureka). Our results are not sensitive to carbon-to-oxygen ratio but do vary strongly when considering lower opaque pressure levels. We also find sensitivity to atmospheric metallicity limits when we consider different methods for removing the effect of the NRS1/NRS2 offset. 

Next, we use the simple three-gas mixture model composed of either \ce{H2O}, \ce{CO2}, or \ce{CH4}, each with backgrounds of \ce{H2}+He, to determine approximate ranges in mean molecular weight limits. We find that we can rule out approximately $<$ 10\% abundance levels of \ce{H2O} and \ce{CO2}. For \ce{H2O} $<$ 10\% can only be ruled out for opaque pressure levels $>10^{-2}$.  For \ce{CH4}, we can rule out up to 100\% abundance levels except for cases where the opaque pressure level is below 10$^{-4}$~bar. Considering these pressure limits as well this corresponds to mean molecular weight limits of about $<$7, $<$4, and $<$16~g/mol for \ce{CO2}, \ce{H2O}, and \ce{CH4}, respectively.

Lastly, we look at how our data fits into the context of current and future data sets. When using the \eureka\ data in combination with HST data from \citet{Bennett2025AJ....169..111B}, we find that we are able to rule out metallicities greater than 520$\times$Solar, which is an increase from 380$\times$Solar ( \eureka\ reduction alone). With the addition of the JWST emission spectrum, we find that when considering pure chemical equilibrium scenarios, surface pressures greater than $10^{-2}$~bar can also be ruled out for metallicities greater than 500$\times$Solar. However, this analysis must be expanded to include broader atmospheric scenarios in order to truly understand the limits we can place on atmospheric pressure. Lastly, we also consider a simulated spectrum of all the allocated JWST transits that will be available for \ltt. In total this includes 2 visits of NIRSS/SOSS, 2 visits of NIRSpec/G395H, 4 visits of NIRCam/F322W2, and 4 visits of NIRCam/F444W from JWST Programs \#2512, \#7073, and \#7251.  We find that \ce{CO2} dominant atmospheres nearing 100\% will be the most challenging case to detect and could still result in non-detections. However, many other scenarios could be detectable including pure \ce{H2O} atmospheres and metal-rich atmospheres up to 1000 $\times$ Solar. We look forward to future observations of \ltt\ that could provide clues to the nature of this closest, M-dwarf hosted rocky world. 

The light curves, final spectra, modeled spectra, and metallicity limits can all be found in the Zenodo posting: \citet{batalha_2026_19957166}.

%%%%%%%%%%%%%%%%%%%%%%%%%%%%%%%%%%%%%%%%%%%%%%%%%%%%%%%%%%%

\begin{acknowledgments}
This work is based on observations made with the NASA/ESA/CSA James Webb Space Telescope. The JWST data were obtained from the Mikulski Archive for Space Telescopes at the Space Telescope Science Institute, which is operated by the Association of Universities for Research in Astronomy, Inc., under NASA contract NAS 5-03127 for JWST. These observations are associated with program \#2512. The specific observations analyzed can be accessed via \dataset[doi: 10.17909/p4zm-qr95]{https://doi.org/10.17909/p4zm-qr95}. Support for program \#2512 was provided by NASA through a grant from the Space Telescope Science Institute, which is operated by the Association of Universities for Research in Astronomy, Inc., under NASA contract NAS 5-03127.

This work is funded in part by the Alfred P. Sloan Foundation under grant G202114194. Support for this work was provided by NASA through grant 80NSSC19K0290 to JT and NW. SM is supported by NASA through the NASA Hubble Fellowship grant HST-HF2-51563 awarded by the Space Telescope Science Institute, which is operated by the Association of Universities for Research in Astronomy, Inc., for NASA, under contract NAS5-26555. TG and NB acknowledges support from NASA’s Interdisciplinary Consortia for Astrobiology Research (grant No. NNH19ZDA001N-ICAR) under award number 19-ICAR19\_2-0041.
\end{acknowledgments}

%%remember our collaboration agreement and make sure everyone's contributions are acknowledged 
%https://compass-jwst.github.io/collab.html
\begin{contribution}
Co-Author contributions are as follows: NB led the overall study, and conducted the theoretical interpretation. NW conducted the \eureka\ reduction and synthesized the data reduction results. TG conducted the \jedi\ reduction. NW conducted the thermal emission study of \ltt . KB provided simulated data for the NIRCam \ltt\ program. JA, MLM, JT, and JF contributed to observation planning for \ltt. All authors contributed comments and technical expertise to this study. 
\end{contribution}

\software{PICASO \citep{batalha19}, PandExo \citep{2017PASP..129f4501B}, Virga \citep{2026AJ....171...98B}, Photochem \citep{photochem2025PSJ.....6..256W}, \jedi\ \citep{2022zndo...7185855A}, \eureka \citep{2022JOSS....7.4503B}, 
SciPy \citep{2020SciPy-NMeth}, UltraNest \citep{UltraNest}, NumPy \citep{harris2020array}}

%Need to add Zenodo and MAST datasets 
%MAST:  DOI: 10.17909/p4zm-qr95

\bibliography{references}{}

@ARTICLE{Gordon2026,
       author = {{Gordon}, Tyler A. and {Batalha}, Natalie M. and {Batalha}, Natasha E. and {Aguichine}, Artyom and {Gagnebin}, Anna and {Kirk}, James and {L{\'o}pez-Morales}, Mercedes and {Meech}, Annabella and {Scarsdale}, Nicholas and {Teske}, Johanna and {Wallack}, Nicole L. and {Wogan}, Nicholas},
        title = "{JWST COMPASS: Insights into the Systematic Noise Properties of NIRSpec/G395H from a Uniform Reanalysis of Seven Transmission Spectra}",
      journal = {\aj},
         year = 2026,
        month = mar,
       volume = {171},
       number = {3},
          eid = {178},
        pages = {178},
          doi = {10.3847/1538-3881/ae3de9},
archivePrefix = {arXiv},
       eprint = {2511.18196},
 primaryClass = {astro-ph.EP},
       adsurl = {https://ui.adsabs.harvard.edu/abs/2026AJ....171..178G}
}

@ARTICLE{Moran2022,
       author = {{Moran}, Sarah E. and {H{\"o}rst}, Sarah M. and {He}, Chao and {Radke}, Michael J. and {Sebree}, Joshua A. and {Izenberg}, Noam R. and {Vuitton}, V{\'e}ronique and {Flandinet}, Laur{\`e}ne and {Orthous-Daunay}, Fran{\c{c}}ois-R{\'e}gis and {Wolters}, C{\'e}dric},
        title = "{Triton Haze Analogs: The Role of Carbon Monoxide in Haze Formation}",
      journal = {Journal of Geophysical Research (Planets)},
         year = 2022,
        month = jan,
       volume = {127},
       number = {1},
          eid = {e06984},
        pages = {e06984},
          doi = {10.1029/2021JE006984},
archivePrefix = {arXiv},
       eprint = {2112.11627},
 primaryClass = {astro-ph.EP},
       adsurl = {https://ui.adsabs.harvard.edu/abs/2022JGRE..12706984M}
}

@ARTICLE{Horst2018,
       author = {{H{\"o}rst}, Sarah M. and {He}, Chao and {Lewis}, Nikole K. and {Kempton}, Eliza M.-R. and {Marley}, Mark S. and {Morley}, Caroline V. and {Moses}, Julianne I. and {Valenti}, Jeff A. and {Vuitton}, V{\'e}ronique},
        title = "{Haze production rates in super-Earth and mini-Neptune atmosphere experiments}",
      journal = {Nature Astronomy},
         year = 2018,
        month = mar,
       volume = {2},
        pages = {303-306},
          doi = {10.1038/s41550-018-0397-0},
archivePrefix = {arXiv},
       eprint = {1801.06512},
 primaryClass = {astro-ph.EP},
       adsurl = {https://ui.adsabs.harvard.edu/abs/2018NatAs...2..303H}
}

@ARTICLE{Mang2026,
       author = {{Mang}, James and {Batalha}, Natasha E. and {Morley}, Caroline V. and {Wogan}, Nicholas F. and {Mukherjee}, Sagnick and {Visscher}, Channon and {Marley}, Mark S. and {Fortney}, Jonathan J. and {Chubb}, Katy L. and {Gao}, Peter and {Malsky}, Isaac},
        title = "{PICASO 4.0: Clouds and Photochemistry in Climate Models of Brown Dwarfs and Exoplanets}",
      journal = {arXiv e-prints},
         year = 2026,
        month = feb,
          eid = {arXiv:2602.22468},
        pages = {arXiv:2602.22468},
          doi = {10.48550/arXiv.2602.22468},
archivePrefix = {arXiv},
       eprint = {2602.22468},
 primaryClass = {astro-ph.EP},
       adsurl = {https://ui.adsabs.harvard.edu/abs/2026arXiv260222468M}
}

@ARTICLE{Meech2026,
       author = {{Meech}, Annabella and {Gao}, Peter and {Wallack}, Nicole L. and {L{\'o}pez-Morales}, Mercedes and {Oddo}, Dominic and {Teske}, Johanna and {Dragomir}, Diana and {Wolfgang}, Angie and {Wogan}, Nicholas and {Wakeford}, Hannah R. and {Moran}, Sarah E. and {Kirk}, James and {Gordon}, Tyler A. and {Gagnebin}, Anna and {Batalha}, Natasha E. and {Batalha}, Natalie M. and {Alderson}, Lili and {Alam}, Munazza K. and {Aguichine}, Artyom},
        title = "{JWST COMPASS: A NIRSpec G395H Transmission Spectrum of Radius Valley-Dweller TOI-260 b}",
      journal = {arXiv e-prints},
         year = 2026,
        month = feb,
          eid = {arXiv:2602.22329},
        pages = {arXiv:2602.22329},
          doi = {10.48550/arXiv.2602.22329},
archivePrefix = {arXiv},
       eprint = {2602.22329},
 primaryClass = {astro-ph.EP},
       adsurl = {https://ui.adsabs.harvard.edu/abs/2026arXiv260222329M}
}

@ARTICLE{Wallack2026,
       author = {{Wallack}, Nicole L. and {Gao}, Peter and {Greklek-McKeon}, Michael and {Meech}, Annabella and {Aguichine}, Artyom and {Alam}, Munazza K. and {Alderson}, Lili and {Batalha}, Natasha E. and {Batalha}, Natalie M. and {Gagnebin}, Anna and {Gordon}, Tyler A. and {Kirk}, James and {L{\'o}pez-Morales}, Mercedes and {Moran}, Sarah E. and {Iyanla Redai}, Jea and {Scarsdale}, Nicholas and {Teske}, Johanna and {Wakeford}, Hannah R. and {Wogan}, Nicholas F. and {Wolfgang}, Angie},
        title = "{JWST COMPASS: NIRSpec/G395H Transmission Observations of the Sub-Neptune HD 15337 c}",
      journal = {arXiv e-prints},
         year = 2026,
        month = feb,
          eid = {arXiv:2602.22327},
        pages = {arXiv:2602.22327},
          doi = {10.48550/arXiv.2602.22327},
archivePrefix = {arXiv},
       eprint = {2602.22327},
 primaryClass = {astro-ph.EP},
       adsurl = {https://ui.adsabs.harvard.edu/abs/2026arXiv260222327W}
}

@ARTICLE{Gressier2022T1h,
       author = {{Gressier}, A. and {Mori}, M. and {Changeat}, Q. and {Edwards}, B. and {Beaulieu}, J.~P. and {Marcq}, E. and {Charnay}, B.},
        title = "{Near-infrared transmission spectrum of TRAPPIST-1 h using Hubble WFC3 G141 observations}",
      journal = {\aap},
         year = 2022,
        month = feb,
       volume = {658},
          eid = {A133},
        pages = {A133},
          doi = {10.1051/0004-6361/202142140},
archivePrefix = {arXiv},
       eprint = {2112.05510},
 primaryClass = {astro-ph.EP},
       adsurl = {https://ui.adsabs.harvard.edu/abs/2022A&A...658A.133G}
}

@ARTICLE{Barclay2025L9859c,
       author = {{Barclay}, Thomas and {Sheppard}, Kyle B. and {Latouf}, Natasha and {Mandell}, Avi M. and {Quintana}, Elisa V. and {Gilbert}, Emily A. and {Liuzzi}, Giuliano and {Villanueva}, Geronimo L. and {Arney}, Giada and {Brande}, Jonathan and {Col{\'o}n}, Knicole D. and {Covone}, Giovanni and {Crossfield}, Ian J.~M. and {Damiano}, Mario and {Domagal-Goldman}, Shawn D. and {Fauchez}, Thomas J. and {Fiscale}, Stefano and {Gallo}, Francesco and {Hedges}, Christina L. and {Hu}, Renyu and {Kite}, Edwin S. and {Koll}, Daniel and {Kopparapu}, Ravi K. and {Kostov}, Veselin B. and {Kreidberg}, Laura and {Lopez}, Eric D. and {Mang}, James and {Morley}, Caroline V. and {Mullally}, Fergal and {Mullally}, Susan E. and {Pidhorodetska}, Daria and {Schlieder}, Joshua E. and {Vega}, Laura D. and {Youngblood}, Allison and {Zieba}, Sebastian},
        title = "{The Transmission Spectrum of the Potentially Rocky Planet L 98-59 c}",
      journal = {\aj},
         year = 2025,
        month = may,
       volume = {169},
       number = {5},
          eid = {241},
        pages = {241},
          doi = {10.3847/1538-3881/ada5f6},
archivePrefix = {arXiv},
       eprint = {2301.10866},
 primaryClass = {astro-ph.EP},
       adsurl = {https://ui.adsabs.harvard.edu/abs/2025AJ....169..241B}
}

@ARTICLE{2019AJ....157...11W,
       author = {{Wakeford}, H.~R. and {Lewis}, N.~K. and {Fowler}, J. and {Bruno}, G. and {Wilson}, T.~J. and {Moran}, S.~E. and {Valenti}, J. and {Batalha}, N.~E. and {Filippazzo}, J. and {Bourrier}, V. and {H{\"o}rst}, S.~M. and {Lederer}, S.~M. and {de Wit}, J.},
        title = "{Disentangling the Planet from the Star in Late-Type M Dwarfs: A Case Study of TRAPPIST-1g}",
      journal = {\aj},
         year = 2019,
        month = jan,
       volume = {157},
       number = {1},
          eid = {11},
        pages = {11},
          doi = {10.3847/1538-3881/aaf04d},
archivePrefix = {arXiv},
       eprint = {1811.04877},
 primaryClass = {astro-ph.EP},
       adsurl = {https://ui.adsabs.harvard.edu/abs/2019AJ....157...11W}
}

@ARTICLE{Winters2019,
       author = {{Winters}, Jennifer G. and {Medina}, Amber A. and {Irwin}, Jonathan M. and {Charbonneau}, David and {Astudillo-Defru}, Nicola and {Horch}, Elliott P. and {Eastman}, Jason D. and {Vrijmoet}, Eliot Halley and {Henry}, Todd J. and {Diamond-Lowe}, Hannah and {Winston}, Elaine and {Barclay}, Thomas and {Bonfils}, Xavier and {Ricker}, George R. and {Vanderspek}, Roland and {Latham}, David W. and {Seager}, Sara and {Winn}, Joshua N. and {Jenkins}, Jon M. and {Udry}, St{\'e}phane and {Twicken}, Joseph D. and {Teske}, Johanna K. and {Tenenbaum}, Peter and {Pepe}, Francesco and {Murgas}, Felipe and {Muirhead}, Philip S. and {Mink}, Jessica and {Lovis}, Christophe and {Levine}, Alan M. and {L{\'e}pine}, S{\'e}bastien and {Jao}, Wei-Chun and {Henze}, Christopher E. and {Fur{\'e}sz}, G{\'a}bor and {Forveille}, Thierry and {Figueira}, Pedro and {Esquerdo}, Gilbert A. and {Dressing}, Courtney D. and {D{\'\i}az}, Rodrigo F. and {Delfosse}, Xavier and {Burke}, Christopher J. and {Bouchy}, Fran{\c{c}}ois and {Berlind}, Perry and {Almenara}, Jose-Manuel},
        title = "{Three Red Suns in the Sky: A Transiting, Terrestrial Planet in a Triple M-dwarf System at 6.9 pc}",
      journal = {\aj},
         year = 2019,
        month = oct,
       volume = {158},
       number = {4},
          eid = {152},
        pages = {152},
          doi = {10.3847/1538-3881/ab364d},
archivePrefix = {arXiv},
       eprint = {1906.10147},
 primaryClass = {astro-ph.EP},
       adsurl = {https://ui.adsabs.harvard.edu/abs/2019AJ....158..152W}
}

@ARTICLE{Henry2024,
       author = {{Henry}, Todd J. and {Jao}, Wei-Chun},
        title = "{The Character of M Dwarfs}",
      journal = {\araa},
         year = 2024,
        month = sep,
       volume = {62},
       number = {1},
        pages = {593-633},
          doi = {10.1146/annurev-astro-052722-102740},
       adsurl = {https://ui.adsabs.harvard.edu/abs/2024ARA&A..62..593H}
}

@ARTICLE{2009LanB...4B..712L,
       author = {{Lodders}, K. and {Palme}, H. and {Gail}, H. -P.},
        title = "{Abundances of the Elements in the Solar System}",
      journal = {Landolt B{\"o}rnstein},
         year = 2009,
        month = jan,
       volume = {4B},
        pages = {712},
          doi = {10.1007/978-3-540-88055-4_34},
archivePrefix = {arXiv},
       eprint = {0901.1149},
 primaryClass = {astro-ph.EP},
       adsurl = {https://ui.adsabs.harvard.edu/abs/2009LanB...4B..712L}
}

@article{Foreman-Mackey2013,
doi = {10.1086/670067},
url = {https://dx.doi.org/10.1086/670067},
year = {2013},
month = {feb},
publisher = {University of Chicago Press},
volume = {125},
number = {925},
pages = {306},
author = {Daniel Foreman-Mackey and David W. Hogg and Dustin Lang and Jonathan Goodman},
title = {emcee: The MCMC Hammer},
journal = {Publications of the Astronomical Society of the Pacific}
}

@ARTICLE{Grant2024,
       author = {{Grant}, David and {Wakeford}, Hannah},
        title = "{ExoTiC-LD: thirty seconds to stellar limb-darkening coefficients}",
      journal = {The Journal of Open Source Software},
         year = 2024,
        month = aug,
       volume = {9},
       number = {100},
          eid = {6816},
        pages = {6816},
          doi = {10.21105/joss.06816},
archivePrefix = {arXiv},
       eprint = {2408.10341},
 primaryClass = {astro-ph.IM},
       adsurl = {https://ui.adsabs.harvard.edu/abs/2024JOSS....9.6816G}
}

@software{Grant2022,
       author = {{Grant}, David and {Wakeford}, Hannah R.},
        title = "{Exo-TiC/ExoTiC-LD: ExoTiC-LD v3.0.0}",
 howpublished = {Zenodo},
         year = 2022,
        month = dec,
          eid = {10.5281/zenodo.7437681},
          doi = {10.5281/zenodo.7437681},
      version = {v3.0.0},
    publisher = {Zenodo},
       adsurl = {https://ui.adsabs.harvard.edu/abs/2022zndo...7437681G}
}

@article{Kreidberg2015,
doi = {10.1086/683602},
url = {https://dx.doi.org/10.1086/683602},
year = {2015},
month = {nov},
publisher = {University of Chicago Press},
volume = {127},
number = {957},
pages = {1161},
author = {Laura Kreidberg},
title = {batman: BAsic Transit Model cAlculatioN in Python},
journal = {Publications of the Astronomical Society of the Pacific},
}

@ARTICLE{Batalha2017,
       author = {{Batalha}, Natasha E. and {Mandell}, Avi and {Pontoppidan}, Klaus and {Stevenson}, Kevin B. and {Lewis}, Nikole K. and {Kalirai}, Jason and {Earl}, Nick and {Greene}, Thomas and {Albert}, Lo{\"\i}c and {Nielsen}, Louise D.},
        title = "{PandExo: A Community Tool for Transiting Exoplanet Science with JWST \& HST}",
      journal = {\pasp},
         year = 2017,
        month = jun,
       volume = {129},
       number = {976},
        pages = {064501},
          doi = {10.1088/1538-3873/aa65b0},
archivePrefix = {arXiv},
       eprint = {1702.01820},
 primaryClass = {astro-ph.IM},
       adsurl = {https://ui.adsabs.harvard.edu/abs/2017PASP..129f4501B}
}

@article{underwood2016exomol,
  title={ExoMol molecular line lists--XIV. The rotation--vibration spectrum of hot SO2},
  author={Underwood, Daniel S and Tennyson, Jonathan and Yurchenko, Sergei N and Huang, Xinchuan and Schwenke, David W and Lee, Timothy J and Clausen, S{\o}nnik and Fateev, Alexander},
  journal={Monthly Notices of the Royal Astronomical Society},
  volume={459},
  number={4},
  pages={3890--3899},
  year={2016},
  publisher={Oxford University Press}
}

@ARTICLE{hitran2012,
	author= {Rothman, L. S. and I. E. Gordon and Y. Babikov and A. Barbe and D. Chris Benner and P. F. Bernath and M. Birk and L. Bizzocchi and V. Boudon and L. R. Brown and A. Campargue and K. Chance and E. A. Cohen and L. H. Coudert and V. M. Devi and B. J. Drouin and A. Fayt and J.-M. Flaud and R. R. Gamache and J. J. Harrison and J.-M. Hartmann and C. Hill and J. T. Hodges and D. Jacquemart and A. Jolly and J. Lamouroux and R. J. Le Roy and G. Li and D. A. Long and O. M. Lyulin and C. J. Mackie and S. T. Massie and S. Mikhailenko and H. S. P. Müller and O. V. Naumenko and A. V. Nikitin and J. Orphal and V. Perevalov and A. Perrin and E. R. Polovtseva and C. Richard and M. A. H. Smith and E. Starikova and K. Sung and S. Tashkun and J. Tennyson and G. C. Toon and V. G. Tyuterev and  G. Wagner},
	title= {The HITRAN2012 molecular spectroscopic database},
	journal= {Journal of Quantitative Spectroscopy and Radiative Transfer},
	volume= {130},
	year= {2013},
	doi= {10.1016/j.jqsrt.2013.07.002},
}

@INPROCEEDINGS{HITRAN2016,
       author = {{Gordon}, Iouli E. and {Rothman}, Laurence S. and {Tan}, Yan and {Kochanov}, Roman V. and {Hill}, Christian},
        title = "{HITRAN2016: Part I. Line lists for H\_2O, CO\_2, O\_3, N\_2O, CO, CH\_4, and O\_2}",
    booktitle = {72nd International Symposium on Molecular Spectroscopy},
         year = 2017,
        month = jun,
          eid = {TJ08},
        pages = {TJ08},
          doi = {10.15278/isms.2017.TJ08},
       adsurl = {https://ui.adsabs.harvard.edu/abs/2017isms.confETJ08G}
}

@ARTICLE{Saumon12,
   author = {{Saumon}, D. and {Marley}, M.~S. and {Abel}, M. and {Frommhold}, L. and 
	{Freedman}, R.~S.},
    title = "{New H$_{2}$ Collision-induced Absorption and NH$_{3}$ Opacity and the Spectra of the Coolest Brown Dwarfs}",
  journal = {\apj},
archivePrefix = "arXiv",
   eprint = {1202.6293},
 primaryClass = "astro-ph.SR",
     year = 2012,
    month = may,
   volume = 750,
      eid = {74},
    pages = {74},
      doi = {10.1088/0004-637X/750/1/74},
   adsurl = {http://adsabs.harvard.edu/abs/2012ApJ...750...74S}
}

@ARTICLE{Lenzuni1991h2h2,
       author = {{Lenzuni}, Paolo and {Chernoff}, David F. and {Salpeter}, Edwin E.},
        title = "{Rosseland and Planck Mean Opacities of a Zero-Metallicity Gas}",
      journal = {\apjs},
         year = 1991,
        month = jun,
       volume = {76},
        pages = {759},
          doi = {10.1086/191580},
       adsurl = {https://ui.adsabs.harvard.edu/abs/1991ApJS...76..759L}
}

@ARTICLE{Polyansky2018H2O,
       author = {{Polyansky}, Oleg L. and {Kyuberis}, Aleksandra A. and {Zobov}, Nikolai F. and {Tennyson}, Jonathan and {Yurchenko}, Sergei N. and {Lodi}, Lorenzo},
        title = "{ExoMol molecular line lists XXX: a complete high-accuracy line list for water}",
      journal = {\mnras},
         year = 2018,
        month = oct,
       volume = {480},
       number = {2},
        pages = {2597-2608},
          doi = {10.1093/mnras/sty1877},
archivePrefix = {arXiv},
       eprint = {1807.04529},
 primaryClass = {astro-ph.EP},
       adsurl = {https://ui.adsabs.harvard.edu/abs/2018MNRAS.480.2597P}
}

@article{azzam16exomol,
    author = {Azzam, Ala'a A. A. and Tennyson, Jonathan and Yurchenko, Sergei N. and Naumenko, Olga V.},
    title = "{ExoMol molecular line lists – XVI. The rotation–vibration spectrum of hot H2S}",
    journal = {Monthly Notices of the Royal Astronomical Society},
    volume = {460},
    number = {4},
    pages = {4063-4074},
    year = {2016},
    month = {05},
    issn = {0035-8711},
    doi = {10.1093/mnras/stw1133},
    url = {https://doi.org/10.1093/mnras/stw1133},
    eprint = {https://academic.oup.com/mnras/article-pdf/460/4/4063/13773124/stw1133.pdf},
}

@ARTICLE{GharibNezhad2021,
       author = {{Gharib-Nezhad}, Ehsan and {Iyer}, Aishwarya R. and {Line}, Michael R. and {Freedman}, Richard S. and {Marley}, Mark S. and {Batalha}, Natasha E.},
        title = "{EXOPLINES: Molecular Absorption Cross-section Database for Brown Dwarf and Giant Exoplanet Atmospheres}",
      journal = {\apjs},
         year = 2021,
        month = jun,
       volume = {254},
       number = {2},
          eid = {34},
        pages = {34},
          doi = {10.3847/1538-4365/abf504},
archivePrefix = {arXiv},
       eprint = {2104.00264},
 primaryClass = {astro-ph.EP},
       adsurl = {https://ui.adsabs.harvard.edu/abs/2021ApJS..254...34G}
}

@ARTICLE{li15rovibrational,
       author = {{Li}, Gang and {Gordon}, Iouli E. and {Rothman}, Laurence S. and
         {Tan}, Yan and {Hu}, Shui-Ming and {Kassi}, Samir and
         {Campargue}, Alain and {Medvedev}, Emile S.},
        title = "{Rovibrational Line Lists for Nine Isotopologues of the CO Molecule in the X $^{1}${\ensuremath{\Sigma}}$^{+}$ Ground Electronic State}",
      journal = {\apjs},
         year = 2015,
        month = jan,
       volume = {216},
       number = {1},
          eid = {15},
        pages = {15},
          doi = {10.1088/0067-0049/216/1/15},
       adsurl = {https://ui.adsabs.harvard.edu/abs/2015ApJS..216...15L}
}

@article{HUANG2014reliable,
title = "Reliable infrared line lists for 13 CO2 isotopologues up to E′=18,000cm−1 and 1500K, with line shape parameters",
journal = "Journal of Quantitative Spectroscopy and Radiative Transfer",
volume = "147",
pages = "134 - 144",
year = "2014",
issn = "0022-4073",
doi = "https://doi.org/10.1016/j.jqsrt.2014.05.015",
url = "http://www.sciencedirect.com/science/article/pii/S0022407314002246",
author = "Xinchuan {Huang} and Robert R. Gamache and Richard S. Freedman and David W. Schwenke and Timothy J. Lee"
}

@ARTICLE{2026MNRAS.546ag143C,
       author = {{Claringbold}, Alastair B. and {Fisher}, Chloe E. and {Kirk}, James and {Ahrer}, Eva-Maria and {Penzlin}, Anna B.~T. and {Thorngren}, Daniel P. and {L{\'o}pez-Morales}, Mercedes and {Wheatley}, Peter J. and {Alderson}, Lili and {Booth}, Richard A. and {Christie}, Duncan A. and {Fairman}, Charlotte and {Mayne}, Nathan J. and {McCormack}, Mason and {Meech}, Annabella and {Owen}, James E. and {Panwar}, Vatsal and {Sergeev}, Denis E. and {Valentine}, Daniel and {Wakeford}, Hannah R. and {Zamyatina}, Maria},
        title = "{BOWIE-ALIGN: Sub-solar C/O ratio and metallicity atmosphere of the misaligned hot Jupiter HAT-P-30 b}",
      journal = {\mnras},
         year = 2026,
        month = mar,
       volume = {546},
       number = {4},
          eid = {stag143},
        pages = {stag143},
          doi = {10.1093/mnras/stag143},
archivePrefix = {arXiv},
       eprint = {2601.13104},
 primaryClass = {astro-ph.EP},
       adsurl = {https://ui.adsabs.harvard.edu/abs/2026MNRAS.546ag143C}
}

@dataset{batalha_2026_19957166,
  author       = {Batalha, Natasha and
                  Wallack, Nicole and
                  Gordon, Tyler},
  title        = {Data for: "JWST COMPASS Program: The 3--5$\mu$m
                   transmission spectrum of LTT 1445 A b"
                  },
  month        = may,
  year         = 2026,
  publisher    = {Zenodo},
  doi          = {10.5281/zenodo.19957166},
  url          = {https://doi.org/10.5281/zenodo.19957166},
}

@ARTICLE{2025MNRAS.537.3027K,
       author = {{Kirk}, James and {Ahrer}, Eva-Maria and {Claringbold}, Alastair B. and {Zamyatina}, Maria and {Fisher}, Chloe and {McCormack}, Mason and {Panwar}, Vatsal and {Powell}, Diana and {Taylor}, Jake and {Thorngren}, Daniel P. and {Christie}, Duncan A. and {Esparza-Borges}, Emma and {Tsai}, Shang-Min and {Alderson}, Lili and {Booth}, Richard A. and {Fairman}, Charlotte and {L{\'o}pez-Morales}, Mercedes and {Mayne}, N.~J. and {Meech}, Annabella and {Molli{\`e}re}, Paul and {Owen}, James E. and {Penzlin}, Anna B.~T. and {Sergeev}, Denis E. and {Valentine}, Daniel and {Wakeford}, Hannah R. and {Wheatley}, Peter J.},
        title = "{BOWIE-ALIGN: JWST reveals hints of planetesimal accretion and complex sulphur chemistry in the atmosphere of the misaligned hot Jupiter WASP-15b}",
      journal = {\mnras},
         year = 2025,
        month = mar,
       volume = {537},
       number = {4},
        pages = {3027-3052},
          doi = {10.1093/mnras/staf208},
archivePrefix = {arXiv},
       eprint = {2410.08116},
 primaryClass = {astro-ph.EP},
       adsurl = {https://ui.adsabs.harvard.edu/abs/2025MNRAS.537.3027K}
}

@ARTICLE{2025arXiv250902128G,
       author = {{Gillon}, Micha{\"e}l and {Ducrot}, Elsa and {Bell}, Taylor J. and {Huang}, Ziyu and {Lincowski}, Andrew and {Lyu}, Xintong and {Maurel}, Alice and {Revol}, Alexandre and {Agol}, Eric and {Bolmont}, Emeline and {Dong}, Chuanfei and {Fauchez}, Thomas J. and {Koll}, Daniel D.~B. and {Leconte}, J{\'e}r{\'e}my and {Meadows}, Victoria S. and {Selsis}, Franck and {Turbet}, Martin and {Charnay}, Benjamin and {Delre}, Laetita and {Demory}, Brice-Olivier and {Householder}, Aaron and {Zieba}, Sebastian and {Berardo}, David and {Dyrek}, Achr{\`e}ne and {Edwards}, Billy and {de Wit}, Julien and {Greene}, Thomas P. and {Hu}, Renyu and {Iro}, Nicolas and {Kreidberg}, Laura and {Lagage}, Pierre-Olivier and {Lustig-Yaeger}, Jacob and {Iyer}, Aishwarya},
        title = "{First JWST thermal phase curves of temperate terrestrial exoplanets reveal no thick atmosphere around TRAPPIST-1 b and c}",
      journal = {arXiv e-prints},
         year = 2025,
        month = sep,
          eid = {arXiv:2509.02128},
        pages = {arXiv:2509.02128},
          doi = {10.48550/arXiv.2509.02128},
archivePrefix = {arXiv},
       eprint = {2509.02128},
 primaryClass = {astro-ph.EP},
       adsurl = {https://ui.adsabs.harvard.edu/abs/2025arXiv250902128G}
}

@ARTICLE{2025AJ....170..226C,
       author = {{Coulombe}, Louis-Philippe and {Benneke}, Bj{\"o}rn and {Krissansen-Totton}, Joshua and {L'Heureux}, Alexandrine and {Piaulet-Ghorayeb}, Caroline and {Radica}, Michael and {Roy}, Pierre-Alexis and {Ahrer}, Eva-Maria and {Cadieux}, Charles and {Miguel}, Yamila and {Schlichting}, Hilke E. and {Delgado-Mena}, Elisa and {Monaghan}, Christopher and {Adamski}, Hanna and {Raul}, Eshan and {Cloutier}, Ryan and {Komacek}, Thaddeus D. and {Taylor}, Jake and {Gapp}, Cyril and {Allart}, Romain and {Bouchy}, Fran{\c{c}}ois and {Canto Martins}, Bruno L. and {Cook}, Neil J. and {Doyon}, Ren{\'e} and {Evans-Soma}, Thomas M. and {Larue}, Pierre and {Su{\'a}rez Mascare{\~n}o}, Alejandro and {Wardenier}, Joost P.},
        title = "{Possible Evidence for the Presence of Volatiles on the Warm Super-Earth TOI-270 b}",
      journal = {\aj},
         year = 2025,
        month = oct,
       volume = {170},
       number = {4},
          eid = {226},
        pages = {226},
          doi = {10.3847/1538-3881/adfc6a},
archivePrefix = {arXiv},
       eprint = {2509.14224},
 primaryClass = {astro-ph.EP},
       adsurl = {https://ui.adsabs.harvard.edu/abs/2025AJ....170..226C}
}

@ARTICLE{2025ApJ...979L...5R,
       author = {{Radica}, Michael and {Piaulet-Ghorayeb}, Caroline and {Taylor}, Jake and {Coulombe}, Louis-Philippe and {Benneke}, Bj{\"o}rn and {Albert}, Loic and {Artigau}, {\'E}tienne and {Cowan}, Nicolas B. and {Doyon}, Ren{\'e} and {Lafreni{\`e}re}, David and {L'Heureux}, Alexandrine and {Lim}, Olivia},
        title = "{Promise and Peril: Stellar Contamination and Strict Limits on the Atmosphere Composition of TRAPPIST-1 c from JWST NIRISS Transmission Spectra}",
      journal = {\apjl},
         year = 2025,
        month = jan,
       volume = {979},
       number = {1},
          eid = {L5},
        pages = {L5},
          doi = {10.3847/2041-8213/ada381},
archivePrefix = {arXiv},
       eprint = {2409.19333},
 primaryClass = {astro-ph.EP},
       adsurl = {https://ui.adsabs.harvard.edu/abs/2025ApJ...979L...5R}
}

@ARTICLE{2025AJ....170...49L,
       author = {{Luque}, Rafael and {Coy}, Brandon Park and {Xue}, Qiao and {Feinstein}, Adina D. and {Ahrer}, Eva-Maria and {Changeat}, Quentin and {Zhang}, Michael and {Moran}, Sarah E. and {Bean}, Jacob L. and {Kite}, Edwin and {Weiner Mansfield}, Megan and {Pall{\'e}}, Enric},
        title = "{A Dark, Bare Rock for TOI-1685 b from a JWST NIRSpec G395H Phase Curve}",
      journal = {\aj},
         year = 2025,
        month = jul,
       volume = {170},
       number = {1},
          eid = {49},
        pages = {49},
          doi = {10.3847/1538-3881/addb40},
archivePrefix = {arXiv},
       eprint = {2412.03411},
 primaryClass = {astro-ph.EP},
       adsurl = {https://ui.adsabs.harvard.edu/abs/2025AJ....170...49L}
}

@Article{         harris2020array,
 title         = {Array programming with {NumPy}},
 author        = {Charles R. Harris and K. Jarrod Millman and St{\'{e}}fan J.
                 van der Walt and Ralf Gommers and Pauli Virtanen and David
                 Cournapeau and Eric Wieser and Julian Taylor and Sebastian
                 Berg and Nathaniel J. Smith and Robert Kern and Matti Picus
                 and Stephan Hoyer and Marten H. van Kerkwijk and Matthew
                 Brett and Allan Haldane and Jaime Fern{\'{a}}ndez del
                 R{\'{i}}o and Mark Wiebe and Pearu Peterson and Pierre
                 G{\'{e}}rard-Marchant and Kevin Sheppard and Tyler Reddy and
                 Warren Weckesser and Hameer Abbasi and Christoph Gohlke and
                 Travis E. Oliphant},
 year          = {2020},
 month         = sep,
 journal       = {Nature},
 volume        = {585},
 number        = {7825},
 pages         = {357--362},
 doi           = {10.1038/s41586-020-2649-2},
 publisher     = {Springer Science and Business Media {LLC}},
 url           = {https://doi.org/10.1038/s41586-020-2649-2}
}

@ARTICLE{2020SciPy-NMeth,
  author  = {Virtanen, Pauli and Gommers, Ralf and Oliphant, Travis E. and
            Haberland, Matt and Reddy, Tyler and Cournapeau, David and
            Burovski, Evgeni and Peterson, Pearu and Weckesser, Warren and
            Bright, Jonathan and {van der Walt}, St{\'e}fan J. and
            Brett, Matthew and Wilson, Joshua and Millman, K. Jarrod and
            Mayorov, Nikolay and Nelson, Andrew R. J. and Jones, Eric and
            Kern, Robert and Larson, Eric and Carey, C J and
            Polat, {\.I}lhan and Feng, Yu and Moore, Eric W. and
            {VanderPlas}, Jake and Laxalde, Denis and Perktold, Josef and
            Cimrman, Robert and Henriksen, Ian and Quintero, E. A. and
            Harris, Charles R. and Archibald, Anne M. and
            Ribeiro, Ant{\^o}nio H. and Pedregosa, Fabian and
            {van Mulbregt}, Paul and {SciPy 1.0 Contributors}},
  title   = {{{SciPy} 1.0: Fundamental Algorithms for Scientific
            Computing in Python}},
  journal = {Nature Methods},
  year    = {2020},
  volume  = {17},
  pages   = {261--272},
  adsurl  = {https://rdcu.be/b08Wh},
  doi     = {10.1038/s41592-019-0686-2},
}

@ARTICLE{2025ApJ...990L..53G,
       author = {{Glidden}, Ana and {Ranjan}, Sukrit and {Seager}, Sara and {Espinoza}, N{\'e}stor and {MacDonald}, Ryan J. and {Allen}, Natalie H. and {Ca{\~n}as}, Caleb I. and {Grant}, David and {Gressier}, Am{\'e}lie and {Stevenson}, Kevin B. and {Batalha}, Natasha E. and {Lewis}, Nikole K. and {Long}, Douglas and {Wakeford}, Hannah R. and {Alderson}, Lili and {Challener}, Ryan C. and {Col{\'o}n}, Knicole and {Huang}, Jingcheng and {Lin}, Zifan and {Louie}, Dana R. and {Mullens}, Elijah and {Sotzen}, Kristin S. and {Valenti}, Jeff A. and {Valentine}, Daniel and {Clampin}, Mark and {Mountain}, C. Matt and {Perrin}, Marshall and {van der Marel}, Roeland P.},
        title = "{JWST-TST DREAMS: Secondary Atmosphere Constraints for the Habitable Zone Planet TRAPPIST-1 e}",
      journal = {\apjl},
         year = 2025,
        month = sep,
       volume = {990},
       number = {2},
          eid = {L53},
        pages = {L53},
          doi = {10.3847/2041-8213/adf62e},
archivePrefix = {arXiv},
       eprint = {2509.05407},
 primaryClass = {astro-ph.EP},
       adsurl = {https://ui.adsabs.harvard.edu/abs/2025ApJ...990L..53G}
}

@ARTICLE{2025ApJ...990L..52E,
       author = {{Espinoza}, N{\'e}stor and {Allen}, Natalie H. and {Glidden}, Ana and {Lewis}, Nikole K. and {Seager}, Sara and {Ca{\~n}as}, Caleb I. and {Grant}, David and {Gressier}, Am{\'e}lie and {Courreges}, Shelby and {Stevenson}, Kevin B. and {Ranjan}, Sukrit and {Col{\'o}n}, Knicole and {Morris}, Brett M. and {MacDonald}, Ryan J. and {Long}, Douglas and {Wakeford}, Hannah R. and {Valenti}, Jeff A. and {Alderson}, Lili and {Batalha}, Natasha E. and {Challener}, Ryan C. and {Huang}, Jingcheng and {Lin}, Zifan and {Louie}, Dana R. and {Mullens}, Elijah and {Valentine}, Daniel and {Mountain}, C. Matt and {Pueyo}, Laurent and {Perrin}, Marshall D. and {Bellini}, Andrea and {Kammerer}, Jens and {Libralato}, Mattia and {Rebollido}, Isabel and {Rickman}, Emily and {Sohn}, Sangmo Tony and {van der Marel}, Roeland P.},
        title = "{JWST-TST DREAMS: NIRSpec/PRISM Transmission Spectroscopy of the Habitable Zone Planet TRAPPIST-1 e}",
      journal = {\apjl},
         year = 2025,
        month = sep,
       volume = {990},
       number = {2},
          eid = {L52},
        pages = {L52},
          doi = {10.3847/2041-8213/adf42e},
archivePrefix = {arXiv},
       eprint = {2509.05414},
 primaryClass = {astro-ph.EP},
       adsurl = {https://ui.adsabs.harvard.edu/abs/2025ApJ...990L..52E}
}

@ARTICLE{2026AJ....171..105A,
       author = {{Allen}, Natalie H. and {Espinoza}, N{\'e}stor and {Boehm}, V.~A. and {Ca{\~n}as}, Caleb I. and {Stevenson}, Kevin B. and {Lewis}, Nikole K. and {MacDonald}, Ryan J. and {Morris}, Brett M. and {Agol}, Eric and {Col{\'o}n}, Knicole and {Diamond-Lowe}, Hannah and {Glidden}, Ana and {Gressier}, Am{\'e}lie and {Huang}, Jingcheng and {Lin}, Zifan and {Long}, Douglas and {Louie}, Dana R. and {MacGregor}, Meredith A. and {Pueyo}, Laurent and {Rackham}, Benjamin V. and {Ranjan}, Sukrit and {Seager}, Sara and {Tovar Mendoza}, Guadalupe and {Valenti}, Jeff A. and {Valentine}, Daniel and {van der Marel}, Roeland P. and {Wakeford}, Hannah R.},
        title = "{JWST TRAPPIST-1 e/b Program: Motivation and First Observations}",
      journal = {\aj},
         year = 2026,
        month = feb,
       volume = {171},
       number = {2},
          eid = {105},
        pages = {105},
          doi = {10.3847/1538-3881/ae28cb},
archivePrefix = {arXiv},
       eprint = {2512.07695},
 primaryClass = {astro-ph.EP},
       adsurl = {https://ui.adsabs.harvard.edu/abs/2026AJ....171..105A}
}

@ARTICLE{2024ApJ...973L...8X,
       author = {{Xue}, Qiao and {Bean}, Jacob L. and {Zhang}, Michael and {Mahajan}, Alexandra and {Ih}, Jegug and {Eastman}, Jason D. and {Lunine}, Jonathan and {Mansfield}, Megan Weiner and {Coy}, Brandon Park and {Kempton}, Eliza M.-R. and {Koll}, Daniel and {Kite}, Edwin},
        title = "{JWST Thermal Emission of the Terrestrial Exoplanet GJ 1132b}",
      journal = {\apjl},
         year = 2024,
        month = sep,
       volume = {973},
       number = {1},
          eid = {L8},
        pages = {L8},
          doi = {10.3847/2041-8213/ad72e9},
archivePrefix = {arXiv},
       eprint = {2408.13340},
 primaryClass = {astro-ph.EP},
       adsurl = {https://ui.adsabs.harvard.edu/abs/2024ApJ...973L...8X}
}

@ARTICLE{Pass2023AJ....166..171P,
       author = {{Pass}, Emily K. and {Winters}, Jennifer G. and {Charbonneau}, David and {Balkanski}, Aurelia and {Lewis}, Nikole and {Lally}, Maura and {Bean}, Jacob L. and {Cloutier}, Ryan and {Eastman}, Jason D.},
        title = "{HST/WFC3 Light Curve Supports a Terrestrial Composition for the Closest Exoplanet to Transit an M Dwarf}",
      journal = {\aj},
         year = 2023,
        month = oct,
       volume = {166},
       number = {4},
          eid = {171},
        pages = {171},
          doi = {10.3847/1538-3881/acf561},
archivePrefix = {arXiv},
       eprint = {2307.02970},
 primaryClass = {astro-ph.EP},
       adsurl = {https://ui.adsabs.harvard.edu/abs/2023AJ....166..171P}
}

@ARTICLE{Kopparapu2013ApJ...767L...8K,
       author = {{Kopparapu}, Ravi Kumar},
        title = "{A Revised Estimate of the Occurrence Rate of Terrestrial Planets in the Habitable Zones around Kepler M-dwarfs}",
      journal = {\apjl},
         year = 2013,
        month = apr,
       volume = {767},
       number = {1},
          eid = {L8},
        pages = {L8},
          doi = {10.1088/2041-8205/767/1/L8},
archivePrefix = {arXiv},
       eprint = {1303.2649},
 primaryClass = {astro-ph.EP},
       adsurl = {https://ui.adsabs.harvard.edu/abs/2013ApJ...767L...8K}
}

@ARTICLE{Allard2019,
       author = {{Allard}, N.~F. and {Spiegelman}, F. and {Leininger}, T. and {Molliere}, P.},
        title = "{New study of the line profiles of sodium perturbed by H$_{2}$}",
      journal = {\aap},
         year = 2019,
        month = aug,
       volume = {628},
          eid = {A120},
        pages = {A120},
          doi = {10.1051/0004-6361/201935593},
archivePrefix = {arXiv},
       eprint = {1908.01989},
 primaryClass = {astro-ph.SR},
       adsurl = {https://ui.adsabs.harvard.edu/abs/2019A&A...628A.120A}
}

@ARTICLE{Koll2022ApJ...924..134K,
       author = {{Koll}, Daniel D.~B.},
        title = "{A Scaling for Atmospheric Heat Redistribution on Tidally Locked Rocky Planets}",
      journal = {\apj},
         year = 2022,
        month = jan,
       volume = {924},
       number = {2},
          eid = {134},
        pages = {134},
          doi = {10.3847/1538-4357/ac3b48},
archivePrefix = {arXiv},
       eprint = {1907.13145},
 primaryClass = {astro-ph.EP},
       adsurl = {https://ui.adsabs.harvard.edu/abs/2022ApJ...924..134K}
}

@article{libby2022featureless,
  title={The featureless HST/WFC3 transmission spectrum of the rocky exoplanet GJ 1132b: No evidence for a cloud-free primordial atmosphere and constraints on starspot contamination},
  author={Libby-Roberts, Jessica E and Berta-Thompson, Zachory K and Diamond-Lowe, Hannah and Gully-Santiago, Michael A and Irwin, Jonathan M and Kempton, Eliza M-R and Rackham, Benjamin V and Charbonneau, David and D{\'e}sert, Jean-Michel and Dittmann, Jason A and others},
  journal={The Astronomical Journal},
  volume={164},
  number={2},
  pages={59},
  year={2022},
  publisher={The American Astronomical Society}
}

@article{damiano2022transmission,
  title={A Transmission Spectrum of the Sub-Earth Planet L98-59 b in 1.1--1.7 $\mu$ m},
  author={Damiano, Mario and Hu, Renyu and Barclay, Thomas and Zieba, Sebastian and Kreidberg, Laura and Brande, Jonathan and Colon, Knicole D and Covone, Giovanni and Crossfield, Ian and Domagal-Goldman, Shawn D and others},
  journal={The Astronomical Journal},
  volume={164},
  number={5},
  pages={225},
  year={2022},
  publisher={The American Astronomical Society}
}

@article{garcia2022hst,
  title={HST/WFC3 transmission spectroscopy of the cold rocky planet TRAPPIST-1h},
  author={Garcia, LJ and Moran, SE and Rackham, BV and Wakeford, HR and Gillon, Micha{\"e}l and de Wit, Julien and Lewis, NK},
  journal={Astronomy \& Astrophysics},
  volume={665},
  pages={A19},
  year={2022},
  publisher={EDP Sciences}
}

@article{zhang2018near,
  title={The Near-infrared Transmission Spectra of TRAPPIST-1 Planets b, c, d, e, f, and g and Stellar Contamination in Multi-epoch Transit Spectra},
  author={Zhang, Zhanbo and Zhou, Yifan and Rackham, Benjamin V and Apai, D{\'a}niel},
  journal={The Astronomical Journal},
  volume={156},
  number={4},
  pages={178},
  year={2018},
  publisher={The American Astronomical Society}
}

@article{dewit2016combined,
  title={A combined transmission spectrum of the Earth-sized exoplanets TRAPPIST-1 b and c},
  author={de Wit, Julien and Wakeford, Hannah R and Gillon, Micha{\"e}l and Lewis, Nikole K and Valenti, Jeff A and Demory, Brice-Olivier and Burgasser, Adam J and Burdanov, Artem and Delrez, Laetitia and Jehin, Emmanu{\"e}l and others},
  journal={Nature},
  volume={537},
  number={7618},
  pages={69--72},
  year={2016},
  publisher={Nature Publishing Group UK London}
}

@ARTICLE{2026AJ....171...98B,
       author = {{Batalha}, Natasha E. and {Rooney}, Caoimhe M. and {Visscher}, Channon and {Moran}, Sarah E. and {Marley}, Mark S. and {Sengupta}, Aditya R. and {Kiefer}, Sven and {Lodge}, Matt G. and {Mang}, James and {Morley}, Caroline V. and {Mukherjee}, Sagnick and {Fortney}, Jonathan J. and {Gao}, Peter and {Lewis}, Nikole K. and {Mayorga}, L.~C. and {Pearce}, Logan A. and {Wakeford}, Hannah R.},
        title = "{Condensation Clouds in Substellar Atmospheres with Virga}",
      journal = {\aj},
         year = 2026,
        month = feb,
       volume = {171},
       number = {2},
          eid = {98},
        pages = {98},
          doi = {10.3847/1538-3881/ae29e5},
archivePrefix = {arXiv},
       eprint = {2508.15102},
 primaryClass = {astro-ph.EP},
       adsurl = {https://ui.adsabs.harvard.edu/abs/2026AJ....171...98B}
}

@ARTICLE{2017PASP..129f4501B,
       author = {{Batalha}, Natasha E. and {Mandell}, Avi and {Pontoppidan}, Klaus and {Stevenson}, Kevin B. and {Lewis}, Nikole K. and {Kalirai}, Jason and {Earl}, Nick and {Greene}, Thomas and {Albert}, Lo{\"\i}c and {Nielsen}, Louise D.},
        title = "{PandExo: A Community Tool for Transiting Exoplanet Science with JWST \& HST}",
      journal = {\pasp},
         year = 2017,
        month = jun,
       volume = {129},
       number = {976},
        pages = {064501},
          doi = {10.1088/1538-3873/aa65b0},
archivePrefix = {arXiv},
       eprint = {1702.01820},
 primaryClass = {astro-ph.IM},
       adsurl = {https://ui.adsabs.harvard.edu/abs/2017PASP..129f4501B}
}

@ARTICLE{batalha19,
       author = {{Batalha}, Natasha E. and {Marley}, Mark S. and {Lewis}, Nikole K. and {Fortney}, Jonathan J.},
        title = "{Exoplanet Reflected-light Spectroscopy with PICASO}",
      journal = {\apj},
         year = 2019,
        month = jun,
       volume = {878},
       number = {1},
          eid = {70},
        pages = {70},
          doi = {10.3847/1538-4357/ab1b51},
archivePrefix = {arXiv},
       eprint = {1904.09355},
 primaryClass = {astro-ph.EP},
       adsurl = {https://ui.adsabs.harvard.edu/abs/2019ApJ...878...70B}
}

@ARTICLE{2017PASP..129a5001S,
       author = {{Schlawin}, E. and {Rieke}, M. and {Leisenring}, J. and {Walker}, L.~M. and {Fraine}, J. and {Kelly}, D. and {Misselt}, K. and {Greene}, T. and {Line}, M. and {Lewis}, N. and {Stansberry}, J.},
        title = "{Two NIRCam Channels are Better than One: How JWST Can Do More Science with NIRCam{\textquoteright}s Short-wavelength Dispersed Hartmann Sensor}",
      journal = {\pasp},
         year = 2017,
        month = jan,
       volume = {129},
       number = {971},
        pages = {015001},
          doi = {10.1088/1538-3873/129/971/015001},
archivePrefix = {arXiv},
       eprint = {1610.02026},
 primaryClass = {astro-ph.IM},
       adsurl = {https://ui.adsabs.harvard.edu/abs/2017PASP..129a5001S}
}

@ARTICLE{2023Natur.614..664A,
       author = {{Alderson}, Lili and {Wakeford}, Hannah R. and {Alam}, Munazza K. and {Batalha}, Natasha E. and {Lothringer}, Joshua D. and {Adams Redai}, Jea and {Barat}, Saugata and {Brande}, Jonathan and {Damiano}, Mario and {Daylan}, Tansu and {Espinoza}, N{\'e}stor and {Flagg}, Laura and {Goyal}, Jayesh M. and {Grant}, David and {Hu}, Renyu and {Inglis}, Julie and {Lee}, Elspeth K.~H. and {Mikal-Evans}, Thomas and {Ramos-Rosado}, Lakeisha and {Roy}, Pierre-Alexis and {Wallack}, Nicole L. and {Batalha}, Natalie M. and {Bean}, Jacob L. and {Benneke}, Bj{\"o}rn and {Berta-Thompson}, Zachory K. and {Carter}, Aarynn L. and {Changeat}, Quentin and {Col{\'o}n}, Knicole D. and {Crossfield}, Ian J.~M. and {D{\'e}sert}, Jean-Michel and {Foreman-Mackey}, Daniel and {Gibson}, Neale P. and {Kreidberg}, Laura and {Line}, Michael R. and {L{\'o}pez-Morales}, Mercedes and {Molaverdikhani}, Karan and {Moran}, Sarah E. and {Morello}, Giuseppe and {Moses}, Julianne I. and {Mukherjee}, Sagnick and {Schlawin}, Everett and {Sing}, David K. and {Stevenson}, Kevin B. and {Taylor}, Jake and {Aggarwal}, Keshav and {Ahrer}, Eva-Maria and {Allen}, Natalie H. and {Barstow}, Joanna K. and {Bell}, Taylor J. and {Blecic}, Jasmina and {Casewell}, Sarah L. and {Chubb}, Katy L. and {Crouzet}, Nicolas and {Cubillos}, Patricio E. and {Decin}, Leen and {Feinstein}, Adina D. and {Fortney}, Joanthan J. and {Harrington}, Joseph and {Heng}, Kevin and {Iro}, Nicolas and {Kempton}, Eliza M.-R. and {Kirk}, James and {Knutson}, Heather A. and {Krick}, Jessica and {Leconte}, J{\'e}r{\'e}my and {Lendl}, Monika and {MacDonald}, Ryan J. and {Mancini}, Luigi and {Mansfield}, Megan and {May}, Erin M. and {Mayne}, Nathan J. and {Miguel}, Yamila and {Nikolov}, Nikolay K. and {Ohno}, Kazumasa and {Palle}, Enric and {Parmentier}, Vivien and {Petit dit de la Roche}, Dominique J.~M. and {Piaulet}, Caroline and {Powell}, Diana and {Rackham}, Benjamin V. and {Redfield}, Seth and {Rogers}, Laura K. and {Rustamkulov}, Zafar and {Tan}, Xianyu and {Tremblin}, P. and {Tsai}, Shang-Min and {Turner}, Jake D. and {de Val-Borro}, Miguel and {Venot}, Olivia and {Welbanks}, Luis and {Wheatley}, Peter J. and {Zhang}, Xi},
        title = "{Early Release Science of the exoplanet WASP-39b with JWST NIRSpec G395H}",
      journal = {\nat},
         year = 2023,
        month = feb,
       volume = {614},
       number = {7949},
        pages = {664-669},
          doi = {10.1038/s41586-022-05591-3},
archivePrefix = {arXiv},
       eprint = {2211.10488},
 primaryClass = {astro-ph.EP},
       adsurl = {https://ui.adsabs.harvard.edu/abs/2023Natur.614..664A}
}

@dataset{natasha_batalha_2025_14861730,
  author       = {Natasha Batalha and
                  Richard Freedman and
                  Ehsan Gharib-Nezhad and
                  Roxana Lupu},
  title        = {Resampled Opacity Database for PICASO},
  month        = feb,
  year         = 2025,
  publisher    = {Zenodo},
  doi          = {10.5281/zenodo.14861730},
  url          = {https://doi.org/10.5281/zenodo.14861730},
}

@ARTICLE{2020ApJS..247...55H,
       author = {{Hargreaves}, Robert J. and {Gordon}, Iouli E. and {Rey}, Michael and {Nikitin}, Andrei V. and {Tyuterev}, Vladimir G. and {Kochanov}, Roman V. and {Rothman}, Laurence S.},
        title = "{An Accurate, Extensive, and Practical Line List of Methane for the HITEMP Database}",
      journal = {\apjs},
         year = 2020,
        month = apr,
       volume = {247},
       number = {2},
          eid = {55},
        pages = {55},
          doi = {10.3847/1538-4365/ab7a1a},
archivePrefix = {arXiv},
       eprint = {2001.05037},
 primaryClass = {astro-ph.EP},
       adsurl = {https://ui.adsabs.harvard.edu/abs/2020ApJS..247...55H}
}

@ARTICLE{2017PASP..129d4402K,
       author = {{Kempton}, Eliza M.-R. and {Lupu}, Roxana and {Owusu-Asare}, Albert and {Slough}, Patrick and {Cale}, Bryson},
        title = "{Exo-Transmit: An Open-Source Code for Calculating Transmission Spectra for Exoplanet Atmospheres of Varied Composition}",
      journal = {\pasp},
         year = 2017,
        month = apr,
       volume = {129},
       number = {974},
        pages = {044402},
          doi = {10.1088/1538-3873/aa61ef},
archivePrefix = {arXiv},
       eprint = {1611.03871},
 primaryClass = {astro-ph.EP},
       adsurl = {https://ui.adsabs.harvard.edu/abs/2017PASP..129d4402K}
}

@ARTICLE{2018AJ....156..252M,
       author = {{Moran}, Sarah E. and {H{\"o}rst}, Sarah M. and {Batalha}, Natasha E. and {Lewis}, Nikole K. and {Wakeford}, Hannah R.},
        title = "{Limits on Clouds and Hazes for the TRAPPIST-1 Planets}",
      journal = {\aj},
         year = 2018,
        month = dec,
       volume = {156},
       number = {6},
          eid = {252},
        pages = {252},
          doi = {10.3847/1538-3881/aae83a},
archivePrefix = {arXiv},
       eprint = {1810.05210},
 primaryClass = {astro-ph.EP},
       adsurl = {https://ui.adsabs.harvard.edu/abs/2018AJ....156..252M}
}

@ARTICLE{2025NatAs.tmp..256R,
       author = {{Roy}, Pierre-Alexis and {Benneke}, Bj{\"o}rn and {Fournier-Tondreau}, Marylou and {Coulombe}, Louis-Philippe and {Piaulet-Ghorayeb}, Caroline and {Lafreni{\`e}re}, David and {Allart}, Romain and {Cowan}, Nicolas B. and {Dang}, Lisa and {Johnstone}, Doug and {Langeveld}, Adam B. and {Pelletier}, Stefan and {Radica}, Michael and {Taylor}, Jake and {Albert}, Lo{\"\i}c and {Doyon}, Ren{\'e} and {Flagg}, Laura and {Jayawardhana}, Ray and {MacDonald}, Ryan J. and {Turner}, Jake D.},
        title = "{Diversity in the haziness and chemistry of temperate sub-Neptunes}",
      journal = {Nature Astronomy},
         year = 2025,
        month = dec,
          doi = {10.1038/s41550-025-02723-3},
archivePrefix = {arXiv},
       eprint = {2512.10876},
 primaryClass = {astro-ph.EP},
       adsurl = {https://ui.adsabs.harvard.edu/abs/2025NatAs.tmp..256R}
}

@ARTICLE{photochem2025PSJ.....6..256W,
       author = {{Wogan}, Nicholas F. and {Batalha}, Natasha E. and {Zahnle}, Kevin and {Krissansen-Totton}, Joshua and {Catling}, David C. and {Wolf}, Eric T. and {Robinson}, Tyler D. and {Meadows}, Victoria and {Arney}, Giada and {Domagal-Goldman}, Shawn},
        title = "{The Open-source Photochem Code: A General Chemical and Climate Model for Interpreting (Exo)Planet Observations}",
      journal = {\psj},
         year = 2025,
        month = nov,
       volume = {6},
       number = {11},
          eid = {256},
        pages = {256},
          doi = {10.3847/PSJ/ae0e1c},
archivePrefix = {arXiv},
       eprint = {2509.25578},
 primaryClass = {astro-ph.EP},
       adsurl = {https://ui.adsabs.harvard.edu/abs/2025PSJ.....6..256W}
}

@ARTICLE{2010A&A...520A..27G,
       author = {{Guillot}, T.},
        title = "{On the radiative equilibrium of irradiated planetary atmospheres}",
      journal = {\aap},
         year = 2010,
        month = sep,
       volume = {520},
          eid = {A27},
        pages = {A27},
          doi = {10.1051/0004-6361/200913396},
archivePrefix = {arXiv},
       eprint = {1006.4702},
 primaryClass = {astro-ph.EP},
       adsurl = {https://ui.adsabs.harvard.edu/abs/2010A&A...520A..27G}
}

@ARTICLE{UltraNest,
       author = {{Buchner}, Johannes},
        title = "{UltraNest - a robust, general purpose Bayesian inference engine}",
      journal = {The Journal of Open Source Software},
         year = 2021,
        month = apr,
       volume = {6},
       number = {60},
          eid = {3001},
        pages = {3001},
          doi = {10.21105/joss.03001},
archivePrefix = {arXiv},
       eprint = {2101.09604},
 primaryClass = {stat.CO},
       adsurl = {https://ui.adsabs.harvard.edu/abs/2021JOSS....6.3001B}
}

@ARTICLE{Teske2025AJ....169..249T,
       author = {{Teske}, Johanna and {Batalha}, Natasha E. and {Wallack}, Nicole L. and {Kirk}, James and {Wogan}, Nicholas F. and {Gordon}, Tyler A. and {Alam}, Munazza K. and {Aguichine}, Artyom and {Wolfgang}, Angie and {Wakeford}, Hannah R. and {Scarsdale}, Nicholas and {Adams Redai}, Jea and {Moran}, Sarah E. and {L{\'o}pez-Morales}, Mercedes and {Meech}, Annabella and {Gao}, Peter and {Batalha}, Natalie M. and {Alderson}, Lili and {Gagnebin}, Anna},
        title = "{JWST COMPASS: NIRSpec/G395H Transmission Observations of TOI-776 c, a 2 R$_{{\ensuremath{\oplus}}}$ M Dwarf Planet}",
      journal = {\aj},
         year = 2025,
        month = may,
       volume = {169},
       number = {5},
          eid = {249},
        pages = {249},
          doi = {10.3847/1538-3881/adb975},
archivePrefix = {arXiv},
       eprint = {2502.20501},
 primaryClass = {astro-ph.EP},
       adsurl = {https://ui.adsabs.harvard.edu/abs/2025AJ....169..249T},
      shorthand = {Teske, Batalha et al. 2025}
}

@ARTICLE{Teske2025AJ_paren,
       author = {{Teske}, Johanna and {Batalha}, Natasha E. and {Wallack}, Nicole L. and {Kirk}, James and {Wogan}, Nicholas F. and {Gordon}, Tyler A. and {Alam}, Munazza K. and {Aguichine}, Artyom and {Wolfgang}, Angie and {Wakeford}, Hannah R. and {Scarsdale}, Nicholas and {Adams Redai}, Jea and {Moran}, Sarah E. and {L{\'o}pez-Morales}, Mercedes and {Meech}, Annabella and {Gao}, Peter and {Batalha}, Natalie M. and {Alderson}, Lili and {Gagnebin}, Anna},
        title = "{JWST COMPASS: NIRSpec/G395H Transmission Observations of TOI-776 c, a 2 R$_{{\ensuremath{\oplus}}}$ M Dwarf Planet}",
      journal = {\aj},
         year = 2025,
        month = may,
       volume = {169},
       number = {5},
          eid = {249},
        pages = {249},
          doi = {10.3847/1538-3881/adb975},
archivePrefix = {arXiv},
       eprint = {2502.20501},
 primaryClass = {astro-ph.EP},
       adsurl = {https://ui.adsabs.harvard.edu/abs/2025AJ....169..249T},
      shorthand = {Teske, Batalha et al. 2025}
}

@ARTICLE{Wallack2024AJ....168...77W,
       author = {{Wallack}, Nicole L. and {Batalha}, Natasha E. and {Alderson}, Lili and {Scarsdale}, Nicholas and {Adams Redai}, Jea I. and {Aguichine}, Artyom and {Alam}, Munazza K. and {Gao}, Peter and {Wolfgang}, Angie and {Batalha}, Natalie M. and {Kirk}, James and {L{\'o}pez-Morales}, Mercedes and {Moran}, Sarah E. and {Teske}, Johanna and {Wakeford}, Hannah R. and {Wogan}, Nicholas F.},
        title = "{JWST COMPASS: A NIRSpec/G395H Transmission Spectrum of the Sub-Neptune TOI-836c}",
      journal = {\aj},
         year = 2024,
        month = aug,
       volume = {168},
       number = {2},
          eid = {77},
        pages = {77},
          doi = {10.3847/1538-3881/ad3917},
archivePrefix = {arXiv},
       eprint = {2404.01264},
 primaryClass = {astro-ph.EP},
       adsurl = {https://ui.adsabs.harvard.edu/abs/2024AJ....168...77W}
}

@ARTICLE{Diamond2023AJ....165..169D,
       author = {{Diamond-Lowe}, Hannah and {Mendon{\c{c}}a}, Jo{\~a}o M. and {Charbonneau}, David and {Buchhave}, Lars A.},
        title = "{Ground-based Optical Transmission Spectroscopy of the Nearby Terrestrial Exoplanet LTT 1445Ab}",
      journal = {\aj},
         year = 2023,
        month = apr,
       volume = {165},
       number = {4},
          eid = {169},
        pages = {169},
          doi = {10.3847/1538-3881/acbf39},
archivePrefix = {arXiv},
       eprint = {2210.11809},
 primaryClass = {astro-ph.EP},
       adsurl = {https://ui.adsabs.harvard.edu/abs/2023AJ....165..169D}
}

@ARTICLE{Zahnle2017ApJ...843..122Z,
       author = {{Zahnle}, Kevin J. and {Catling}, David C.},
        title = "{The Cosmic Shoreline: The Evidence that Escape Determines which Planets Have Atmospheres, and what this May Mean for Proxima Centauri B}",
      journal = {\apj},
         year = 2017,
        month = jul,
       volume = {843},
       number = {2},
          eid = {122},
        pages = {122},
          doi = {10.3847/1538-4357/aa7846},
archivePrefix = {arXiv},
       eprint = {1702.03386},
 primaryClass = {astro-ph.EP},
       adsurl = {https://ui.adsabs.harvard.edu/abs/2017ApJ...843..122Z}
}

@ARTICLE{BertaThompson2025arXiv250702136B,
       author = {{Berta-Thompson}, Zach K. and {Wachiraphan}, Patcharapol and {Murray}, Catriona},
        title = "{The 3D Cosmic Shoreline for Nurturing Planetary Atmospheres}",
      journal = {arXiv e-prints},
         year = 2025,
        month = jul,
          eid = {arXiv:2507.02136},
        pages = {arXiv:2507.02136},
          doi = {10.48550/arXiv.2507.02136},
archivePrefix = {arXiv},
       eprint = {2507.02136},
 primaryClass = {astro-ph.EP},
       adsurl = {https://ui.adsabs.harvard.edu/abs/2025arXiv250702136B}
}

@ARTICLE{Winters2022AJ....163..168W,
       author = {{Winters}, Jennifer G. and {Cloutier}, Ryan and {Medina}, Amber A. and {Irwin}, Jonathan M. and {Charbonneau}, David and {Astudillo-Defru}, Nicola and {Bonfils}, Xavier and {Howard}, Andrew W. and {Isaacson}, Howard and {Bean}, Jacob L. and {Seifahrt}, Andreas and {Teske}, Johanna K. and {Eastman}, Jason D. and {Twicken}, Joseph D. and {Collins}, Karen A. and {Jensen}, Eric L.~N. and {Quinn}, Samuel N. and {Payne}, Matthew J. and {Kristiansen}, Martti H. and {Spencer}, Alton and {Vanderburg}, Andrew and {Zechmeister}, Mathias and {Weiss}, Lauren M. and {Wang}, Sharon Xuesong and {Wang}, Gavin and {Udry}, St{\'e}phane and {Terentev}, Ivan A. and {St{\"u}rmer}, Julian and {Stef{\'a}nsson}, Gudmundur and {Shporer}, Avi and {Shectman}, Stephen and {Sefako}, Ramotholo and {Schwengeler}, Hans Martin and {Schwarz}, Richard P. and {Scarsdale}, Nicholas and {Rubenzahl}, Ryan A. and {Roy}, Arpita and {Rosenthal}, Lee J. and {Robertson}, Paul and {Petigura}, Erik A. and {Pepe}, Francesco and {Omohundro}, Mark and {Murphy}, Joseph M. Akana and {Murgas}, Felipe and {Mo{\v{c}}nik}, Teo and {Montet}, Benjamin T. and {Mennickent}, Ronald and {Mayo}, Andrew W. and {Massey}, Bob and {Lubin}, Jack and {Lovis}, Christophe and {Lewin}, Pablo and {Kasper}, David and {Kane}, Stephen R. and {Jenkins}, Jon M. and {Huber}, Daniel and {Horne}, Keith and {Hill}, Michelle L. and {Gorrini}, Paula and {Giacalone}, Steven and {Fulton}, Benjamin and {Forveille}, Thierry and {Figueira}, Pedro and {Fetherolf}, Tara and {Dressing}, Courtney and {D{\'\i}az}, Rodrigo F. and {Delfosse}, Xavier and {Dalba}, Paul A. and {Dai}, Fei and {Cort{\'e}s}, C.~C. and {Crossfield}, Ian J.~M. and {Crane}, Jeffrey D. and {Conti}, Dennis M. and {Collins}, Kevin I. and {Chontos}, Ashley and {Butler}, R. Paul and {Brown}, Peyton and {Brady}, Madison and {Behmard}, Aida and {Beard}, Corey and {Batalha}, Natalie M. and {Almenara}, Jose-Manuel},
        title = "{A Second Planet Transiting LTT 1445A and a Determination of the Masses of Both Worlds}",
      journal = {\aj},
         year = 2022,
        month = apr,
       volume = {163},
       number = {4},
          eid = {168},
        pages = {168},
          doi = {10.3847/1538-3881/ac50a9},
archivePrefix = {arXiv},
       eprint = {2107.14737},
 primaryClass = {astro-ph.EP},
       adsurl = {https://ui.adsabs.harvard.edu/abs/2022AJ....163..168W}
}

@ARTICLE{Wachiraphan2025AJ....169..311W,
       author = {{Wachiraphan}, Patcharapol and {Berta-Thompson}, Zachory K. and {Diamond-Lowe}, Hannah and {Winters}, Jennifer G. and {Murray}, Catriona and {Zhang}, Michael and {Xue}, Qiao and {Morley}, Caroline V. and {Rosario-Franco}, Marialis and {Duvvuri}, Girish M.},
        title = "{The Thermal Emission Spectrum of the Nearby Rocky Exoplanet LTT 1445A b from JWST MIRI/LRS}",
      journal = {\aj},
         year = 2025,
        month = jun,
       volume = {169},
       number = {6},
          eid = {311},
        pages = {311},
          doi = {10.3847/1538-3881/adc990},
archivePrefix = {arXiv},
       eprint = {2410.10987},
 primaryClass = {astro-ph.EP},
       adsurl = {https://ui.adsabs.harvard.edu/abs/2025AJ....169..311W}
}

@ARTICLE{Damiano2024ApJ...968L..22D,
       author = {{Damiano}, Mario and {Bello-Arufe}, Aaron and {Yang}, Jeehyun and {Hu}, Renyu},
        title = "{LHS 1140 b Is a Potentially Habitable Water World}",
      journal = {\apjl},
         year = 2024,
        month = jun,
       volume = {968},
       number = {2},
          eid = {L22},
        pages = {L22},
          doi = {10.3847/2041-8213/ad5204},
archivePrefix = {arXiv},
       eprint = {2403.13265},
 primaryClass = {astro-ph.EP},
       adsurl = {https://ui.adsabs.harvard.edu/abs/2024ApJ...968L..22D}
}

@ARTICLE{Alderson2025AJ....169..142A,
       author = {{Alderson}, Lili and {Moran}, Sarah E. and {Wallack}, Nicole L. and {Batalha}, Natasha E. and {Wogan}, Nicholas F. and {Dattilo}, Anne and {Wakeford}, Hannah R. and {Redai}, Jea Adams and {Alam}, Munazza K. and {Aguichine}, Artyom and {Batalha}, Natalie M. and {Gagnebin}, Anna and {Gao}, Peter and {Kirk}, James and {L{\'o}pez-Morales}, Mercedes and {Meech}, Annabella and {Teske}, Johanna and {Wolfgang}, Angie},
        title = "{JWST COMPASS: NIRSpec/G395H Transmission Observations of the Super-Earth TOI-776 b}",
      journal = {\aj},
         year = 2025,
        month = mar,
       volume = {169},
       number = {3},
          eid = {142},
        pages = {142},
          doi = {10.3847/1538-3881/adad64},
archivePrefix = {arXiv},
       eprint = {2501.14596},
 primaryClass = {astro-ph.EP},
       adsurl = {https://ui.adsabs.harvard.edu/abs/2025AJ....169..142A}
}

@ARTICLE{Alderson2024AJ....167..216A,
       author = {{Alderson}, Lili and {Batalha}, Natasha E. and {Wakeford}, Hannah R. and {Wallack}, Nicole L. and {Aguichine}, Artyom and {Teske}, Johanna and {Adams Redai}, Jea and {Alam}, Munazza K. and {Batalha}, Natalie M. and {Gao}, Peter and {Kirk}, James and {L{\'o}pez-Morales}, Mercedes and {Moran}, Sarah E. and {Scarsdale}, Nicholas and {Wogan}, Nicholas F. and {Wolfgang}, Angie},
        title = "{JWST COMPASS: NIRSpec/G395H Transmission Observations of the Super-Earth TOI-836b}",
      journal = {\aj},
         year = 2024,
        month = may,
       volume = {167},
       number = {5},
          eid = {216},
        pages = {216},
          doi = {10.3847/1538-3881/ad32c9},
archivePrefix = {arXiv},
       eprint = {2404.00093},
 primaryClass = {astro-ph.EP},
       adsurl = {https://ui.adsabs.harvard.edu/abs/2024AJ....167..216A}
}

@ARTICLE{Fisher2026MNRAS.545f2187F,
       author = {{Fisher}, Chloe E. and {Hooton}, Matthew J. and {Gressier}, Am{\'e}lie and {Zgraggen}, Merlin and {Tian}, Meng and {Heng}, Kevin and {Allen}, Natalie H. and {Chatterjee}, Richard D. and {Morris}, Brett M. and {Borsato}, Nicholas W. and {Espinoza}, N{\'e}stor and {Kitzmann}, Daniel and {Meier}, Tobias G. and {Buchhave}, Lars A. and {Burgasser}, Adam J. and {Demory}, Brice-Olivier and {Fortune}, Mark and {Hoeijmakers}, H. Jens and {Luque}, Raphael and {Meier Vald{\'e}s}, Erik A. and {Mendon{\c{c}}a}, Jo{\~a}o M. and {Prinoth}, Bibiana and {Rathcke}, Alexander D. and {Taylor}, Jake},
        title = "{JWST NIRSpec finds no clear signs of an atmosphere on TOI-1685 b}",
      journal = {\mnras},
         year = 2026,
        month = feb,
       volume = {545},
       number = {4},
          eid = {staf2187},
        pages = {staf2187},
          doi = {10.1093/mnras/staf2187},
archivePrefix = {arXiv},
       eprint = {2512.15338},
 primaryClass = {astro-ph.EP},
       adsurl = {https://ui.adsabs.harvard.edu/abs/2026MNRAS.545f2187F}
}

@ARTICLE{Gressier2024ApJ...975L..10G,
       author = {{Gressier}, Am{\'e}lie and {Espinoza}, N{\'e}stor and {Allen}, Natalie H. and {Sing}, David K. and {Banerjee}, Agnibha and {Barstow}, Joanna K. and {Valenti}, Jeff A. and {Lewis}, Nikole K. and {Birkmann}, Stephan M. and {Challener}, Ryan C. and {Manjavacas}, Elena and {Alves de Oliveira}, Catarina and {Crouzet}, Nicolas and {Beck}, Tracy. L.},
        title = "{Hints of a Sulfur-rich Atmosphere around the 1.6 R $_{{\ensuremath{\oplus}}}$ Super-Earth L98-59 d from JWST NIRspec G395H Transmission Spectroscopy}",
      journal = {\apjl},
         year = 2024,
        month = nov,
       volume = {975},
       number = {1},
          eid = {L10},
        pages = {L10},
          doi = {10.3847/2041-8213/ad73d1},
archivePrefix = {arXiv},
       eprint = {2408.15855},
 primaryClass = {astro-ph.EP},
       adsurl = {https://ui.adsabs.harvard.edu/abs/2024ApJ...975L..10G}
}

@ARTICLE{Alam2025AJ....169...15A,
       author = {{Alam}, Munazza K. and {Gao}, Peter and {Adams Redai}, Jea and {Wallack}, Nicole L. and {Wogan}, Nicholas F. and {Aguichine}, Artyom and {Dattilo}, Anne and {Alderson}, Lili and {Batalha}, Natasha E. and {Batalha}, Natalie M. and {Kirk}, James and {L{\'o}pez-Morales}, Mercedes and {Meech}, Annabella and {Moran}, Sarah E. and {Teske}, Johanna and {Wakeford}, Hannah R. and {Wolfgang}, Angie},
        title = "{JWST COMPASS: The First Near- to Mid-infrared Transmission Spectrum of the Hot Super-Earth L 168-9 b}",
      journal = {\aj},
         year = 2025,
        month = jan,
       volume = {169},
       number = {1},
          eid = {15},
        pages = {15},
          doi = {10.3847/1538-3881/ad8eb5},
archivePrefix = {arXiv},
       eprint = {2411.03154},
 primaryClass = {astro-ph.EP},
       adsurl = {https://ui.adsabs.harvard.edu/abs/2025AJ....169...15A}
}

@ARTICLE{Weiner2024ApJ...975L..22W,
       author = {{Weiner Mansfield}, Megan and {Xue}, Qiao and {Zhang}, Michael and {Mahajan}, Alexandra S. and {Ih}, Jegug and {Koll}, Daniel and {Bean}, Jacob L. and {Coy}, Brandon Park and {Eastman}, Jason D. and {Kempton}, Eliza M.-R. and {Kite}, Edwin S.},
        title = "{No Thick Atmosphere on the Terrestrial Exoplanet Gl 486b}",
      journal = {\apjl},
         year = 2024,
        month = nov,
       volume = {975},
       number = {1},
          eid = {L22},
        pages = {L22},
          doi = {10.3847/2041-8213/ad8161},
archivePrefix = {arXiv},
       eprint = {2408.15123},
 primaryClass = {astro-ph.EP},
       adsurl = {https://ui.adsabs.harvard.edu/abs/2024ApJ...975L..22W}
}

@ARTICLE{Fortune2025A&A...701A..25F,
       author = {{Fortune}, Mark and {Gibson}, Neale P. and {Diamond-Lowe}, Hannah and {Mendon{\c{c}}a}, Jo{\~a}o M. and {Gressier}, Am{\'e}lie and {Kitzmann}, Daniel and {Allen}, Natalie H. and {August}, Prune C. and {Ih}, Jegug and {Meier Vald{\'e}s}, Erik and {Zgraggen}, Merlin and {Buchhave}, Lars A. and {Demory}, Brice-Olivier and {Espinoza}, N{\'e}stor and {Heng}, Kevin and {Jones}, Kathryn and {Rathcke}, Alexander D.},
        title = "{Hot Rocks Survey: III. A deep eclipse for LHS 1140c and a new Gaussian process method to account for correlated noise in individual pixels}",
      journal = {\aap},
         year = 2025,
        month = sep,
       volume = {701},
          eid = {A25},
        pages = {A25},
          doi = {10.1051/0004-6361/202554198},
archivePrefix = {arXiv},
       eprint = {2505.22186},
 primaryClass = {astro-ph.EP},
       adsurl = {https://ui.adsabs.harvard.edu/abs/2025A&A...701A..25F}
}

@ARTICLE{Adams2025AJ....170..219A,
       author = {{Adams Redai}, Jea and {Wogan}, Nicholas and {Wallack}, Nicole L. and {Alam}, Munazza K. and {Aguichine}, Artyom and {Wolfgang}, Angie and {Wakeford}, Hannah R. and {Teske}, Johanna and {Scarsdale}, Nicholas and {Moran}, Sarah E. and {L{\'o}pez-Morales}, Mercedes and {Meech}, Annabella and {Gao}, Peter and {Gagnebin}, Anna and {Batalha}, Natasha E. and {Batalha}, Natalie M. and {Alderson}, Lili},
        title = "{JWST COMPASS: A NIRSpec G395H Transmission Spectrum of the Super-Earth GJ 357 b}",
      journal = {\aj},
         year = 2025,
        month = oct,
       volume = {170},
       number = {4},
          eid = {219},
        pages = {219},
          doi = {10.3847/1538-3881/adee92},
archivePrefix = {arXiv},
       eprint = {2507.07165},
 primaryClass = {astro-ph.EP},
       adsurl = {https://ui.adsabs.harvard.edu/abs/2025AJ....170..219A}
}

@ARTICLE{Taylor2025MNRAS.540.3677T,
       author = {{Taylor}, Jake and {Radica}, Michael and {Chatterjee}, Richard D. and {Hammond}, Mark and {Meier}, Tobias and {Aigrain}, Suzanne and {MacDonald}, Ryan J. and {Albert}, Loic and {Benneke}, Bj{\"o}rn and {Coulombe}, Louis-Philippe and {Cowan}, Nicolas B. and {Dang}, Lisa and {Doyon}, Ren{\'e} and {Flagg}, Laura and {Johnstone}, Doug and {Kaltenegger}, Lisa and {Lafreni{\`e}re}, David and {Pelletier}, Stefan and {Piaulet-Ghorayeb}, Caroline and {Rowe}, Jason F. and {Roy}, Pierre-Alexis},
        title = "{JWST NIRISS transmission spectroscopy of the super-Earth GJ 357b, a favourable target for atmospheric retention}",
      journal = {\mnras},
         year = 2025,
        month = jul,
       volume = {540},
       number = {4},
        pages = {3677-3692},
          doi = {10.1093/mnras/staf894},
archivePrefix = {arXiv},
       eprint = {2505.24462},
 primaryClass = {astro-ph.EP},
       adsurl = {https://ui.adsabs.harvard.edu/abs/2025MNRAS.540.3677T}
}

@ARTICLE{Bennett2025AJ....169..111B,
       author = {{Bennett}, Katherine A. and {Sing}, David K. and {Stevenson}, Kevin B. and {Wakeford}, Hannah R. and {Rustamkulov}, Zafar and {Allen}, Natalie H. and {Lothringer}, Joshua D. and {MacDonald}, Ryan J. and {Mayne}, Nathan J. and {Fu}, Guangwei},
        title = "{An HST Transmission Spectrum of the Closest M Dwarf Transiting Rocky Planet LTT 1445Ab}",
      journal = {\aj},
         year = 2025,
        month = feb,
       volume = {169},
       number = {2},
          eid = {111},
        pages = {111},
          doi = {10.3847/1538-3881/ad9dd1},
archivePrefix = {arXiv},
       eprint = {2410.11054},
 primaryClass = {astro-ph.EP},
       adsurl = {https://ui.adsabs.harvard.edu/abs/2025AJ....169..111B}
}

@ARTICLE{Bennett2025AJ....170..205B,
       author = {{Bennett}, Katherine A. and {MacDonald}, Ryan J. and {Peacock}, Sarah and {Perez}, Junellie and {May}, E.~M. and {Moran}, Sarah E. and {Alderson}, Lili and {Lustig-Yaeger}, Jacob and {Wakeford}, Hannah R. and {Sing}, David K. and {Stevenson}, Kevin B. and {Batalha}, Natasha E. and {L{\'o}pez-Morales}, Mercedes and {Alam}, Munazza K. and {Lothringer}, Joshua D. and {Fu}, Guangwei and {Kirk}, James and {Valenti}, Jeff A. and {Mayorga}, L.~C. and {Sotzen}, Kristin S.},
        title = "{Additional JWST/NIRSpec Transits of the Rocky M Dwarf Exoplanet GJ 1132 b Reveal a Featureless Spectrum}",
      journal = {\aj},
         year = 2025,
        month = oct,
       volume = {170},
       number = {4},
          eid = {205},
        pages = {205},
          doi = {10.3847/1538-3881/adf198},
archivePrefix = {arXiv},
       eprint = {2508.10579},
 primaryClass = {astro-ph.EP},
       adsurl = {https://ui.adsabs.harvard.edu/abs/2025AJ....170..205B}
}

@ARTICLE{May2023ApJ...959L...9M,
       author = {{May}, E.~M. and {MacDonald}, Ryan J. and {Bennett}, Katherine A. and {Moran}, Sarah E. and {Wakeford}, Hannah R. and {Peacock}, Sarah and {Lustig-Yaeger}, Jacob and {Highland}, Alicia N. and {Stevenson}, Kevin B. and {Sing}, David K. and {Mayorga}, L.~C. and {Batalha}, Natasha E. and {Kirk}, James and {L{\'o}pez-Morales}, Mercedes and {Valenti}, Jeff A. and {Alam}, Munazza K. and {Alderson}, Lili and {Fu}, Guangwei and {Gonzalez-Quiles}, Junellie and {Lothringer}, Joshua D. and {Rustamkulov}, Zafar and {Sotzen}, Kristin S.},
        title = "{Double Trouble: Two Transits of the Super-Earth GJ 1132 b Observed with JWST NIRSpec G395H}",
      journal = {\apjl},
         year = 2023,
        month = dec,
       volume = {959},
       number = {1},
          eid = {L9},
        pages = {L9},
          doi = {10.3847/2041-8213/ad054f},
archivePrefix = {arXiv},
       eprint = {2310.10711},
 primaryClass = {astro-ph.EP},
       adsurl = {https://ui.adsabs.harvard.edu/abs/2023ApJ...959L...9M}
}

@ARTICLE{Greene2023Natur.618...39G,
       author = {{Greene}, Thomas P. and {Bell}, Taylor J. and {Ducrot}, Elsa and {Dyrek}, Achr{\`e}ne and {Lagage}, Pierre-Olivier and {Fortney}, Jonathan J.},
        title = "{Thermal emission from the Earth-sized exoplanet TRAPPIST-1 b using JWST}",
      journal = {\nat},
         year = 2023,
        month = jun,
       volume = {618},
       number = {7963},
        pages = {39-42},
          doi = {10.1038/s41586-023-05951-7},
archivePrefix = {arXiv},
       eprint = {2303.14849},
 primaryClass = {astro-ph.EP},
       adsurl = {https://ui.adsabs.harvard.edu/abs/2023Natur.618...39G}
}

@ARTICLE{Zhang2024ApJ...961L..44Z,
       author = {{Zhang}, Michael and {Hu}, Renyu and {Inglis}, Julie and {Dai}, Fei and {Bean}, Jacob L. and {Knutson}, Heather A. and {Lam}, Kristine and {Goffo}, Elisa and {Gandolfi}, Davide},
        title = "{GJ 367b Is a Dark, Hot, Airless Sub-Earth}",
      journal = {\apjl},
         year = 2024,
        month = feb,
       volume = {961},
       number = {2},
          eid = {L44},
        pages = {L44},
          doi = {10.3847/2041-8213/ad1a07},
archivePrefix = {arXiv},
       eprint = {2401.01400},
 primaryClass = {astro-ph.EP},
       adsurl = {https://ui.adsabs.harvard.edu/abs/2024ApJ...961L..44Z}
}

@ARTICLE{Zieba2023Natur.620..746Z,
       author = {{Zieba}, Sebastian and {Kreidberg}, Laura and {Ducrot}, Elsa and {Gillon}, Micha{\"e}l and {Morley}, Caroline and {Schaefer}, Laura and {Tamburo}, Patrick and {Koll}, Daniel D.~B. and {Lyu}, Xintong and {Acu{\~n}a}, Lorena and {Agol}, Eric and {Iyer}, Aishwarya R. and {Hu}, Renyu and {Lincowski}, Andrew P. and {Meadows}, Victoria S. and {Selsis}, Franck and {Bolmont}, Emeline and {Mandell}, Avi M. and {Suissa}, Gabrielle},
        title = "{No thick carbon dioxide atmosphere on the rocky exoplanet TRAPPIST-1 c}",
      journal = {\nat},
         year = 2023,
        month = aug,
       volume = {620},
       number = {7975},
        pages = {746-749},
          doi = {10.1038/s41586-023-06232-z},
archivePrefix = {arXiv},
       eprint = {2306.10150},
 primaryClass = {astro-ph.EP},
       adsurl = {https://ui.adsabs.harvard.edu/abs/2023Natur.620..746Z}
}

@ARTICLE{Lustig2023NatAs...7.1317L,
       author = {{Lustig-Yaeger}, Jacob and {Fu}, Guangwei and {May}, E.~M. and {Ceballos}, Kevin N. Ortiz and {Moran}, Sarah E. and {Peacock}, Sarah and {Stevenson}, Kevin B. and {Kirk}, James and {L{\'o}pez-Morales}, Mercedes and {MacDonald}, Ryan J. and {Mayorga}, L.~C. and {Sing}, David K. and {Sotzen}, Kristin S. and {Valenti}, Jeff A. and {Redai}, J{\'e}a I. Adams and {Alam}, Munazza K. and {Batalha}, Natasha E. and {Bennett}, Katherine A. and {Gonzalez-Quiles}, Junellie and {Kruse}, Ethan and {Lothringer}, Joshua D. and {Rustamkulov}, Zafar and {Wakeford}, Hannah R.},
        title = "{A JWST transmission spectrum of the nearby Earth-sized exoplanet LHS 475 b}",
      journal = {Nature Astronomy},
         year = 2023,
        month = nov,
       volume = {7},
        pages = {1317-1328},
          doi = {10.1038/s41550-023-02064-z},
archivePrefix = {arXiv},
       eprint = {2301.04191},
 primaryClass = {astro-ph.EP},
       adsurl = {https://ui.adsabs.harvard.edu/abs/2023NatAs...7.1317L}
}

@ARTICLE{Kirk2024AJ....167...90K,
       author = {{Kirk}, James and {Stevenson}, Kevin B. and {Fu}, Guangwei and {Lustig-Yaeger}, Jacob and {Moran}, Sarah E. and {Peacock}, Sarah and {Alam}, Munazza K. and {Batalha}, Natasha E. and {Bennett}, Katherine A. and {Gonzalez-Quiles}, Junellie and {L{\'o}pez-Morales}, Mercedes and {Lothringer}, Joshua D. and {MacDonald}, Ryan J. and {May}, E.~M. and {Mayorga}, L.~C. and {Rustamkulov}, Zafar and {Sing}, David K. and {Sotzen}, Kristin S. and {Valenti}, Jeff A. and {Wakeford}, Hannah R.},
        title = "{JWST/NIRCam Transmission Spectroscopy of the Nearby Sub-Earth GJ 341b}",
      journal = {\aj},
         year = 2024,
        month = mar,
       volume = {167},
       number = {3},
          eid = {90},
        pages = {90},
          doi = {10.3847/1538-3881/ad19df},
archivePrefix = {arXiv},
       eprint = {2401.06043},
 primaryClass = {astro-ph.EP},
       adsurl = {https://ui.adsabs.harvard.edu/abs/2024AJ....167...90K}
}

@ARTICLE{BelloArufe2025ApJ...980L..26B,
       author = {{Bello-Arufe}, Aaron and {Damiano}, Mario and {Bennett}, Katherine A. and {Hu}, Renyu and {Welbanks}, Luis and {MacDonald}, Ryan J. and {Seligman}, Darryl Z. and {Sing}, David K. and {Tokadjian}, Armen and {Oza}, Apurva V. and {Yang}, Jeehyun},
        title = "{Evidence for a Volcanic Atmosphere on the Sub-Earth L 98-59 b}",
      journal = {\apjl},
         year = 2025,
        month = feb,
       volume = {980},
       number = {2},
          eid = {L26},
        pages = {L26},
          doi = {10.3847/2041-8213/adaf22},
archivePrefix = {arXiv},
       eprint = {2501.18680},
 primaryClass = {astro-ph.EP},
       adsurl = {https://ui.adsabs.harvard.edu/abs/2025ApJ...980L..26B}
}

@ARTICLE{Scarsdale2024AJ....168..276S,
       author = {{Scarsdale}, Nicholas and {Wogan}, Nicholas and {Wakeford}, Hannah R. and {Wallack}, Nicole L. and {Batalha}, Natasha E. and {Alderson}, Lili and {Aguichine}, Artyom and {Wolfgang}, Angie and {Teske}, Johanna and {Moran}, Sarah E. and {L{\'o}pez-Morales}, Mercedes and {Kirk}, James and {Gordon}, Tyler and {Gao}, Peter and {Batalha}, Natalie M. and {Alam}, Munazza K. and {Adams Redai}, Jea},
        title = "{JWST COMPASS: The 3{\textendash}5 {\ensuremath{\mu}}m Transmission Spectrum of the Super-Earth L 98-59 c}",
      journal = {\aj},
         year = 2024,
        month = dec,
       volume = {168},
       number = {6},
          eid = {276},
        pages = {276},
          doi = {10.3847/1538-3881/ad73cf},
archivePrefix = {arXiv},
       eprint = {2409.07552},
 primaryClass = {astro-ph.EP},
       adsurl = {https://ui.adsabs.harvard.edu/abs/2024AJ....168..276S}
}

@ARTICLE{2023ApJ...948L..11M,
       author = {{Moran}, Sarah E. and {Stevenson}, Kevin B. and {Sing}, David K. and {MacDonald}, Ryan J. and {Kirk}, James and {Lustig-Yaeger}, Jacob and {Peacock}, Sarah and {Mayorga}, L.~C. and {Bennett}, Katherine A. and {L{\'o}pez-Morales}, Mercedes and {May}, E.~M. and {Rustamkulov}, Zafar and {Valenti}, Jeff A. and {Adams Redai}, J{\'e}a I. and {Alam}, Munazza K. and {Batalha}, Natasha E. and {Fu}, Guangwei and {Gonzalez-Quiles}, Junellie and {Highland}, Alicia N. and {Kruse}, Ethan and {Lothringer}, Joshua D. and {Ortiz Ceballos}, Kevin N. and {Sotzen}, Kristin S. and {Wakeford}, Hannah R.},
        title = "{High Tide or Riptide on the Cosmic Shoreline? A Water-rich Atmosphere or Stellar Contamination for the Warm Super-Earth GJ 486b from JWST Observations}",
      journal = {\apjl},
         year = 2023,
        month = may,
       volume = {948},
       number = {1},
          eid = {L11},
        pages = {L11},
          doi = {10.3847/2041-8213/accb9c},
archivePrefix = {arXiv},
       eprint = {2305.00868},
 primaryClass = {astro-ph.EP},
       adsurl = {https://ui.adsabs.harvard.edu/abs/2023ApJ...948L..11M}
}

@ARTICLE{2022JOSS....7.4503B,
       author = {{Bell}, Taylor and {Ahrer}, Eva-Maria and {Brande}, Jonathan and {Carter}, Aarynn and {Feinstein}, Adina and {Caloca}, Giannina and {Mansfield}, Megan and {Zieba}, Sebastian and {Piaulet}, Caroline and {Benneke}, Bj{\"o}rn and {Filippazzo}, Joseph and {May}, Erin and {Roy}, Pierre-Alexis and {Kreidberg}, Laura and {Stevenson}, Kevin},
        title = "{Eureka!: An End-to-End Pipeline for JWST Time-Series Observations}",
      journal = {The Journal of Open Source Software},
         year = 2022,
        month = nov,
       volume = {7},
       number = {79},
          eid = {4503},
        pages = {4503},
          doi = {10.21105/joss.04503},
archivePrefix = {arXiv},
       eprint = {2207.03585},
 primaryClass = {astro-ph.IM},
       adsurl = {https://ui.adsabs.harvard.edu/abs/2022JOSS....7.4503B}
}

@MISC{2022zndo...7185855A,
       author = {{Alderson}, Lili and {Grant}, David and {Wakeford}, Hannah},
        title = "{Exo-TiC/ExoTiC-JEDI: v0.1-beta-release}",
 howpublished = {Zenodo},
         year = 2022,
        month = oct,
          eid = {10.5281/zenodo.7185855},
          doi = {10.5281/zenodo.7185855},
      version = {v0.1},
    publisher = {Zenodo},
       adsurl = {https://ui.adsabs.harvard.edu/abs/2022zndo...7185855A}
}

@MISC{2021jwst.prop.2512B,
       author = {{Batalha}, Natasha and {Teske}, Johanna and {Alam}, Munazza and {Alderson}, Lili and {Batalha}, Natalie and {Gao}, Peter and {Lopez-Morales}, Mercedes and {Marley}, Mark S. and {Shahar}, Anat and {Wakeford}, Hannah and {Wolfgang}, Angie},
        title = "{Seeing the Forest and the Trees: Unveiling Small Planet Atmospheres with a Population-Level Framework}",
 howpublished = {JWST Proposal. Cycle 1, ID. \#2512},
         year = 2021,
        month = mar,
        pages = {2512},
       adsurl = {https://ui.adsabs.harvard.edu/abs/2021jwst.prop.2512B}
}

@ARTICLE{2019ApJ...878...70B,
       author = {{Batalha}, Natasha E. and {Marley}, Mark S. and {Lewis}, Nikole K. and {Fortney}, Jonathan J.},
        title = "{Exoplanet Reflected-light Spectroscopy with PICASO}",
      journal = {\apj},
         year = 2019,
        month = jun,
       volume = {878},
       number = {1},
          eid = {70},
        pages = {70},
          doi = {10.3847/1538-4357/ab1b51},
archivePrefix = {arXiv},
       eprint = {1904.09355},
 primaryClass = {astro-ph.EP},
       adsurl = {https://ui.adsabs.harvard.edu/abs/2019ApJ...878...70B}
}

@ARTICLE{2010ApJ...716.1060V,
       author = {{Visscher}, Channon and {Lodders}, Katharina and {Fegley}, Bruce, Jr.},
        title = "{Atmospheric Chemistry in Giant Planets, Brown Dwarfs, and Low-mass Dwarf Stars. III. Iron, Magnesium, and Silicon}",
      journal = {\apj},
         year = 2010,
        month = jun,
       volume = {716},
       number = {2},
        pages = {1060-1075},
          doi = {10.1088/0004-637X/716/2/1060},
archivePrefix = {arXiv},
       eprint = {1001.3639},
 primaryClass = {astro-ph.EP},
       adsurl = {https://ui.adsabs.harvard.edu/abs/2010ApJ...716.1060V}
}

@ARTICLE{Dai2022,
       author = {{Dai}, Longkang and {Zhang}, Xi and {Shao}, Wencheng D. and {Bierson}, Carver J. and {Cui}, Jun},
        title = "{A Simple Condensation Model for the H$_{2}$SO$_{4}$-H$_{2}$O Gas-Cloud System on Venus}",
      journal = {Journal of Geophysical Research (Planets)},
         year = 2022,
        month = mar,
       volume = {127},
       number = {3},
          eid = {e07060},
        pages = {e07060},
          doi = {10.1029/2021JE007060},
archivePrefix = {arXiv},
       eprint = {2207.10238},
 primaryClass = {astro-ph.EP},
       adsurl = {https://ui.adsabs.harvard.edu/abs/2022JGRE..12707060D}
}

@ARTICLE{Gao2017,
       author = {{Gao}, Peter and {Marley}, Mark S. and {Zahnle}, Kevin and {Robinson}, Tyler D. and {Lewis}, Nikole K.},
        title = "{Sulfur Hazes in Giant Exoplanet Atmospheres: Impacts on Reflected Light Spectra}",
      journal = {\aj},
         year = 2017,
        month = mar,
       volume = {153},
       number = {3},
          eid = {139},
        pages = {139},
          doi = {10.3847/1538-3881/aa5fab},
archivePrefix = {arXiv},
       eprint = {1701.00318},
 primaryClass = {astro-ph.EP},
       adsurl = {https://ui.adsabs.harvard.edu/abs/2017AJ....153..139G}
}
\bibliographystyle{aasjournal}

\end{document}